\documentclass[twocolumn,footinbib,superscriptaddress,floatfix,longbibliography]{revtex4-2}

\usepackage{times}
\usepackage{float}
\usepackage{multirow}
\usepackage[dvipsnames]{xcolor}
\usepackage{amsmath}
\usepackage{amsthm, mathrsfs}
\usepackage{amssymb}
\usepackage{amsbsy}
\usepackage{enumitem}
\usepackage{wasysym}
\usepackage[english]{babel}
\usepackage[T1]{fontenc}
\usepackage[utf8]{inputenc} 
\usepackage{graphicx}
\usepackage[colorlinks,bookmarks=false,citecolor=blue,linkcolor=red,urlcolor=blue]{hyperref}
\usepackage{pstricks}
\usepackage{tabularx,hhline}	
\usepackage[caption=false]{subfig}		
\newcolumntype{P}[1]{>{\centering\arraybackslash}p{#1}}

\usepackage{pdfpages}

\makeatletter
\AtBeginDocument{\let\LS@rot\@undefined}
\makeatother

\newcommand{\mi}[1]{\textcolor{black}{#1}}

\definecolor{indigo}{rgb}{0.5,0,0.7}
\newcommand{\yy}[1]{\textcolor{black}{#1}}

\newcommand{\rp}{\textcolor{black}}

\newcommand{\eqsa}[1]{\begin{eqnarray} #1 \end{eqnarray}}

\usepackage{pifont}
\DeclareFontFamily{U}{cryst}{}
\DeclareFontShape{U}{cryst}{m}{n}{<-> cryst}{}

\makeatletter
\newcommand*{\rom}[1]{\expandafter\@slowromancap\romannumeral #1@}
\makeatother

\begin{document}
\graphicspath{{./Images/}}

\title{Ground state of the $S$=1/2 pyrochlore Heisenberg antiferromagnet: \\
A quantum spin liquid emergent from dimensional reduction}

\author{Rico Pohle}
\affiliation{Waseda Research Institute for Science and Engineering,
Waseda University, 3-4-1 Okubo, Shinjuku-ku, Tokyo, 169-8555, Japan}
\affiliation{Department of Applied Physics, University of Tokyo, Hongo,  Bunkyo-ku,
Tokyo, 113-8656, Japan}
\affiliation{Graduate School of Science and Technology, Keio University, Yokohama 223-8522, Japan}
\author{Youhei Yamaji}
\affiliation{Center for Green Research on Energy and Environmental Materials (GREEN), 
and Research Center for Materials Nanoarchitectonics (MANA), National 
Institute for Materials Science (NIMS), Namiki, Tsukuba-shi, Ibaraki 305-0044, Japan}

\author{Masatoshi Imada}
\affiliation{Waseda Research Institute for Science and Engineering,
Waseda University, 3-4-1 Okubo, Shinjuku-ku, Tokyo, 169-8555, Japan}
\affiliation{Toyota Physical and Chemical Research Institute, Yokomichi, Nagakute, 
Aichi, 480-1192, Japan}
\affiliation{Sophia University, Kioicho, Chiyoda-ku, Tokyo, Japan}

\date{\today} 

%
%
\begin{abstract}
%

The quantum antiferromagnet on the pyrochlore lattice offers an archetypal frustrated system,
which potentially realizes a quantum spin liquid characterized by the absence of standard 
spontaneous symmetry breaking even at zero temperature, unusually as an isotropic 3D system.
Despite tremendous progress in the literature, however, 
the nature of the ground state of the fully quantum-mechanical
spin Hamiltonian on the pyrochlore lattice still remains elusive.
Here, we show that an unconventional type of quantum spin liquid is born out from the 
pyrochlore system after the self-organized dimensional reduction leading to confined states in 2D layers. 
This conclusion is obtained from state-of-the-art variational Monte Carlo (VMC) simulations 
at zero temperature. 
Quantum spin liquids triggered by the emergent dimensional reduction is an unexplored 
route of the spin-liquid formation.
The dimensional reduction from 3D to 2D is a consequence of a conventional spontaneous 
symmetry breaking, while the resultant decoupling of layers enables the emergence of a 
2D quantum spin liquid that is adiabatically disconnected from trivial product states and exhibits strong quantum entanglement.
The stabilized quantum spin liquid exhibits 
an algebraic decay of correlations and vanishing excitation 
gap in the thermodynamic limit. 
The wave-function structure supports the fractionalization of the spin into spinons.
This spin-liquid ground state persists in the presence of spin-orbit interactions, 
which expands the possibilities of realizing
quantum spin liquids in real pyrochlore-structured materials.

\end{abstract}

\maketitle

%
%
\section{Introduction}									
\label{sec:Intro}
%

Quantum entanglement represents the holistic nature of 
an ensemble of particles at the heart of quantum mechanics.
The entangled state cannot be described as
an assembly of mutually interacting but essentially individual and isolated components,
which is called the product state, as the classical state also belongs to.
Such entangled states offer the possibility towards conceptually unexplored and innovative 
types of information transport~\cite{Bennett1993, Bouwmeester1997, Duan2001, Barrett2004, Zhang2006}, 
quantum computation~\cite{QuantumComp2010}, and cryptography~\cite{Gisin2002}, 
beyond the accessibility from any classical principles of physics.
However, in nature, most macroscopic systems tend to favor less entangled 
states that can essentially be described by the product state. 
This state is decomposed into local and microscopic subsystems despite 
the existence of mutual interactions.
The reduction to product states is usually a consequence of either classical 
dissipation generated by couplings to random and extensive degrees 
of freedom~\cite{CaldeiraLeggett,PhysRevA.65.012101,
PhysRevA.69.052105} or by spontaneous symmetry breaking, as observed 
in magnetically ordered states of conventional magnets.

Originally inspired by quantum resonance in molecules, such as resonating
valence bond states of benzene~\cite{Pauling1933}, a quantum spin liquid (QSL)
was proposed in a Heisenberg model on the triangular 
lattice~\cite{Anderson1973,fazekas1974ground} to shed light 
on such a quantum entangled state persistent in bulk magnets 
against the general trend~\cite{PhysRevLett.99.170502,Vedral2008}.
QSLs are indeed characterized as states of matter
that exhibit long-ranged entanglement of spins far apart 
without falling into the product states or conventional magnetically ordered states 
described by the framework of the conventional Landau paradigm of magnetism~\cite{Landau1, Landau2}.  
In fact, QSLs are believed to exhibit
exotic properties such as topological entanglement~\cite{Kitaev2006b,Isakov2011}, 
emergent gauge fields, and fractionalized excitations~\cite{Lee2008, Balents2010, Savary2017, Zhou2017},
providing us with a platform to discover new physics of quantum matter, including candidates of future 
quantum information devices.

Despite such proposals for intriguing features, QSLs still remain elusive in terms 
of both experimental characterization and theoretical understanding since the 
first proposal more than 50 years ago~\cite{Anderson1973}.
A key idea to induce QSLs is provided by the concept of geometrical frustration: When 
magnetic interactions compete with each other hindering simple magnetic order due to the lattice geometry 
or orbital configurations of electrons, such a system is called a frustrated magnet. 
Frustration works to suppress spontaneous symmetry breaking down to 
temperatures much lower than the energy scale of spin-spin 
interactions~\cite{Intro.Frustr.Mag2011, Diep2013, Broholm2020}.

Some of not comprehensive but typical candidate materials of QSL 
are found in quasi-two-dimensional systems:
Molecular solids with anisotropic triangular lattice structures 
provide us with such examples~\cite{Kanoda2011,Zhou2017}, and their 
theoretical aspects suggesting the fractionalization of spins have been 
elucidated in {\it ab initio} studies~\cite{Ido2022}. 
Other examples include the Herbertsmithite compound ZnCu$_3$(OH)$_6$Cl$_2$, 
proposed as a good experimental realization of the spin-1/2 
Heisenberg model on the Kagome lattice~\cite{Shores2005}.
The ground state of this model was theoretically 
proposed to be a gapped $\mathbb{Z}_2$ QSL through extensive density matrix 
renormalization group (DMRG)~\cite{Yan2011} and exact diagonalization 
studies~\cite{Lauchli2019} (see for a review Ref.~[\onlinecite{Norman2016}]). 
Ca$_{10}$Cr$_7$O$_{28}$ was modeled by a distorted kagome-bilayer lattice showing 
dynamical properties consistent with a QSL at low temperatures~\cite{Balz2016}. 
It was claimed that the experimental indications are accounted for 
by a gapless $\mathbb{Z}_2$ QSL represented by spinon pairing with $f$-wave 
symmetry~\cite{Sonnenschein2019}, where the pairing exhibits characteristics 
of a heavy fermion superconductor~\cite{Pohle2021}. 
Another intriguing class includes Kitaev honeycomb materials, as 
reviewed in Refs.~[\onlinecite{Takagi2019}] and [\onlinecite{Trebst2022}]. 
In these systems, the frustration arises from spin anisotropic 
interactions~\cite{Kitaev2006a}, leading to the emergence of fractionalized 
Majorana fermions, with ongoing efforts to identify them in 
experiments~\cite{Motome2020}.

In contrast to these 2D candidates, 3D systems are believed to have much stronger tendency to 
some type of symmetry breaking, which hampers the QSL ground state and 
leaves the realization of QSLs in 3D challenging.
Among them, the Heisenberg model on the pyrochlore lattice offers a widely 
studied theoretical playground~\cite{Harris1997, Moessner1998a, Moessner1998b}
and has been proposed to mimic the essence of many materials in 
nature~\cite{Subramanian1983, Castelnovo2012, SpinIce2021, Gingras2014, Yan2017, Rau2019}, 
making it a good platform for the search of QSLs in 3D.
Experimentally, rare-earth pyrochlore oxides have been
proposed as potential realizations of classical spin liquids, modeled by the
XXZ Heisenberg model on the pyrochlore lattice. 
Illustrative examples like Ho$_2$Ti$_2$O$_7$~\cite{Fennell2009} and 
Dy$_2$Ti$_2$O$_7$~\cite{Morris2009}, exhibit unique ground states constrained by
the ``ice rules''~\cite{Bernal1933, Pauling1935, SpinIce2021}, giving them the
name ``spin ice''. 
Departing from classical spin ice, Pr$_2$Hf$_2$O$_7$~\cite{Sibille2018} 
has been proposed to induce a QSL phase driven by enhanced quantum fluctuations.
However, its underlying physics can be understood as a perturbative
extension of the classical spin ice.
In contrast, a wide range of materials with lighter magnetic
ions such as $5d$ iridium pyrochlores $R_2$Ir$_2$O$_7$ ($R$ is a rare earth element)
remain to be understood, since they require full quantum mechanical
treatments.

Despite 30 years of extensive research, the ground state 
of the full quantum $S$=1/2 Heisenberg antiferromagnet on the pyrochlore lattice remains 
controversial, with both 
positive~\cite{Canals1998, Canals2000, Kim2008,  Burnell2009, Mueller2019,
Chandra2018, Iqbal2019} and 
negative ~\cite{Harris1991, Sobral1997, Isoda1998, Tsunetsugu2001a, Tsunetsugu2001b, Moessner2006,  
Tchernyshyov2006}
indications of a QSL ground state.
Recent advancements in numerical techniques have made a significant step forward. 
Studies utilizing SU(2) density matrix renormalization group (DMRG)~\cite{Hagymasi2021}, 
the variational Monte Carlo method~\cite{Astrakhantsev2021}, 
and numerical linked cluster expansion~\cite{Schafer2023} have reported the 
presence of spontaneous symmetry breaking, and suggested the absence 
of a QSL ground state. 
However, to reach a convincing and conclusive understanding, theoretical 
analyses are required to satisfy high accuracy and perform proper finite-size scaling to 
make reliable estimates in the thermodynamic limit.
Without such an analysis, the nature of the ground state remains an 
open question.


In this paper, after careful analysis of size dependence on accurate simulation results, we 
clarify that the ground state of the
$S$=1/2 pyrochlore Heisenberg model is a QSL. Furthermore, the QSL persists under perturbations 
such as Dzyaloshinskii-Moriya (DM) interactions, thereby expanding the possibilities for realizing
QSL in real materials.

To solve the full quantum many-body problem, we employ a state-of-the-art VMC method by 
incorporating various symmetry projections. We utilize the open source software 
mVMC ~\cite{Tahara2008, Misawa2019, mVMC_GitHub}.
This quantum solver has already demonstrated its accuracy and has identified 
the existence of QSL ground states in 2D strongly correlated 
systems~\cite{Kaneko2014, Morita2015, Nomura2021, Ido2022}.
This standard method and its accuracy in the current context are summarized in 
Appendices~\ref{app:Method} and \ref{app:pyro3D.energy}, respectively and 
are further supplemented in Supplemental Materials (SM)~\cite{SM}.

\begin{figure}[t]
	\centering
	\includegraphics[width=0.47\textwidth]{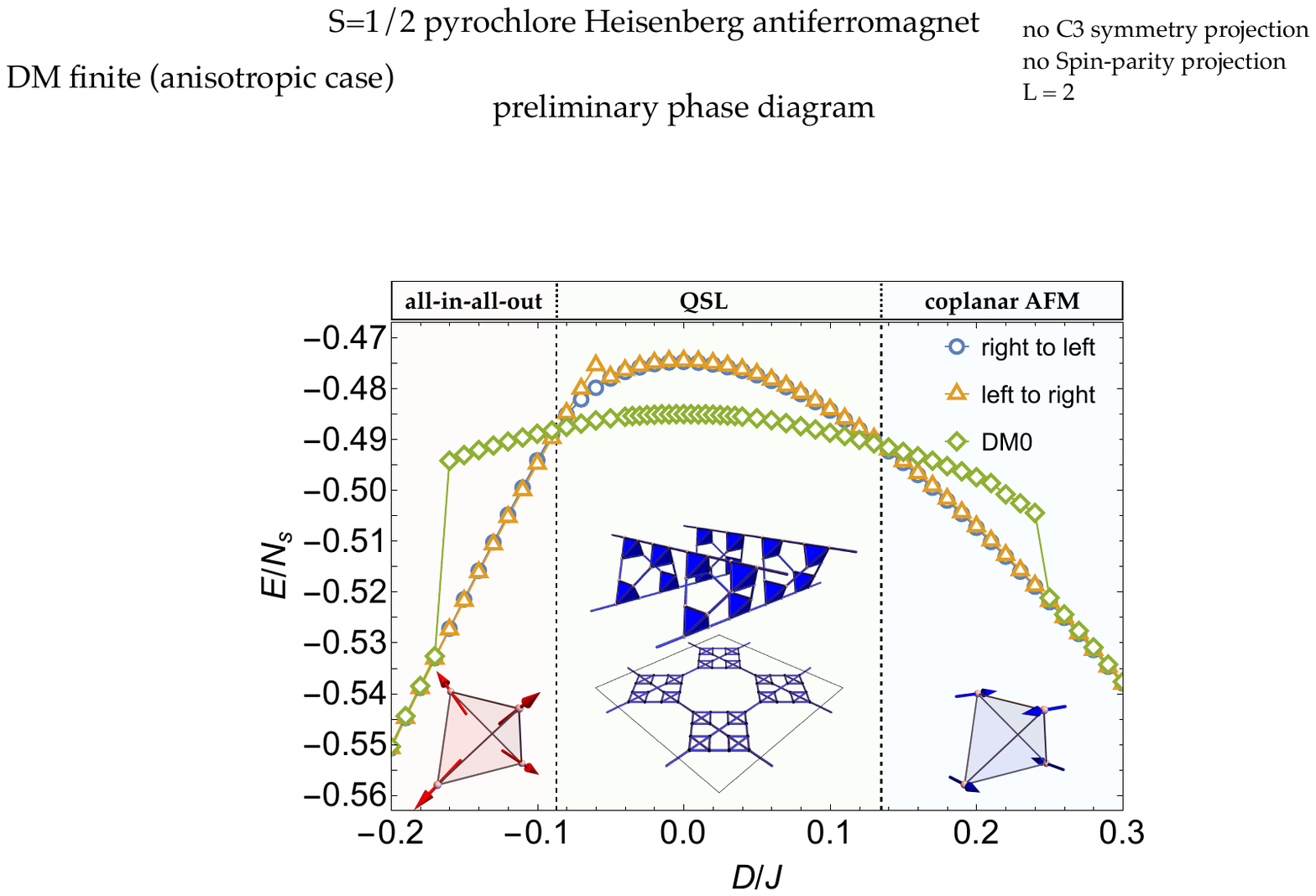}
	\caption{
         Ground state phase diagram of the $S$=1/2 pyrochlore Heisenberg 
         antiferromagnet in presence of Dzyaloshinskii-Moriya (DM) 
         interactions $D/J$ [Eq.~(\ref{eq:Ham})].
	Here, $D$ is the spin-orbit coupling defined in Appendix \ref{app:DM.def} 
	and $J$ is the nearest-neighbor Heisenberg exchange coupling defined in 
	Sec.~\ref{sec:model}.
	The model reveals a QSL ground state in an extended parameter 
	region \mbox{$-0.085(5) < D/J < 0.135(5)$}, surrounded by   
	all-in/all-out and a coplanar antiferromagnetic (AFM) ordered phases.
	The energy per site, $\mi{E/N_s}$, has been obtained from mVMC calculations 
        for a cubic cluster of the linear dimension $L = 2$ 
	($N_s$ = 128 total sites), by energy-optimization sweeps from 
        right to left (blue circles), left to right (orange triangles) and 
        by initializing the mVMC optimization with
        the maximally flippable dimer state
        [see Appendix~\ref{app:max.flip.state}] from $D = 0$ (green diamonds).
	In the QSL phase, spin correlations are confined within a 2D subspace 
        on the super-tetrahedron square lattice (STSL) as is illustrated in the central inset 
        (see details in Sec.~\ref{sec:dim.reduction}).
	}
	\label{fig:PhaseDiagram.DM}
\end{figure}
%


\begin{figure*}[htb]
	\centering
	\subfloat[ 3D -- pyrochlore lattice ]{
		\includegraphics[width=0.28\textwidth]{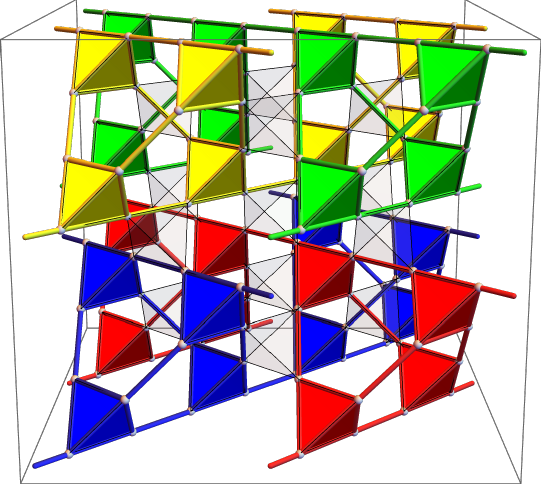} }	
	\hspace*{0.8cm}
	\subfloat[ 2D -- single layer pyrochlore  ]{
		\includegraphics[width=0.32\textwidth]{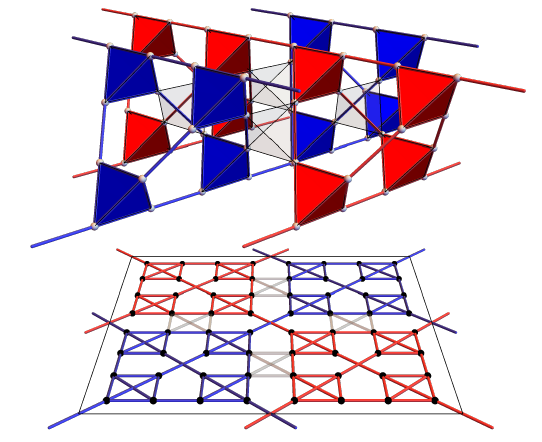} }  	
	\subfloat[ 2D -- STSL ]{
		\includegraphics[width=0.32\textwidth]{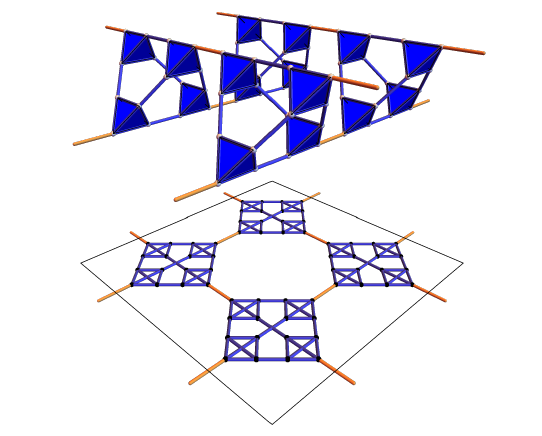} }  
	\caption{	
	Dimensional reduction in the ground state from ${\bf O_h}$ to ${\bf D_{2d}}$
	(to be compared with corresponding numerical results in 
        Fig.~\ref{fig:ground.state.3Dpyrochlore}).
	Each color schematically indicates a complicated network of singlet bonds, 
	made of super-tetrahedra which are strongly correlated within 2D layers
	(here: horizontal $xy$-plane).
	Gray tetrahedra show negligible singlet strength and effectively disconnect 
	networks of different color. 
	(a) shows a finite size cluster containing 128 spins on the 3D pyrochlore lattice with, 
	in total 4 networks (blue, red, green, yellow), which are all effectively disconnected.
	(b) shows a 64 site cluster of the blue and red colored lower layered bonds in (a), 
	and its projection onto the $xy$ plane.
	(c) shows the isolated blue network from (b) as finite-size cluster with 64 spins, 
	and its projection onto the $xy$ plane, forming the super-tetrahedron square lattice (STSL).
	Bonds have been colored blue and orange to indicate definitions of couplings 
	$J_1$ and $J_2$ in ${\mathscr H}_{\sf J_1 J_2}$ of Eq.~\eqref{eq:Ham.J1J2}, respectively.
	}
	\label{fig:3D.to.2D}
\end{figure*}
%

As shown in Fig.~\ref{fig:PhaseDiagram.DM}, we find a non-magnetic phase in a 
region of parameter space, which we propose to be a QSL.
After thorough optimization of the wave function for the SU(2) symmetric Hamiltonian, 
we observe the convergence to the ground state, which breaks the octahedral symmetry 
${\bf O_h}$ of the pyrochlore lattice [Fig.~\ref{fig:PhaseDiagram.DM}(a)].
This symmetry-broken state facilitates a dimensional reduction from isotropic 3D to 
decoupled 2D layers in an emergent fashion. 
Therefore, we are allowed to solve a resultant system confined within each 2D 
layered bond network.
Namely, the ground state shows an enlarged unit cell, involving 16 sites on a super-tetrahedron, 
which are connected via singlets within a 2D plane forming a super-tetrahedron 
square lattice (STSL) lattice, as depicted in the central inset of Fig.~\ref{fig:PhaseDiagram.DM} 
as well as Figs.~\ref{fig:3D.to.2D}(b) and (c), while interlayer correlations essentially vanish 
leading to the confinement in a 2D plane.

By performing 
finite-size scaling, we observe a power-law decay of spin correlations in the 
ground and excited states of the STSL model, with vanishing excitation gap in the 
thermodynamic limit.
By fitting our numerical results to a 16-orbital Hartree-Fock-Bogoliubov (HFB) type mean-field wave function, 
we obtain quadratically dispersing gapless excitations for spinons.
These spinons emerge from the fractionalization of the original spins. Interestingly,
these excitations are gapless not on a single point but on lines in
momentum space.
%

%

The present article is structured as follows:
Section~\ref{sec:model} introduces the \yy{Hamiltonian} and outlines the
mVMC method, with the calculated mVMC results in Sec.~\ref{sec:results}. 
Section~\ref{sec:dim.reduction} demonstrates that the ground state 
exhibits the singlet-bond order on the 3D pyrochlore lattice, 
leading to a dimensional reduction from 3D to 2D  by effectively disconnecting the interlayer 
correlation as a consequence of the symmetry breaking. 
Then, we introduce an effective model on the STSL,
which captures dominant correlations of the obtained QSL ground state. 
Section~\ref{sec:QSL.robustness} discusses the robustness of the 
QSL state in the presence of perturbations, in an example of finite Dzyaloshinskii-Moriya
interactions on the 3D lattice model, and anisotropic exchange interactions on the 
STSL model.
Section~\ref{sec:QSL.2D.pyrochlore} presents finite-size scaling 
on the STSL model for clusters of up to
$N_s=$ 1024 spins corresponding to an effective site number in the full
three-dimensional lattice consisting of approximately $8\times 10^3$ spins, from which 
the presence of a QSL ground state is evidenced from power-law 
correlations between spins and a vanishing gap between
the ground state and excited state energies in the thermodynamic limit.
We propose in Sec.~\ref{sec:BCS.theory} that our numerical findings can be 
interpreted by the fractionalization of an electronic spin into two spinons supported from  
the fitting of the mVMC ground-state wave function to a HFB mean-field theory.
Section~\ref{sec:summary.discussion} summarizes and discusses our results 
and  their implications for future studies.

%
%
\section{Model Hamiltonians}
\label{sec:model}
%

We study the spin-1/2 Heisenberg Hamiltonian
%
\begin{equation}
	{\mathscr H} = J   \sum_{\langle ij \rangle} {\bf S}_i \cdot {\bf S}_j  	\, ,
\label{eq:Ham.iso}
\end{equation}
%
on the pyrochlore lattice, as illustrated in Fig.~\ref{fig:3D.to.2D}(a).
The spin $S=1/2$ vector operator ${\bf S}_i = (S^x_i, S^y_i, S^z_i)$ 
acts on site $i$, with $J$ being the antiferromagnetic isotropic exchange 
interaction between neighboring spins on sites $i$ and $j$.

When the spin-orbit coupling is not negligible,
an asymmetric exchange coupling called Dzyaloshinskii-Moriya (DM) 
interaction~\cite{dzyaloshinsky1958thermodynamic, PhysRevLett.4.228, PhysRev.120.91} 
becomes a relevant perturbation, in addition to the Heisenberg term, as,
%
\begin{equation}
	{\mathscr H}_{\sf DM} = {\mathscr H} + \sum_{\langle ij \rangle} {\bf D}_{ij} \cdot \big( {\bf S}_i \times {\bf S}_j \big) 	\, ,
\label{eq:Ham}
\end{equation}
%
where the vector ${\bf D}_{ij} = D \ {\bf e}_{ij}$
defines the DM interaction with the unit vector ${\bf e}_{ij}$ in the  direction perpendicular to the 
bond bridging the $i$ and $j$ sites~\cite{Canals2008}.
Details can be found in Appendix~\ref{app:DM.def}.
The amplitude of the vector $D=|{\bf D}_{ij}|$ does not depend on the bond.

The ground state of the Hamiltonian ${\mathscr H}$ [Eq.~(\ref{eq:Ham.iso})]
shows a spontaneous dimensional reduction as illustrated in 
Fig.~\ref{fig:3D.to.2D} and discussed later in Sec.~\ref{sec:dim.reduction}.
The effective Hamiltonian after the dimensional reduction is the Heisenberg Hamiltonian
on the STSL, illustrated in the central inset of Fig.~\ref{fig:PhaseDiagram.DM} as 
well as in Fig.~\ref{fig:3D.to.2D}(c).

\begin{figure*}[t]
	\centering
	\includegraphics[width=0.98\textwidth]{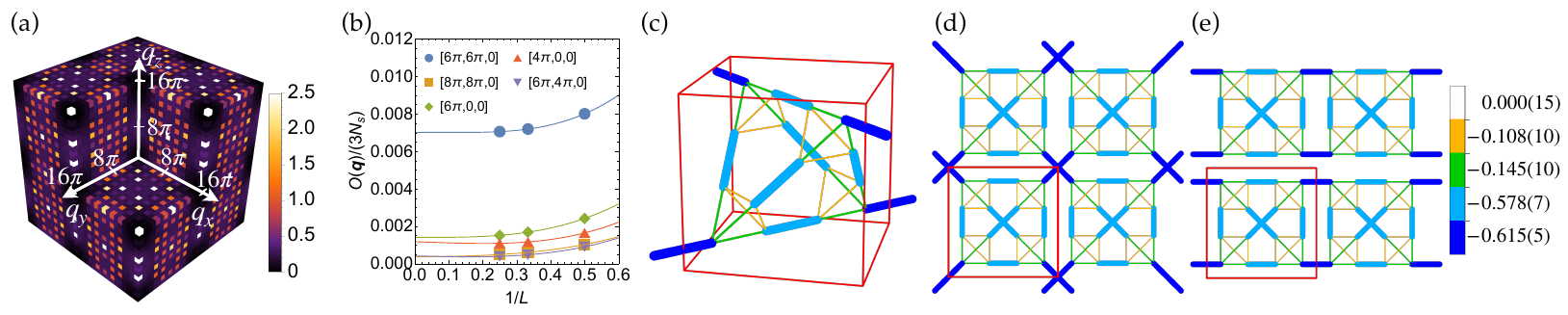}  
	\caption{
	Singlet-bond correlation and symmetry breaking.	
	(a) Momentum-resolved structure factor $O({\bf q})$ [Eq.~\eqref{eq:Oq.with.static}]
	of singlet bonds for a $L=2, N_s=128$ site cubic cluster shows high-intensity points 
	in Brillouin zone centers. 
	(b) Finite-size scaling of selected high-intensity points demonstrating the presence of long-range 
	singlet-bond order in the thermodynamic limit.
	(c)--(e): Real space configuration of singlet-bond intensity [Eq.~\eqref{eq:B_corr}] in 
	the symmetry-broken ground state. 
	Singlets cover the pyrochlore lattice with different intensities (thick = strong, thin = weak)
	on a complex bond network, as shown 
	for (c) the 16-site unit  cell, (d) the top view and (e) the front view of a 
	segment of a $L=4, N_s=1024$ site cubic cluster.
	Singlets with negligibly small intensities are colored white.
	The three-dimensional unit cell is shown as red boxes, in (c), (d) and (e).
	}
	\label{fig:ground.state.3Dpyrochlore}
\end{figure*}
%

To understand the entangled nature of the wave functions intuitively, we further introduce 
an effective Hamiltonian
\begin{equation}
	{\mathscr H}_{\sf J_1 J_2} =  J_1 \sum_{\langle ij \rangle_1} {\bf S}_i \cdot {\bf S}_j 
				     +  J_2 \sum_{\langle ij \rangle_2} {\bf S}_i \cdot {\bf S}_j   \, ,
	\label{eq:Ham.J1J2}
\end{equation}
%
where we classify $J$ into two types $J_1$ and $J_2$, namely $J_1$  corresponds 
to bonds inside the super-tetrahedron and $J_2$ accounts for the inter-super-tetrahedron bond. 
In other words, $\langle ij \rangle_1$ indicates the nearest-neighbor interaction 
between sites within each super-tetrahedron [blue bonds in Fig.~\ref{fig:3D.to.2D}(c)], while 
$\langle ij \rangle_2$ indicates the interaction between sites connecting super-tetrahedra 
[orange bonds in Fig.~\ref{fig:3D.to.2D}(c)].
We start from two extreme limits $J_1=0$ and $J_2=0$, each of which drives the system 
into a different simple product state. 
We then treat interactions perturbatively by gradually switching on nonzero $J_1$ or $J_2$
and examine the growth of the entangled nature in the QSL phase.

The Hamiltonians are solved by VMC, which is outlined in Appendix~\ref{app:Method}  and its 
accuracy is discussed in Appendix~\ref{app:pyro3D.energy}.

%
%
%
\section{Results}
\label{sec:results}  

We find that the ground state of the Hamiltonian in Eq.~\eqref{eq:Ham.iso} is
a spontaneously symmetry-broken phase that cannot be 
represented by a simple product state.
More specifically, the ground state of the three-dimensional pyrochlore Heisenberg 
antiferromagnet emergently breaks down to stacked two-dimensional algebraic 
quantum spin liquids, which is reminiscent of symmetry breaking to a smectic liquid crystal.
This symmetry breaking enlarges the unit cell size to $16$ spins by preserving the 
cubic symmetry of the pyrochlore lattice. 
Technically, this allows the calculation of three system sizes of the cubic lattice with 
the linear dimension $L=2,3$ and 4, corresponding to the total number of sites
$N=128, 432$ and 1024, respectively. 
We later employ an even more efficient method by utilizing the dimensional reduction 
emergent from the symmetry breaking, which allows simulations of effectively much 
larger system sizes.
%

%
%
%
\subsection{Dimensional reduction from $O_h \to {D_{2d}}$}  	
\label{sec:dim.reduction}
%

\subsubsection{Symmetry breaking } 	
\label{sec:sym.breaking} 

We first show, in Fig.~\ref{fig:ground.state.3Dpyrochlore}(a), the Fourier transform of 
the singlet-bond correlation $O({\bf q})$, as defined in Eq.~\eqref{eq:Oq.with.static}, 
which shows high intensities at the ordering vector of \mbox{${\bf q}=(2n\pi,2m\pi)$} 
with $n$ and $m$ being integers. 
Figure~\ref{fig:ground.state.3Dpyrochlore}(b) demonstrates the long-range singlet-bond order after 
the extrapolation to the thermodynamic limit, with especially strong order at $(6\pi,6\pi,0)$. 
The real space configuration of the bond order for a $L=4, N_s=1024$ site cubic cluster
is shown in Figs.~\ref{fig:ground.state.3Dpyrochlore}(c)--(e).
Here, the colors of the bonds represent the singlet strengths, as defined in 
Eq.~\eqref{eq:B_corr}, which is 
$B_m = -0.615(5)$ on dark blue, 
$B_m = -0.578(7)$ on blue, 
$B_m = -0.145(10)$ on green, and 
$B_m = -0.108(10)$ on orange bonds.
Singlet intensities which are negligible within numerical accuracy, $B_m = 0.000(15)$, 
are colored white.

Consistently, the spin structure factor defined by $S^z({\bf q})$ [see Eq.~\eqref{eq:Sq} 
and Fig.~\ref{fig:3D.pyrochlore.spin.singlet}] does not show any 
signature of the order indicating that the ground state of the Heisenberg Hamiltonian in 
Eq.~(\ref{eq:Ham.iso}) is non-magnetic.
Our results for the singlet-bond order is consistent with the previous 
works~\cite{Hagymasi2021,Astrakhantsev2021,Schafer2023} up to the 128 site cubic cluster.

The periodic singlet ordering shown in Figs.~\ref{fig:ground.state.3Dpyrochlore} (c)--(e) 
enlarges the magnetic unit cell to a super-tetrahedron consisting of 16 sites.
Such an ordering induces concomitantly three types of symmetry breaking:
\begin{enumerate}[label=(\roman*)]
	\item Inversion symmetry breaking [(degree of degeneracy) = 2],  which selects 
	the tetrahedral sublattice consisting of only either upward or downward tetrahedra.
	\item Translational symmetry breaking [(degree of degeneracy) = 4 ] that specifies the origin of the super-tetrahedron 
	among one of 4 equivalent tetrahedra within the cubic unit cell.
	\item $C_3$ rotational symmetry breaking [(degree of degeneracy) = 3], where 
	one of 3 possible 2D  planes (namely, $xy$, $xz$ or $yz$) are chosen as the 
	plane of STSL composed of connected super-tetrahedron.
\end{enumerate}
This singles out in total $24 (=2\times 3 \times 4)$ degenerate states in the ground state.
While previous studies identified singlet order in the ground 
state~\cite{Hagymasi2021,Astrakhantsev2021,Schafer2023}, these three types of 
symmetry breaking were not discussed.
%

\subsubsection{decoupling of 2-dimensional layers}  	
\label{sec:2d.layers}
%
 
Among three types of symmetry breaking (i)--(iii), (iii) is especially important because 
it generates the dimensional reduction from 3D to 2D as is illustrated in Figs.~\ref{fig:3D.to.2D}(b) 
and (c), for the example of the symmetry breaking to stacked $xy$ planes.
Since each super-tetrahedron is connected to other super-tetrahedra only via 
strongly coupled inter-tetrahedron bonds colored here by red, blue, green and yellow 
lines in Fig.~\ref{fig:3D.to.2D}(a), different color super-tetrahedra are essentially 
decoupled and form mutually disconnected 4 networks.

A closer look reveals that the 4 subsystems are decomposed to two groups (one 
group colored by blue and red and the other group, green and yellow) as is seen in 
Fig.~\ref{fig:3D.to.2D}(b). 
Two subsystems constituting a group (for instance blue and red) are interpenetrating 
but are not connected by strong bonds with each other.
Therefore we are allowed to treat only one subsystem if one can see that the coupling 
between two subsystems become irrelevant. We will show numerical evidence of this 
decoupling later. Provided that this is the case, in Fig.~\ref{fig:3D.to.2D}(c) we plot 
only one of those sub-systems after  symmetrization and rotation of $\pi/4$.
This lattice forms the minimal network which is necessary to capture the dominant 
correlations in the ground state of the full 3D model on the pyrochlore lattice. 
The STSL refers to this square lattice of super-tetrahedron.

Then, we solve ${\mathscr H}$ [Eq.~(\ref{eq:Ham.iso})] on this effective 
STSL, in order to clarify the nature of the ground state in the full 3D pyrochlore lattice. 
The 2D STSL allows us to access physical quantities for clusters up to 
the linear dimension $L = 8$ in the unit of the unit cell of the STSL, which corresponds to
the number of sites $N_{\sf STS} = 1024$ on the STSL, and an effective system 
size on the 3D pyrochlore lattice of $N_{\sf 3D} \approx 8 \times 10^{3}$ sites.
This gives us freedom to perform a finite-size scaling, and allows
reasonable extrapolations of ground state properties to the thermodynamic limit.

\subsubsection{Evidence for dimensional reduction \\ --- Comparison of correlations}  			
\label{sec:correlations.benchmark}
%

%
\begin{figure}[b]
	\centering
	\includegraphics[width=0.49\textwidth]{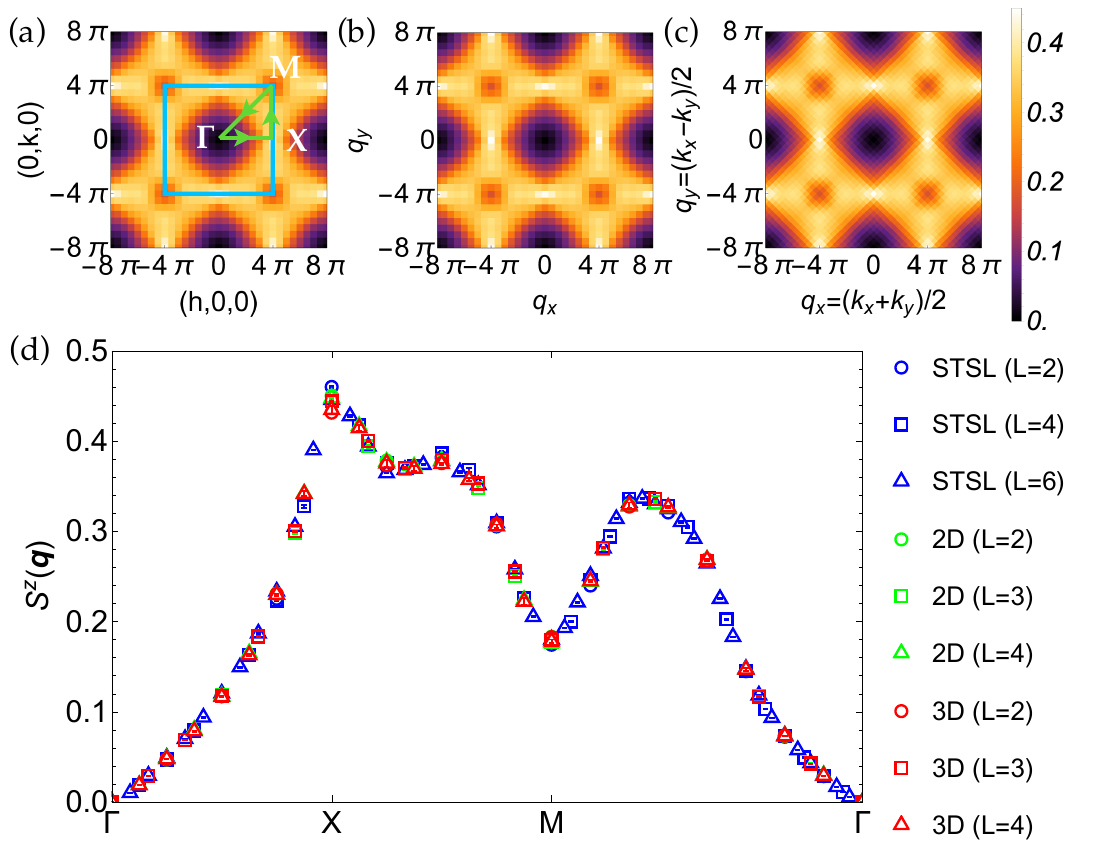} 	
	\caption{
	Comparison of the structure factor $S^z({\bf q})$ [Eq.~\eqref{eq:Sq}] in the ground state of ${\mathscr H}$ 
	[Eq.~(\ref{eq:Ham.iso})] on
	(a) the 3D pyrochlore lattice, 
	(b) the 2D pyrochlore layer, and 
	(c) the STSL (see Fig.\ref{fig:3D.to.2D}) for systems of linear dimension $L=4$
        in (a) and (b), and $L=6$ for (c).
  	The magnetic Brillouin zone is drawn as blue square in (a).
	The momentum coordinates $h$ and $k$ in (a), correspond to 
	$q_x$ and $q_y$ in (b), respectively. 
	Note that the Brillouin zone of the original STSL with reciprocal 
	lattice vectors $k_x$ and $k_y$ (see definition in Appendix~\ref{app:STS.coord})
	is 45 degrees rotated in (c) to make a direct comparison with (a) and (b) easier.
	(d) shows the $S^z({\bf q})$ along the irreducible wedge [see green path in (a)], 
	where results of all three lattice models are plotted together 
	for different system sizes.	
	The comparison between all three models shows negligible difference,
	supporting the reliability of using the STSL to model the ground state 
	of the 3D pyrochlore lattice. 
        }
	\label{fig:comp.Sq.Oq.3D.2D}
\end{figure}
%

To justify the validity of the effective STSL model we explicitly compare in 
Fig.~\ref{fig:comp.Sq.Oq.3D.2D} (a)--(c) the equal-time spin structure factor $S^z({\bf q})$ for the ground 
states of 3D pyrochlore, 2D layered pyrochlore and the STSL, respectively. 
(a) and (b) are shown for $L=4$, while (c) is shown for $L=6$ and rotated by $\pi / 4$, since 
we symmetrized and rotated the STSL in real space [see Fig.~\ref{fig:3D.to.2D}(c)].
$S^z({\bf q})$ shows features which are nearly identical among all three different lattices models. 
For quantitative comparison, we also plot intensities along the irreducible 
wedge for in-plane correlations in Fig.~\ref{fig:comp.Sq.Oq.3D.2D}(d).
We find that all three lattice models give the same result within numerical 
errors, supporting the reliability of using the STSL to capture dominant correlations 
in the ground state of the original 3D pyrochlore lattice. 
A detailed analysis of correlations and their implications on the ground state is demonstrated for 
the STSL model in Sec.~\ref{sec:QSL.2D.pyrochlore}.

\subsubsection{Symmetry of the super-tetraheron-square lattice}  	
\label{sec:Sym.Analysis}
%

%
\begin{table}
	\newcolumntype{C}{>{}c<{}} 
	\renewcommand{\arraystretch}{1.2}
	\setlength{\tabcolsep}{0.3cm} 
	\begin{tabular}{ C  C  C  C  C C}
	\hhline{======}		
		${\bf D_{2d}}$ & E & 2S$_4$ & C$_2$(z) & 2C$'_2$ & 2$\sigma_d$ 	\\
	\hhline{------}
		$A_{1}$ 	& 1 & 1 & 1 & 1 & 1	 \\
		$A_{2}$ 	& 1 & 1 & 1 & -1 & -1	 \\
		$B_{1}$ 	& 1 & -1 & 1 & 1 & -1	 \\
		$B_{2}$ 	& 1 & -1 & 1 & -1 & 1	 \\
		$E$		& 2 & 0 & -2 & 0 & 0	\\
	\hhline{======}
	\end{tabular}  
	\caption{
	Space group of the STSL belonging to the symmetry $ P\bar{4}m2$ (No. 115), 
	with point group symmetry ${\bf D_{2d}}$.
	}
\label{tab:STS.character.table}
\end{table}
%

The STSL is reminiscent of the square-octagon lattice
(also known as Fisher, or bathroom tile (4-8) lattice), which hosts intriguing physical 
properties by itself~\cite{Kargarian2010, Maity2020, Hearth2022, He2022}.
However, differently from the purely 2D square-octagon lattices, the STSL 
possesses a finite height along the $z$-direction, placing it within the tetragonal 
space group $P\bar{4}m2$, (space group No.115 and layer group No.59).
The resulting point group symmetry describing the STSL is ${\bf D_{2d}}$, 
with the character table shown in Table~\ref{tab:STS.character.table}.
Relevant symmetry operations are identity E, 
S$_4$ improper rotations of $\pi/2$,  
C$_2(z)$ rotations of $\pi$ about the $z$-axis,  
C$_2$ rotations of $\pi$, 
and $\sigma_d$ mirror within the $xy$ plane, 
producing the allowed irreducible representations (irreps) 
$A_1$, $A_2$, $B_1$, $B_2$ and $E$. 
Here, symmetry dependent calculations become relevant in order to 
separate different states within mVMC calculations and to characterize their excitations. 
Explicit definitions of point-group projection operators can be found in the 
SM~\cite{SM}.

%
%
\subsection{Existence of Unconventional Phase and its Robustness against Perturbation}  	
\label{sec:QSL.robustness}
%

Before clarifying the nature of the ground state itself, we first reveal the existence of an 
unconventional phase for the Heisenberg antiferromagnet on the pyrochlore lattice and 
show its stability and robustness against perturbations.
In Sec.~\ref{sec:DM} we consider finite Dzyaloshinskii-Moriya (DM) interactions, which 
become relevant as realistic perturbations from the isotropic Heisenberg model in 
real materials such as pyrochlore iridates~\cite{kurita2011topological,PhysRevB.85.045124}.
In Sec.~\ref{sec:J1J2_STS} we analyse the bond-anisotropic STSL by monitoring the 
strength of the inter-super-tetrahedron bonds $J_2$ relative to the intra-super-tetrahedron 
bond $J_1$ to gain insights into the systematic growth of quantum entanglement in the 
intermediate region between two well-defined limits of product wave functions at 
$J_2/J_1=0$ and $J_2/J_1 \to \infty$. 
In both cases we find a robust new phase in an extended region around the isotropic 
2D Heisenberg limit.

%
\subsubsection{ spin orbit coupling }  	
\label{sec:DM}
%

The symmetry of the pyrochlore lattice allows for anisotropic spin exchange in the form of 
DM interaction $D$~\cite{Elhajal2005, Canals2008}.
In real materials such interactions originate from spin-orbit coupling, as relevant in,
e.g., pyrochlore oxides~\cite{Pesin2010, Gardner2010, Arakawa2016}.
Here, we investigate the $S=1/2$ Heisenberg model with $D$ [Eq.~(\ref{eq:Ham})].
In  Fig.~\ref{fig:PhaseDiagram.DM} of Sec.~\ref{sec:Intro} we show the ground-state     
phase diagram of ${\mathscr H}_{\sf DM}$ [see Eq.~\eqref{eq:Ham}] for a $L=2$ ($N_s=128$) 
site cluster, respecting the ${\bf O_h}$ cubic symmetry of the pyrochlore lattice. 
The model shows  an ``all-in / all-out'' magnetic ordered dipolar phase
(2-fold degenerate) for large negative $D$, and a 
so-called $\Psi_3$-coplanar antiferromagnetic (AFM) phase (6-fold degenerate) for 
large positive $D$.
Between those classically ordered phases we obtain an intermediate new phase, 
which will be identified as QSL later in Sec.~\ref{sec:QSL.2D.pyrochlore}, 
over a wide range \mbox{$-0.085(5) < D/J < 0.135(5)$}.
By performing energy-optimization sweeps from right to left (blue circles), 
left to right (orange triangles) and optimization from a maximally flippable dimer 
initial state (green diamonds) [see Appendix~\ref{app:max.flip.state}]
we observe first-order transitions to the intermediate new phase both from the two sides 
of the classically ordered phase as is visible from their energy 
level crossings. 

A previous study on the same model in its classical limit found the same ordered 
magnetic phases as observed in the present $S=1/2$ case for largely negative and 
positive $D$ regions~\cite{Chern2010}.
A cooperative paramagnetic state (classical spin liquid state) is stabilized around $D = 0$ 
at nonzero temperatures, but exists only at $D = 0$ at zero temperature.
Comparing these classical findings to our quantum results,
we interpret that quantum fluctuations play a crucial role in stabilizing the 
intermediate ground state in an extended region of finite $D$ identified as the QSL in 
Sec.~\ref{sec:QSL.2D.pyrochlore}.

Destabilization of the magnetic phases around $D/J=0$ has been
reported in a pseudo-fermion functional renormalization group (PFFRG) study~\cite{Noculak2023},
which suggests the existence of a quantum paramagnetic phase for 
\mbox{$-0.20 \lesssim D/J \lesssim 0.23$} at zero temperature.
While their results on the destabilization of the magnetically ordered phases
are qualitatively consistent with our phase diagram, the phase boundaries are
different from ours.
The first-order nature of the phase transitions found in the present study may be the 
origin of discrepancy in the estimation of the phase boundaries, because the transition 
point inferred from divergence of the susceptibility in the PFFRG study leads to 
overestimate the paramagnetic phase. 
In addition, the PFFRG study did not clearly identify the nature of the QSL phase.

%
\subsubsection{ anisotropic STSL model }  	
\label{sec:J1J2_STS}
%
\begin{figure}[t]
	\centering
	\includegraphics[width=0.42\textwidth]{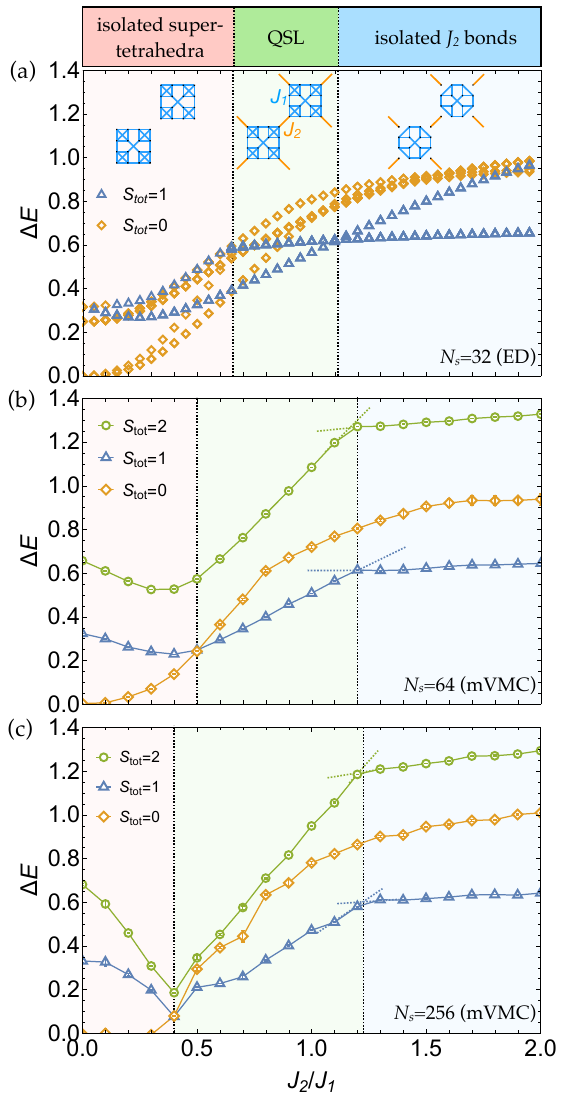} 
	\caption{
	Excitation energy $\Delta E$ for the STSL model [see Eq.~\eqref{eq:Ham.J1J2}
	and inset of (a)], as a function of coupling $J_2/J_1$.
	Results for states with total spin $S_{\rm tot}=0,1,2$ were obtained from (a) 
	exact diagonalization results for a $N_s=32$ site cluster, 
	and from mVMC after variance extrapolation for (b) $N_s=128$ and (c) 
        $N_s=256$ site clusters.
	The two limits of product wave functions of isolated super-tetrahedra ($J_2/J_1 = 0$)
	and isolated $J_2$ bonds ($J_2/J_1 \to \infty$) are separated by an extended  
	quantum spin liquid (QSL) phase. 
	Phase boundaries have been identified by level crossings in the excited states.
	}
	\label{fig:PD.J2}
\end{figure}
%
As discussed in Sec.~\ref{sec:dim.reduction} the ground state of the 3D antiferromagnetic
Heisenberg model on the pyroclore lattice
can be described sufficiently well by the STSL model. 
In the following, we make use of the lattice anisotropy in ${\mathscr H}_{\sf J_1 J_2}$ 
[see Eq.~\eqref{eq:Ham.J1J2}], in order to probe the robustness of the ground state. 
We distinguish the amplitudes of the exchange interactions on bonds inside a super-tetrahedron 
[blue bonds Fig.~\ref{fig:3D.to.2D}(c) and inset of Fig.~\ref{fig:PD.J2}(a)], denoted as 
$J_1$, and interactions which connect two nearest-neighbor super-tetrahedra 
[orange bonds Fig.~\ref{fig:3D.to.2D}(c) and inset of Fig.~\ref{fig:PD.J2}(a)],
denoted as $J_2$.
By controlling the ratio of those two interaction strengths $J_2/J_1$, we are able to tune 
the model between two trivial and well-distinct states. 

For $J_2/ J_1 = 0$ correlations are strictly localized within a super-tetrahedron.
Since an isolated super-tetrahedron respects the tetrahedral point group 
symmetry $T_d$, its ground state belongs to a doubly degenerate irreducible representation $E$,
like the ground state of a simple isolated tetrahedron made of $4$ sites. 
The global wave function becomes a classical product state of these individual 
states with a classical degeneracy of $2^{L_x \times L_y}$, where $L_x$ and 
$L_y$ is the linear system size along $x$ and $y$ direction of the lattice. 
In the other limit of $J_2/ J_1 \to \infty$, the ground state of the model is represented by another 
product wave function, where isolated strong singlets occupy the $J_2$ bonds and also effectively 
eliminate correlations between nearest-neighbour super-tetrahedra. 
Each of the resulting isolated singlets and isolated truncated super-tetrahedra 
has a unique and symmetric ground state, forming a global, 
singly-degenerate product wave function without long-range entanglement.

In Fig.~\ref{fig:PD.J2} we explicitly show the excitation spectrum of 
${\mathscr H}_{\sf J_1 J_2}$ 
for states with total spin $S_{\rm tot}=0,1,2$ as a 
function of the coupling ratio $J_2/J_1$.
We compare system sizes for $N_s=32$, as obtained from   
exact diagonalization via the Lanczos method~\cite{HPhi}, 
with system sizes $L=2$ ($N_s=64$) and $L=4$ ($N_s=256$) as obtained 
from variance extrapolation from  mVMC optimized variational wave functions, 
including full spin-projections (calculation details can be found in SM~\cite{SM}).
Sandwiched by the two well-known limits of product-wave functions, we find an extended  
region of an intermediate phase in the range between 
$J_2/J_1 \approx 0.4$ and $J_2/J_1 \approx 1.2$ for $L=4$ ($N=256$).
We estimate those phase boundaries from the established technique of level spectroscopy, 
which tells us that phase boundaries of the ground state can be estimated from level crossings 
in the energy spectrum of the lowest excited states \cite{Suwa2016, Wang2018, Nomura2021}.
In the following, we present a systematic 
study to reveal the nature of the new intermediate phase.

%
%
\subsection{The QSL on the super-tetraheron-square lattice}  	
\label{sec:QSL.2D.pyrochlore}
%

Now, we show numerical evidence that the intermediate and unconventional phase 
in the region $0.4\lesssim J_2/J_1 \lesssim 1.2$ or $0.085 \lesssim D/J \lesssim 0.135$ 
really has the nature of a QSL on the STSL. Inside this region, we take a typical example 
at $J_2/J_1=0.6$ and $D=0$ to elucidate the universal feature of the spin correlation 
and excitation spectra to characterize this QSL phase.

\begin{figure*}[t]
	\centering
	\includegraphics[width=0.99\textwidth]{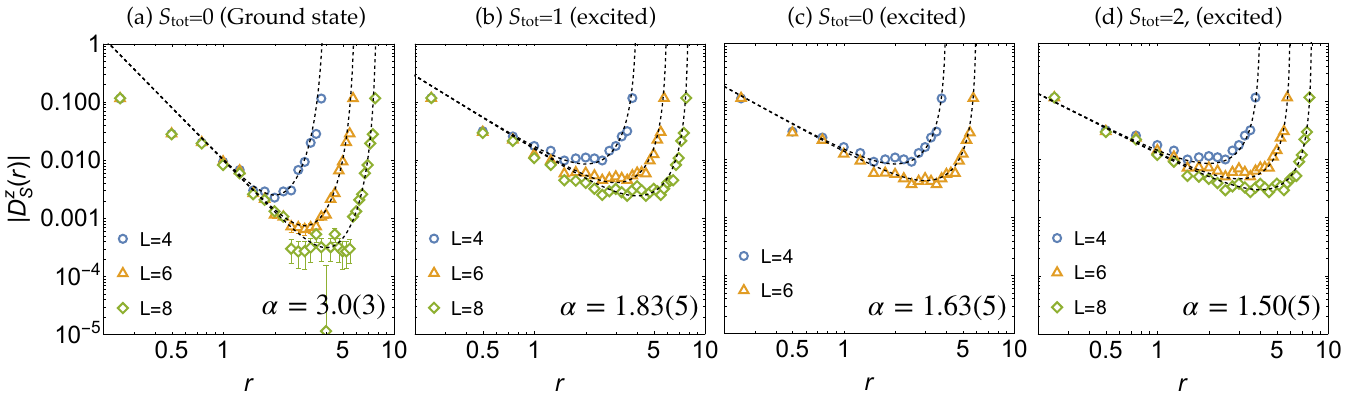}
	\caption{
	Spin-spin correlations in real space for the anisotropic 
	STSL model [see Eq.~\eqref{eq:Ham.J1J2}] for $J_2/J_1 = 0.6$.
	Correlations are shown for their $z$-components, where
 	$|D_S^z({\bf r})|$ [see Eq.~(\ref{eq:Drr})] has been measured along 
        $x$ and $y$ directions and averaged over symmetrically-equivalent paths 
        for 5-10 independent bins, each sampled over more than $10^4$ Monte Carlo 
        steps, while the variances between the bins are plotted as the error bars.
	Correlations decay algebraically at long distances, $\sim 1/r^{\alpha}$, with 
	(a) $\alpha = 3.0(3)$ in the $S_{\rm tot} = 0$ ground state,
        and (b) $\alpha = 1.83(5)$ for the $S_{\rm tot} = 1$,
        (c) $\alpha = 1.63(5)$ for the $S_{\rm tot} = 0$,
        (d) $\alpha = 1.50(5)$ for the $S_{\rm tot} = 2$ excited states
	[see dashed black curve for power-law decay fit by Eq.~\eqref{eq:Dr}]. 
 	}
	\label{fig:spin.correlations.real}
\end{figure*}
%

\subsubsection{spin-spin correlations}  	
\label{sec:STS.correlations.spin}
%

\begin{figure*}[t]
	\centering
	\includegraphics[width=0.99\textwidth]{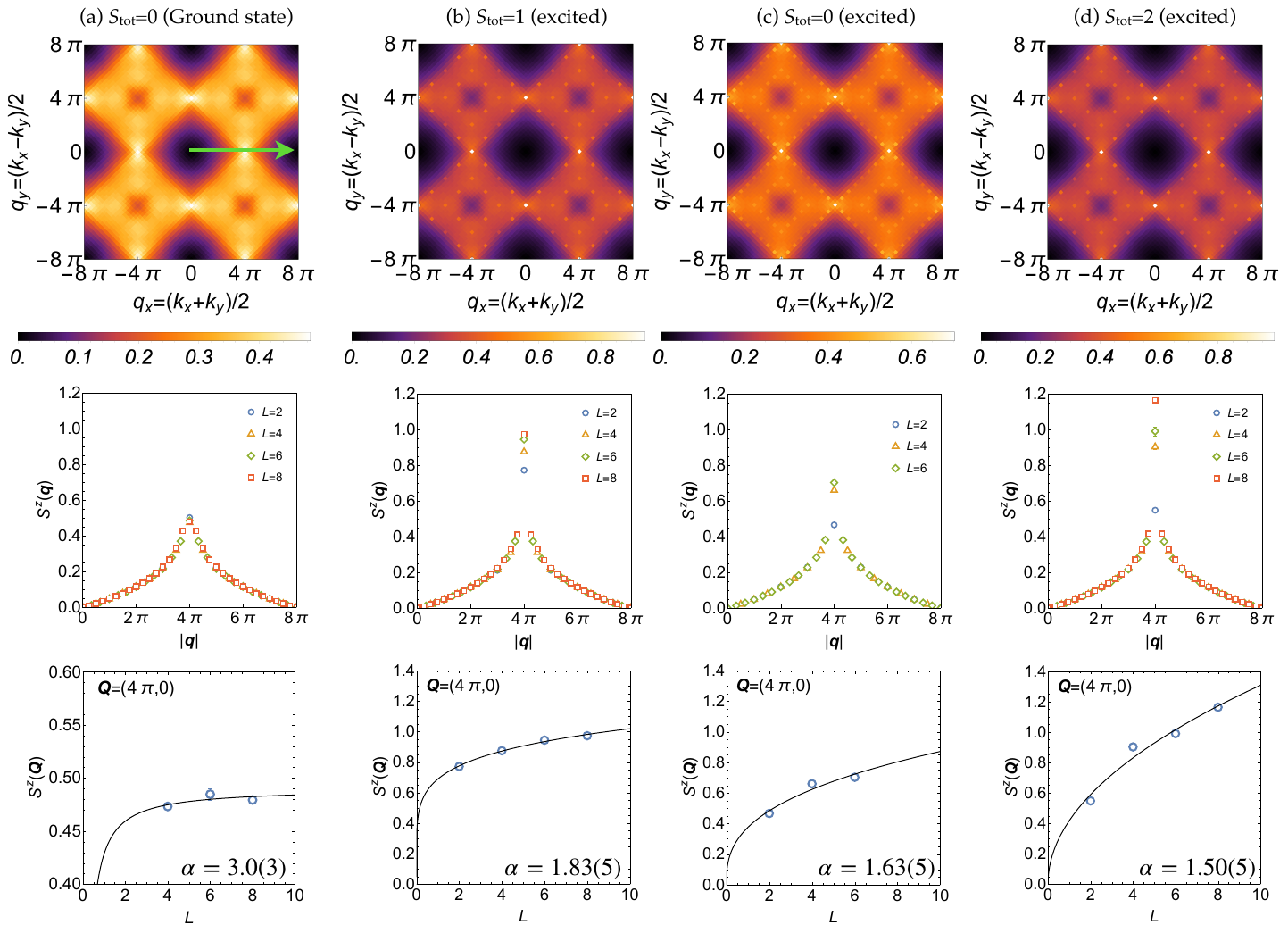}
	\caption{
	Spin-spin correlations in momentum space for (a) the ground state and (b)--(d)
	its excited states in the anisotropic STSL model [see Eq.~\eqref{eq:Ham.J1J2}] 
	for $J_2/J_1 = 0.6$.
	In the top panels, we present the equal-time spin structure factor
	$S^z({\bf q})$ [see Eq.~\eqref{eq:Sq}] for a cluster of size $L=6$.
	The unfolded Brillouin zone has been symmetrized and rotated by 45 degree, 
	to make a comparison to the 3D pyrochlore lattice case in 
	Fig.~\ref{fig:comp.Sq.Oq.3D.2D} easier, where $(k_x, k_y)$ is the momentum 
	of the original STSL (see Appendix~\ref{app:STS.coord}). 
	The scattering shows a diffuse checkerboard pattern and broadened 
	pinch-point-like structures at the corners between the squares composing 
	the checkerboard.
	In the middle panels, we show the $S^z({\bf q})$ along a line cut 
        from ${\bf q} = (0,0)$ to $(8\pi, 0)$ (green arrow in (a) of the top panel) and 
        observe that the broadened pinch-points 
        follow a catenary line shape with a singular cusp at ${\bf Q} =  (4\pi, 0)$.
	In the bottom panels, we show the scaling of the peak value of this cusp with 
	system size, and obtain, by fitting with Eq.~\eqref{eq:Sq.scaling}, power-law 
	exponents which consistently fit the real-space spin correlations
 	in Fig.~\ref{fig:spin.correlations.real}.
	Observables were obtained in the same way as done for 
        Fig.~\ref{fig:spin.correlations.real}.
     	}
	\label{fig:spin.correlations.mom}
\end{figure*}
%

In the following we shall discuss spin-spin correlations for the 
ground state and excited states 
of ${\mathscr H}_{\sf J_1 J_2}$ [Eq.~\eqref{eq:Ham.J1J2}] on the STSL 
for $J_2/J_1 = 0.6$.
In Fig.~\ref{fig:spin.correlations.real}, we show the log-log plot of the 
size-dependent spin-spin correlations defined by Eq.~(\ref{eq:Drr}).
Only the spin correlation in the $z$ component is displayed, because the spin 
correlation must satisfy the spin space symmetry. 
For the total singlet ($S_{\rm tot}=0$) state, the spin correlation must satisfy 
the SU(2) symmetry, which trivially yields the isotropic spin correlation. 
Even for the $S_{\rm tot}=1$ and $S_{\rm tot}=2$ excited states, the total 
spin per site scales to zero in the thermodynamic limit and asymptotically 
satisfies the SU(2) isotropic nature. 

Correlations have been measured along the $x$ and $y$ directions within the STSL
and averaged over symmetrically-equivalent paths.
We observe a power-law decay of correlations for long distances, 
which is well fitted by the form 
%
\begin{equation}
	|D^z_S({\bf r})| = A \left( \frac{1}{r^{\alpha}} + \frac{1}{|L-r|^{\alpha}} \right)\, ,
\label{eq:Dr}
\end{equation}
%
taking into account the periodic boundary condition at the edge of the finite-size cluster.
While the ground state shows a power-law decay with the exponent of  $\alpha = 3.0(3)$, 
its excited states decay weaker with $\alpha = 1.83(5)$,
$\alpha = 1.63(5)$ and $\alpha = 1.50(5)$ for states with $S_{\rm tot} = 1$,
$S_{\rm tot} = 0$ and $S_{\rm tot} = 2$, respectively.
We further plot in SM~\cite{SM} the same data set of 
Fig.~\ref{fig:spin.correlations.real}
on a semi-log scale and confirm an exponential decay only at short distances, 
with a deviation from the exponential fit for long distances. 
The long-range algebraic decay in the ground state is very subtle and only possible to 
distinguish from its short-range exponential decay for system sizes $L \geq 6$.

Such an algebraic decay of correlations must be also visible in the 
momentum resolved spin-spin correlations.
In the first row of Fig.~\ref{fig:spin.correlations.mom}, we 
show the equal-time structure factor
$S^z({\bf q})$ [see Eq.~\eqref{eq:Sq}] on a $L=6$ size cluster for 
the same states as presented in Fig.~\ref{fig:spin.correlations.real}.
$S^z({\bf q})$ shows a checkerboard pattern without 
high-intensity Bragg peaks but with a characteristic ``bow-tie" structure and 
cusp-type singularity at ${\bf Q}=[4\pi(m-n+1),4\pi(m+n)]$,
with integers $m,n$ in the Brillouin zone.
In the second row, we compare intensities along 
the horizontal line cut from \rp{${\bf q} = (0,0)$ to $(8\pi, 0)$} and observe that the broadened 
pinch-points follow a catenary line shape with a singular cusp at 
${\bf Q} = (4\pi, 0 )$.
The asymmetry between $q_x$ and $q_y$ momentum directions can be seen by 
comparison to the vertical line 
cut from ${\bf q} = (4\pi,-4\pi)$ to $(4\pi, 4\pi)$, as shown in SM~\cite{SM}.
While intensities on the ``catenary-line" tails do not show any noticeable size dependence, 
the cusp singularity does scale with linear system size $L$  in the form
%
\begin{equation}
\begin{aligned}
    S^z({\bf Q}) &= \int d{\bf r} 
        \langle S^z({\bf r}) \cdot S^z(\bf r_0) \rangle e^{i {\bf Q}\cdot ({\bf r}-{\bf r}_0)}    \\
        &\propto \int_0^L dr \frac{1}{r^{\alpha - 1}}  
        \sim  \frac{1}{L^{\alpha - 2}}    \, .
\label{eq:Sq.scaling}
\end{aligned}
\end{equation}
%
In the third row of Fig.~\ref{fig:spin.correlations.mom}
we fit measured values with Eq.~\eqref{eq:Sq.scaling}, and 
obtain for the ground state $S^z({\bf Q}) \sim 1/L$, supporting $\alpha = 3.0(3)$,
while for the excited states $S^z({\bf Q}) \sim L^{2-\alpha}$ 
with $\alpha = 1.83(5)$, $\alpha = 1.63(5)$ and $\alpha = 1.50(5)$ for 
$S_{\rm tot} = 1$, $S_{\rm tot} = 0$ and $S_{\rm tot} = 2$, respectively. 
These values are quantitatively consistent with the measured power-law 
decay shown in Fig.~\ref{fig:spin.correlations.real} in real space.

The one-to-one correspondence of the scaling between the real and momentum spaces 
show strong evidence for the existence of a critical phase with power-law correlations of 
form $\sim 1/r^3$ in the ground state. 
At first sight, the cusp-type singularity in the $S^z({\bf q})$ of Fig.~\ref{fig:spin.correlations.mom}
seems to be reminiscent of “pinch-point” singularities as known from the classical Heisenberg 
antiferromagnet on the pyrochlore lattice~\cite{Harris1997, Moessner1998a, Moessner1998b}. 
Pinch-points directly correspond to a local divergence free condition, which impose the sum 
of classical spins sharing the same tetrahedron to vanish. 
This results in an extensive degeneracy in the classical ground state manifold, where 
correlations show a power-law decay with 
$1/r^3$ scaling~\cite{Isakov2004, Henley2005, Henley2010}.
Even though the scaling behaviour appears to coincide, our results for the quantum model 
do not show sharp pinch-points. Instead, the ground state shows  
an angular-shape singularity, while the excited states demonstrate 
cusp-type singularities with a power-law scaling of the 
peak value, both following catenary line-shapes.
Such a signature suggests a different nature of the QSL ground state 
compared to its classical counterpart.

Moreover, pseudo-fermion functional renormalization group (PFFRG) calculations 
also differ on a qualitative level.
The PFFRG result does not exhibit singularities, but rather “rounded pinch points” 
at very low temperature~\cite{Iqbal2019}.
Such rounded signatures suggest exponential decay of correlations, 
which stays in stark contrast to our findings of power-law correlations.

\subsubsection{singlet-singlet correlations}  	
\label{sec:STS.correlations.singlet}
%

\begin{figure*}[t]
	\centering
	\includegraphics[width=0.99\textwidth]{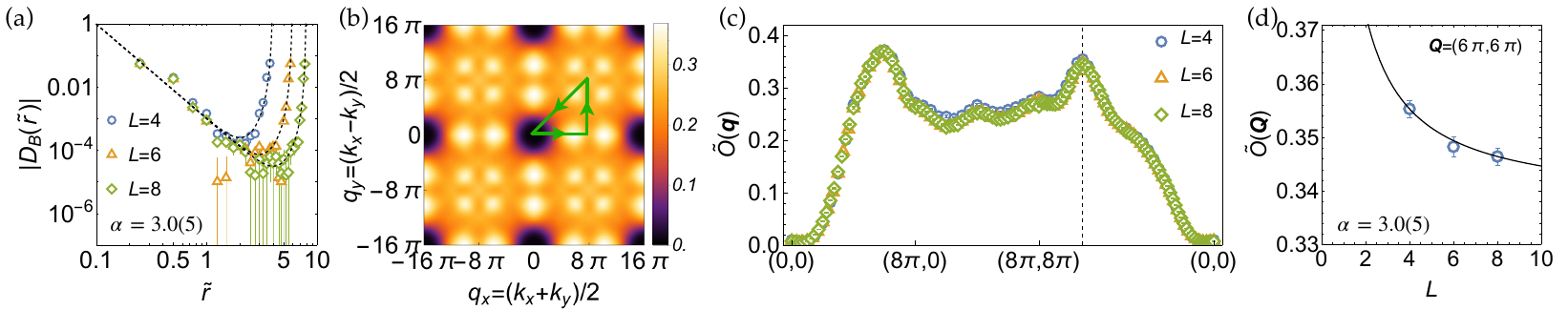}  
	\caption{
	Singlet-singlet correlations in the ground state of the anisotropic STSL 
	model [see Eq.~\eqref{eq:Ham.J1J2}] at $J_2/J_1 = 0.6$, for finite-size systems 
	of linear size $L=4,6,8$.
	(a) Absolute value of real-space correlations, $|D_B({\bf r})|$ [see Eq.~(\ref{eq:DB_no.static})], 
	have been averaged over symmetrically equivalent paths along the 
	$x$ and $y$ directions.
	Correlations decay algebraically at long distances, $\sim 1/r^{\alpha}$, with 
	$\alpha = 3.0(5)$.
	(b) Equal-time singlet structure factor without static contributions, 
	$\tilde{Q}({\bf q})$ [see Eq.~\eqref{eq:Oq.no.static}], for a cluster of size $L=6$, 
	shows a diffuse signal with bright intensities 
	around $(6 \pi, 0)$, and $(6\pi, 6\pi)$, and equivalent momentum points. 
	The Brillouin zone has been symmetrized and rotated by 45 degree, where $(k_x, k_y)$ 
	is the momentum of the original STSL (see Appendix~\ref{app:STS.coord}). 
	The path along the irreducible wedge is drawn in green. 
	(c) The intensity of $\tilde{Q}({\bf q})$ in (b) along the irreducible wedge shows a 
	very small system size dependency. 
	(d) Scaling of the peak value at $(6\pi, 6\pi)$ [dashed line in (c)], fits to a power-law, 
	$1/L^{\alpha}$, consistently to the value measured from real-space spin 
	correlations in (a).
	}
	\label{fig:singlet.correlations}
\end{figure*}
%

The algebraic decay of  spin-spin correlations
is also reflected in the correlations of singlets. 
In Fig.~\ref{fig:ground.state.3Dpyrochlore} we show that the ground state of the pyrochlore 
Heisenberg antiferromagnet breaks the octahedral symmetry of the pyrochlore lattice
by arranging singlets on a 2D layered bond network with an enlarged unit cell of a super-tetrahedron. 
Such an arrangement of singlets on the lattice induces order in the thermodynamic limit 
[see Fig.~\ref{fig:ground.state.3Dpyrochlore}(b)], 
which we confirmed by the extrapolation of diverging Bragg peaks in the equal-time structure factor 
of singlets, $O({\bf q})$ [Eq.~\eqref{eq:Oq.with.static}].

In the following we discuss fluctuations of singlet correlations by subtracting their 
static Bragg-peak contributions.
We measure the equal-time singlet structure factor
%
\begin{equation}
	\tilde{O}({\bf q}) 	= \frac{1}{N_b} \sum_{i,j} e^{i {\bf q} ({\bf \tilde{r}}_m - {\bf \tilde{r}}_n)}
				  		D_B({\bf \tilde{r}}_m - {\bf \tilde{r}}_n )	\, ,
\label{eq:Oq.no.static} 
\end{equation}
%
where $N_b = 3N_s$, with $N_s$ the number of spins, and 
$D_{B}$ the singlet-singlet correlation function in real space
%
\begin{equation}
	D_B({\bf \tilde{r}}_m - {\bf \tilde{r}}_n ) =   
		\langle B_m \ B_n \rangle - \langle B_m \rangle \langle B_n \rangle		\, .
\label{eq:DB_no.static}
\end{equation}
%
Here, the singlet strength $B_m$ on bond $m$ with bond center ${\bf \tilde{r}}_m$
is measured with Eq.~\eqref{eq:B_corr}.

In Fig.~\ref{fig:singlet.correlations} we show the singlet-singlet correlations in the 
variational ground state of the anisotropic STSL model [Eq.~\eqref{eq:Ham.J1J2}] 
at $J_2/J_1 = 0.6$, for finite-size systems of linear size $L=4,6,8$.
We note that excited states show the same type of singlet correlations (not shown here).
Figure~\ref{fig:singlet.correlations}(a) shows the real-space correlations for singlets 
$D_B({\bf \tilde{r}})$, which were measured and averaged over symmetrically-equivalent paths
along the $x$ and $y$ directions within the STSL.
In analogy to correlations for spins, we observe a power-law decay of correlations for singlets 
at long distances, which is well fitted by the form 
%
\begin{equation}
	|D_B({\bf \tilde{r}})| = A \left( \frac{1}{\tilde{r}^{\alpha}} 
                + \frac{1}{|L-\tilde{r}|^{\alpha}} \right)\, ,
\label{eq:Db.fit}
\end{equation}
%
with the power-law exponent $\alpha = 3.0(5)$.
The corresponding equal-time singlet structure factor $\tilde{O}({\bf q})$ [see Eq.~\eqref{eq:Oq.no.static}]
in Fig.~\ref{fig:singlet.correlations}(b) is very diffuse with areas of bright intensities at 
momentum $(6\pi, 0)$, and $(6\pi, 6\pi)$, and equivalent momentum points. 
Figure~\ref{fig:singlet.correlations}(c) presents a quantitative comparison of intensities for $\tilde{O}({\bf q})$
along the irreducible wedge [green line in Fig.~\ref{fig:singlet.correlations}(b)], showing a very small system 
size dependence, somewhat similar to the observed spin correlation function in Fig.~\ref{fig:spin.correlations.mom}(a).
The size dependent scaling for $\tilde{O}({\bf q})$ follows the same relationship as given in 
Eq.~\eqref{eq:Sq.scaling} for spins, resulting in the general form 
$\tilde{O}({\bf q}) \sim 1/L^{\alpha -2}$.
In Fig.~\ref{fig:singlet.correlations}(d) we fit the size-dependent peak intensity $\tilde{O}({\bf Q})$ at 
${\bf Q}=(6\pi , 6\pi)$  [dashed line in Fig.~\ref{fig:singlet.correlations}(c)] with $\tilde{O}({\bf q}) \sim 1/L$, 
supporting $\alpha = 3.0(5)$, which 
is consistent with the measured power-law exponent in real-space, shown in Fig.~\ref{fig:singlet.correlations}(a).

\subsubsection{excitation spectrum}  	
\label{sec:STS.ex.spec}
%

\begin{figure}[b]
	\centering
	\includegraphics[width=0.48 \textwidth]{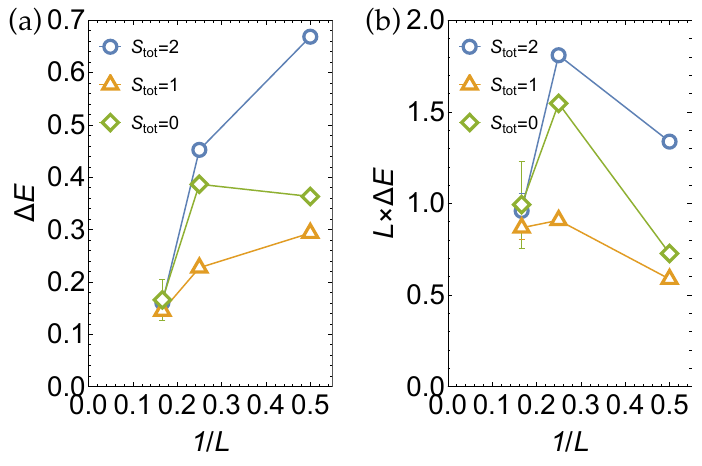}
	\caption{
        Finite-size scaling of energy gap $\Delta E$ to excited states 
        for the STSL model [see Eq.~\eqref{eq:Ham.J1J2}] at $J_2 = 0.6$.
	The data supports that $\Delta E$ decreases as a function of system 
	size and eventually scales to zero in the thermodynamic limit.
	}
\label{fig:Excitation.Energy.Scaling}
\end{figure}
%

Our analysis of the STSL allowed us to effectively access much larger  
cluster sizes than possible in the explicit 3D pyrochlore studies, 
which enables us to distinguish among
remaining long-range order, exponential decay, or algebraic decay of spin correlations, 
not only in the ground state but also in the excited states. 
The fact that excited states also follow a power-law decay of spin 
correlations suggests, that they become degenerate with the ground state
in the thermodynamic limit, which we shall confirm numerically in the following.

In Fig.~\ref{fig:Excitation.Energy.Scaling}(a) we show the energy gap 
$\Delta E$ between the ground and the lowest excited states with the total spin 
$S_{\rm tot} =0,1$ and 2 for $J_2/J_1 = 0.6$ in the spectrum of Fig.~\ref{fig:PD.J2}.
The quasi two-dimensionality of the STSL allows us to treat system sizes with 
$L = 2, 4, 6$ $(N_s = 64,  256,  576)$, with the size extrapolation to $L \to \infty$ for all 
the $S_{\rm tot} =0,1$ and 2 excitations. 
Although the present result is not conclusive because of the limitation of the system size, 
the most plausible case is a vanishing excitation gap not only for the first excited state with 
$S_{\rm tot} = 1$, but also for states with 
$S_{\rm tot} = 0$ and $S_{\rm tot} = 2$ at higher energy, consistently with the power law decay 
in all these excited states revealed in Secs.~\ref{sec:STS.correlations.spin} and \ref{sec:STS.correlations.singlet}.
The absence of the gap is further supported in Fig.~\ref{fig:Excitation.Energy.Scaling}(b),
where $L\cdot \Delta E$ does not seem to have the tendency to diverge at large sizes.
If $L\cdot \Delta E$ stays a nonzero constant in the thermodynamic limit, it implies the Dirac-like 
linear dispersion of the excitation spectra around the zero energy. On the other hand, if $L\cdot \Delta E$ is scaled 
to zero, a higher order dispersion including quadratic one is expected.  
The power-law decay of spin correlation and the consistency with the gapless excitations reported in 
this section support the emergence of the gapless QSL after the dimensional reduction to the STSL. 
We argue in Sec.~\ref{sec:BCS.theory} about the plausibility of quadratic gapless dispersion of fractionalized spins.

%
%
\section{Discussion: nature of fractionalization of spins in quantum spin liquid} 	
\label{sec:BCS.theory}
%

In this section, we further investigate the nature of this QSL,
by elucidating the structure of the mVMC variational wave function. 
Similar analyses on the structure of the variational wave functions have been 
successfully performed to clarify the nature of QSLs~\cite{PhysRevB.93.144411, Nomura2021, Ido2022} 
inspired by the projected BCS ansatz~\cite{Wen-Book}.

\begin{figure*}[t]
	\centering
	\includegraphics[width=0.99\textwidth]{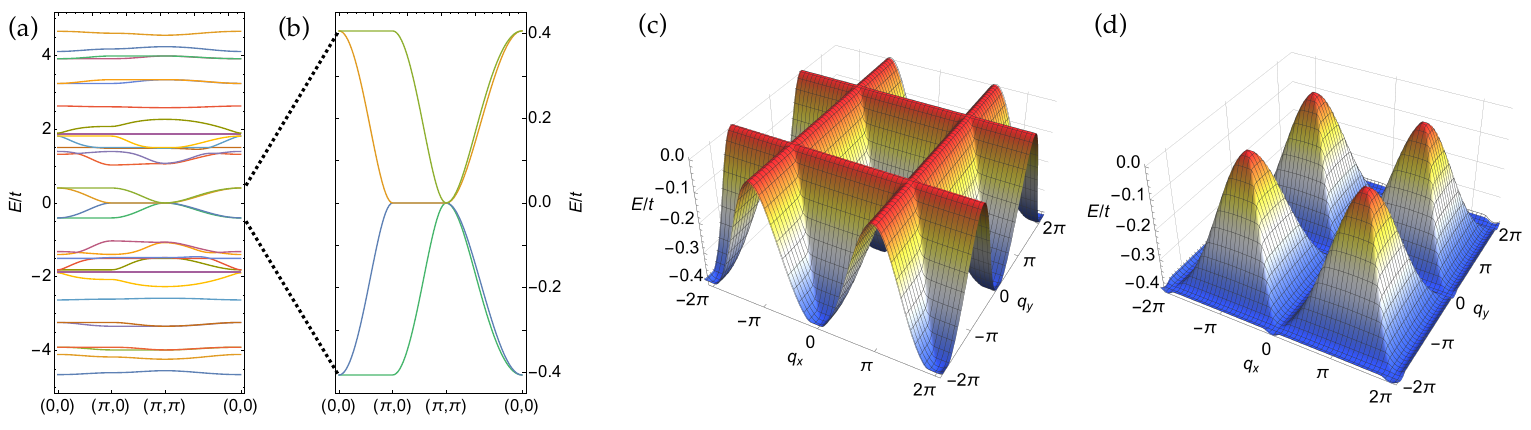}
	\caption{
	Dispersion of the HFB mean-field Hamiltonian ${\mathscr H}_{\rm HFB}$ 
	[see Eq.~\eqref{eq:BCS1} and Eq.~\eqref{eq:Hbcs_Nambu}] along the irreducible wedge, 
	after sufficient minimization of the loss function in Eq.~\eqref{eq:BCS.fij.loss.function}
	(Details and fitting parameters are shown in Table~\ref{tab:BCS.fitted}
	of Appendix~\ref{app:HFB}).
	(a) The dispersion contains 32 bands, with only four bands characterizing the 
	low-energy spectrum.
	(b) Zoom-in of (a) shows that bands at the fermi energy touch quadratically at the fermi level.
	(c) Surprisingly, the first bands near the fermi energy touch on a nodal line in momentum space 
	along $(\pm \pi, q_y)$ and $(q_x, \pm \pi)$, while (d) the second bands
	touch on the points $(\pm \pi, \pm \pi)$.
	}
	\label{fig:BCS.mean-field}
\end{figure*}
%

\subsection{Multipartite Hartree-Fock-Bogoliubov Ansatz}
\label{sec:BCS.fit}
%

We fit the optimized pair-product wave function $\left | \psi_{\sf pair} \right \rangle$ 
in Eq.~\eqref{eq:psi.pair}, 
characterized by the variational parameters
 $f_{ij}$ to the ground-state solution of Hartree-Fock-Bogoliubov (HFB)-type 
fermionic mean-field Hamiltonians~\cite{ring2004nuclear} to gain insight into the nature of the 
QSL. 
Note that $f_{ij}$ is the optimized variational parameters obtained after the VMC calculations 
of the STSL model with Eq.~\eqref{eq:Ham.J1J2}.
Here, for the fitting, we employ the HFB Hamiltonian on the STSL with 16 sites in the unit cell 
in the form
%
\begin{equation}
    \yy{{\mathscr H}_{\rm HFB}} = {\mathscr H}_{t} + {\mathscr H}_{\rm \Delta} \, ,	
\label{eq:BCS1}
\end{equation}
%
where the kinetic energy term is written as
%
\begin{equation}
	{\mathscr H}_{t} = \sum_{{\bf k}, \sigma} \sum_{\mu, \nu} 
					\left(\boldsymbol{\epsilon} (\bf k)\right)_{\mu, \nu}\ 
					  \hat{c}_{{\bf k} , \mu, \sigma}^{\dagger} \hat{c}_{{\bf k}, \nu, \sigma}  \, ,
\label{eq:Ham.BCS.hopping}
\end{equation}
%
and the superconducting BCS mean-field term as
%
\begin{align}
	 {\mathscr H}_{\rm \Delta}  =  \sum_{{\bf k}} \sum_{\mu, \nu} 
	 \left[  \left(\boldsymbol{\Delta} ({\bf k})\right)_{\mu, \nu} \ \hat{c}_{{\bf k}, \mu, \uparrow}^{\dagger} \hat{c}_{-{\bf k}, \nu, \downarrow}^{\dagger} 
			+ {\rm h.c.} \right]  \, .
\label{eq:Ham.BCS.pairing}
\end{align}
%
The fermionic creation $\hat{c}^{\dagger}$ and annihilation $\hat{c}$ operators contain the 
spin $\sigma$ and momentum ${\bf k}$ indices as well as the indices $\mu$ and $\nu$, which denote the 
16-site sublattice degrees of freedom in the unit cell of the STSL (see Fig.~\ref{fig:STSL.BCS}).

After diagonalizing the Hamiltonian in Eq.~\eqref{eq:BCS1},
using the explicit form given in Appendix~\ref{app:HFB}, we obtain the Bogoliubov quasiparticle 
eigenfunctions with coefficients ${\bf u}$ and ${\bf v}$, which are matrices with the 
sublattice site index $\mu$ and $\nu$, and the diagonalized band index $n$.
Here, these coefficient matrices satisfy the following HFB equations,
\eqsa{
\left(
\begin{array}{cc}
\boldsymbol{\epsilon}({\bf k}) & \boldsymbol{\Delta}({\bf k}) \\
\boldsymbol{\Delta}({\bf k}) & -\boldsymbol{\epsilon}({\bf k}) \\
\end{array}
\right)
\left(
\begin{array}{c}
{\bf u}({\bf k}) \\
{\bf v}({\bf k}) \\
\end{array}
\right)
&=&
{\bf E}_d ({\bf k}) \left(
\begin{array}{c}
{\bf u}({\bf k}) \\
{\bf v}({\bf k}) \\
\end{array}
\right)
,
\\
\left(
\begin{array}{cc}
\boldsymbol{\epsilon}({\bf k}) & \boldsymbol{\Delta}({\bf k}) \\
\boldsymbol{\Delta}({\bf k}) & -\boldsymbol{\epsilon}({\bf k}) \\
\end{array}
\right)
\left(
\begin{array}{c}
-{\bf v}({\bf k}) \\
{\bf u}({\bf k}) \\
\end{array}
\right)
&=&
-{\bf E}_d ({\bf k}) \left(
\begin{array}{c}
-{\bf v}({\bf k}) \\
{\bf u}({\bf k}) \\
\end{array}
\right)
,
\nonumber\\
}
where ${\bf E}_d ({\bf k})$ is a diagonal eigenvalue matrix whose $n$th diagonal component is the $n$th
positive eigenvalue $E_n({\bf k})$ of the band index $n$.
Then
%
\begin{equation}
	f_{\bf k}^{\rm HFB} \propto \sum_{\mu, \nu} \left[ \sum_n  \mi{({\bf u}({\bf k}))_{\mu, n}} \left( {\bf v}^{-1} ({\bf k}) \right)_{n,\nu} \right]  \, 
\label{eq:BCS.to.fij}
\end{equation}
%
represents the amplitude of the Cooper pairs in momentum space, 
which can be used to represent the singlet Cooper-pair-wave function in Eq.~\eqref{eq:psi.pair}
after Fourier transformation into real space. 
We obtain the best HFB representation of $\left | \psi_{\sf pair} \right \rangle$ by minimizing the loss function 
for $N_k$ momentum points
%
\begin{equation}
	\chi^2 = \frac{1}{N_k} \sum_{\bf k}^{N_k} \left( f_{\bf k}^{\rm HFB} - f_{\bf k}^{\rm mVMC} \right)^2	\, ,
\label{eq:BCS.fij.loss.function}
\end{equation}
%
between the HFB pair amplitude $f_{\bf k}^{\rm HFB}$ and $f_{\bf k}^{\rm mVMC}$ after optimization by mVMC.
Further details are given in Appendix~\ref{app:HFB}.

In Fig.~\ref{fig:BCS.mean-field}(a) we show the energy dispersion of the obtained HFB mean-field 
solution after minimizing the loss function $\chi^2$ in Eq.~\eqref{eq:BCS.fij.loss.function}.
We plot the energy eigenvalues along the irreducible wedge and obtain, as expected for the STSL 
model, 32 bands.
The low-energy spectrum is characterized by four bands [see zoom-in in Fig.~\ref{fig:BCS.mean-field}(b)],
which quadratically touch at the Fermi level at multiples of $(\pi, 0)$, $(0, \pi)$ and $(\pi, \pi)$.
However, the first band near the Fermi energy touches not at a point, 
but on a nodal line in momentum space, as visualized in 
Fig.~\ref{fig:BCS.mean-field}(c), while the second band, shown in Fig.~\ref{fig:BCS.mean-field}(d), 
touches quadratically at the singular points $(\pm \pi, \pm \pi)$.
This result is consistent with the closing of the excitation gap in the thermodynamic limit observed 
numerically with mVMC in Fig.~\ref{fig:Excitation.Energy.Scaling}.

As detailed in Appendix~\ref{app:DSF_BCS}, the dynamical spin structure factor $S^z ({\bf q},\omega)$ 
defined by the Fourier transform of the spin correlations [Eq.~(\ref{eq:Sqw.spinon})] 
also shows the gapless nature of the spin excitation as illustrated in Fig.~\ref{Fig:SQomega_spinon_MF}.  
In Appendix~\ref{app:DSF_BCS}, we demonstrate that the equal-time spin structure factor $S^z ({\bf q})$ 
defined in Eq.~(\ref{eq:Sq}) [or Eq.~(\ref{eq:Sq.spinon})] supports the power-law decay of the spin 
correlation $D_S^z({\bf r}_i-{\bf r}_j)=\langle S^z_i \ S^z_j \rangle$ defined in Eq.~(\ref{eq:Drr}) as 
$\sim C/r^{\alpha}$ with $\alpha\sim 3.0$, which is remarkably the same within the error bar with the 
mVMC results of the original Hamiltonian [Eq.\eqref{eq:Ham.iso}] shown in Fig.~\ref{fig:spin.correlations.real}(a).
%

%
\begin{figure}[t]
	\begin{center}
		\includegraphics[width=0.5\textwidth]{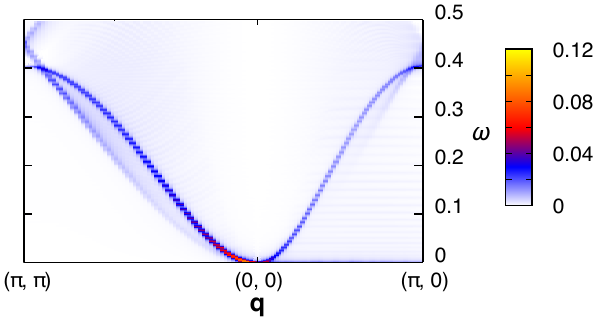}
	\end{center}
	\caption{
	Dynamical spin structure factor $S^z ({\bf q},\omega)$ [Eq.~\eqref{eq:Sqw.spinon}],
	as obtained from spinon mean-field theory.
	The lowest energy branch of the excitation spectrum is shown for $L=128$.
	The spectrum is essentially gapless and quadratic. Here, the broadening 
	factor $\delta = 0.0025$ is used.
	}
\label{Fig:SQomega_spinon_MF}
\end{figure}
%

\section{Summary and conclusions}						
\label{sec:summary.discussion}				

The quest for the ground state of the pyrochlore Heisenberg 
antiferromagnet (HAF) has a long history, yet a conclusive answer remained 
elusive, primarily due to the absence of accurate numerical techniques.
In this article, we present compelling evidence
showing that the ground state of the 
pyrochlore HAF is a quantum spin liquid (QSL), confined within a dimensionally reduced 
subspace. 
This QSL state emerges following the spontaneous breaking of lattice symmetries, 
including inversion, rotation, and translation, achieved by selecting an enlarged unit 
cell comprising 16 sites on a super-tetrahedron.
Our state-of-the-art VMC technique 
reveals dominant correlations within a 2D bond network embedded inside
the 3D pyrochlore lattice, with negligible inter-plane correlations.

To support our findings, we conduct a scaling analysis for a low-energy 
effective model on the super-tetrahedron square lattice (STSL), confirming a
\mbox{$1/r^{\alpha}$}
algebraic decay of spin and singlet correlations
in the ground state with the exponent \mbox{$\alpha\sim 3$}.
Correlations of excited states also decay algebraically, although with a smaller power, 
suggesting the presence of a gapless ground state in the thermodynamic limit.
We validate the gapless nature through numerical finite size scaling of the excitation gap.

To gain deeper insights into the nature of the QSL wave function,
we fit our variational parameters $f_{ij}$ by introducing a general 16-site, multipartite
Hartree-Fock-Bogoliubov (HFB) mean-field Hamiltonian of spinons. 
Our analysis reveals that quadratic bands touch each other at zero energy
in the spinon dispersion, albeit not at a singular point but along
symmetry lines in momentum space.
The spin structure factor based on this spinon HFB 
mean-field dispersion demonstrates a 
gapless, quadratic band dispersion and confirms the \yy{$1/r^{\alpha}$} power law decay of 
spin correlations with $\alpha\sim 3$.

In conclusion, our study of the pyrochlore HAF highlights the interplay 
between nature's preference for less entangled states and the role of 
frustration in generating exotic states of matter.  
Despite historical expectations of a 3D QSL, nature finds a unique compromise
by forming a state where large entanglement persists within a 2D subsystem embedded 
within the 3D lattice.
This unexpected dimensional reduction underscores the remarkable diversity of 
solutions that nature can discover to alleviate frustration.

The persistence of strong correlations within a 2D subsystem suggests the 
possibility of the QSL
being a $\mathbb{Z}_2$ spin liquid similar to examples studied in
several 2D frustrated magnets~\cite{Kitaev2006a,Nomura2021}.
Although definitive proof awaits future investigations, it may be made possible by
a combination of the present results and the symmetry classification of gauge 
degrees of freedom~\cite{Wen-Book, Schneider2022} or effective lattice gauge theories.

As demonstrated in Fig.~\ref{fig:PhaseDiagram.DM},
the QSL stays robust against perturbations, which holds significant implications 
for materials like iridate and molybdate 
pyrochlores~\cite{Gardner2010,PhysRevLett.113.117201,PhysRevMaterials.1.071201}, 
suggesting exciting directions for further exploration.

%
%
\section{Acknowledgements}

\rp{
\yy{RP} is pleased to acknowledge helpful
discussions
with Kota Ido, Tsuyoshi Okubo, Nic Shannon, 
RuQing G. Xu and  Han Yan. 
\yy{YY and RP thank Yong Baek Kim for insightful comments.}
This work was supported by MEXT as “Program for Promoting Researches on the Supercomputer 
Fugaku” (Basic Science for Emergence and Functionality in 
Quantum Matter -- Innovative Strongly-Correlated Electron Science by Integration of “Fugaku” and 
Frontier Experiments -- ) and used computational resources of supercomputer Fugaku provided 
by the RIKEN Center for Computational Science (Project ID: hp200132, No. hp210163, and No. hp220166).
Part of the computation was done using the HPC facilities provided by the 
Supercomputer Center of the Institute for Solid State Physics, the University of Tokyo,
\yy{and computational resources of the supercomputer Fugaku provided
by R-CCS through the HPCI System Research Project (Project ID: hp230169 \rp{and hp120281}) and the 
``Program for Promoting Researches on the Supercomputer Fugaku'' (\mbox{JPMXP1020230410} \rp{and 
\mbox{JPMXP1020230411})}.}
\rp{RP acknowledges the financial support from the JSPS KAKENHI Grant 
No. JP19H05825 (``Quantum Liquid Crystals'').}
\yy{YY acknowledges the support from MANA and World Premier International Research Center Initiative (WPI), MEXT, Japan.}
\mi{MI acknowledges the financial support from the JSPS KAKENHI Grant No. 22H05111 
(``Foundation of Machine Learning Physics'') and 
22H05114 (``Frontiers of Condensed Matter Physics Pioneered by Neural Network'')}.}

%
%
\appendix									

\section{Methods}			
\label{app:Method}

The method used in this paper is called many-variable variational Monte Carlo (mVMC), 
and definitions of relevant 
physical quantities are outlined here, with additional details given in 
Secs. I and III of SM~\cite{SM}.

\subsection{Many-variable variational Monte Carlo}	
\label{sec:variational.MC}	
%

In the present paper, we apply the state-of-the-art variational Monte Carlo 
method~\cite{Tahara2008,Misawa2019} by employing the code of open-source 
software mVMC, which generates variational ground-state 
wave functions $|\Psi\rangle$ after optimization of a large number of variational parameters.   
We express our variational wave function in the form  
%
\begin{equation}
 | \Psi \rangle =  {\mathcal L} {\mathcal P}	\left | \psi_{\sf pair} \right \rangle 	\, ,
\label{eq:var.wave.function}
\end{equation}
%
with the correlation factor ${\mathcal P}$
and the quantum-number projector ${\mathcal L}$. 
In Eq.~\eqref{eq:var.wave.function} we introduce the pair-product form of the many-body 
wave function in their real-space representation as Pfaffian matrix
%
\begin{equation}
	\left | \psi_{\sf pair} \right \rangle  = 	\left( \sum_{ij}^{N_s} f_{ij} 
						c^{\dagger}_{i \uparrow} c^{\dagger}_{j \downarrow}  \right)^{N_e/2}
						\left | 0 \right \rangle \, ,
\label{eq:psi.pair}
\end{equation}
%
an extension to the general Slater determinant.
The amplitude $f_{ij}$ of an electron pair with opposite spin 
serves as the variational parameter which will be optimized.
By introducing the artificial neural-network projector ${\mathcal N}$~\cite{Nomura2017} 
and a first-order power Lanczos step~\cite{Heeb1993} to Eq.~(\ref{eq:var.wave.function}) as  
\eqsa{
| \Psi \rangle =  (1+\alpha_{\rm L}\mathscr{H}){\mathcal L}
 {\mathcal N}  {\mathcal P}	 \left | \psi_{\sf pair} \right \rangle 	\, ,
\label{eq:var.wave.function.RBM.Lanczos}
}
we improve the accuracy of the variational wave function.
Here, the restricted Boltzmann machine (RBM)~\cite{smolensky1986information} 
is used for $\mathcal{N}$, and the parameter $\alpha_{\rm L}$ is optimized by minimizing
the energy expectation value after the other variational parameters in $\mathcal{N}$ and $\left | \psi_{\sf pair} \right \rangle$
are optimized.
In the following sections, the simplest variational wave function 
$| \Psi \rangle =  {\mathcal L} {\mathcal P}\left | \psi_{\sf pair} \right \rangle$
is called the mVMC wave function while
$| \Psi \rangle =  {\mathcal L}
 {\mathcal N}  {\mathcal P}\left | \psi_{\sf pair} \right \rangle$
 $\left(| \Psi \rangle =  (1+\alpha_{\rm L}\mathscr{H}){\mathcal L} {\mathcal P} \left | \psi_{\sf pair} \right \rangle\right)$
is called the mVMC-RBM (mVMC/Lanczos) wave function.
The most accurate variational wave function defined in Eq.~(\ref{eq:var.wave.function.RBM.Lanczos}) is denoted as
the mVMC-RBM/Lanczos wave function.
Further details of mVMC are available in SM~\cite{SM}.

Our target is the quantum $S=1/2$ Heisenberg antiferromagnet on the pyrochlore lattice 
defined by the Hamiltonian in Eq.~\eqref{eq:Ham.iso}.
We have exploited the ground states of finite-size systems up to 1024 lattice sites 
for the original 3D lattice, and for
the 2D effective STSL model corresponding to $\sim 8\times 10^3$ sites of the 3D system, 
with periodic boundary conditions to estimate the thermodynamic limit of physical quantities 
after the size extrapolation.

The pair-product wave function defined in Eq.~(\ref{eq:psi.pair})
only contains the anti-parallel spin pairs, which is used at $D/J = 0$. 
However, it is necessary to use pair-product wave functions with both
anti-parallel and parallel spin pairs for nonzero $D$ (see SM~\cite{SM} for the details of the parallel spin pairs).
We respect the full cubic symmetry of the pyrochlore lattice by the quantum number projection~\cite{Mizusaki2004},
and simulate lattice sizes up to 1024 spins to extrapolate finite-size results to the thermodynamic limit.

The accuracy of the mVMC method is benchmarked in various models in comparison to other 
methods (see Appendix~\ref{app:pyro3D.energy}). In the present case of the Heisenberg 
model on the pyrochlore lattice, the better accuracy and performance has been confirmed in 
comparison to the DMRG result. See also SM Sec. V~\cite{SM} for more details.

\subsection{Correlation functions}

To understand the nature of the wave functions,
we calculate spin-spin correlations in momentum 
space for the $z$-components of the equal-time structure factor $S^z({\bf q})$
%
%
\begin{equation}
	S^z({\bf q}) = \frac{1}{N_s} \sum_{i,j} e^{i {\bf q} ({\bf r}_i - {\bf r}_j)} D_S^z({\bf r}_i-{\bf r}_j), 
 \label{eq:Sq}
 \end{equation}
 \begin{equation}
D_S^z({\bf r}_i-{\bf r}_j) = 		  \langle S^z_i \ S^z_j \rangle 	\, ,
\label{eq:Drr}
\end{equation}
%
where ${\bf r}_i$ is the position of the site $i$ and ${\bf q}$ is the momentum.
We also study the correlations of singlet bonds to understand the nature of 
the nonmagnetic phase. 
Here, the strength of the singlet bond is measured by
%
\begin{align}
	B_m 			  &= {\bf S}_{m_1} \cdot {\bf S}_{m_2}  	\, ,
\label{eq:B_corr}
\end{align}
%
where ${\bf S}_{m_1}$ and ${\bf S}_{m_2}$ are the two vectors of the spin operators
on sites $m_1$ and $m_2$, respectively, which are connected via bond $m$.
The correlation of a singlet bond can be measured by
%
\begin{equation}
	O({\bf q}) 	= \frac{1}{N_b} \sum_{m,n} e^{i {\bf q} ({\bf \tilde{r}}_m - {\bf \tilde{r}}_n)}
			      \langle B_m \ B_n \rangle 		 
\label{eq:Oq.with.static}   \, ,
\end{equation}
%
where the total number of bonds in the pyrochlore lattice is $N_b = 3 N_{s}$, 
with $N_{s}$ being the number of spins,
and ${\bf \tilde{r}}_m = ({\bf r}_{m_1} + {\bf r}_{m_2})/2$  is the vector to the center of the bond $m$.

\section{Accuracy of the present variational wave function and 
comparison to previous work}
\label{app:pyro3D.energy}

\begin{figure}[tb]
	\centering
	\includegraphics[width=0.45\textwidth]{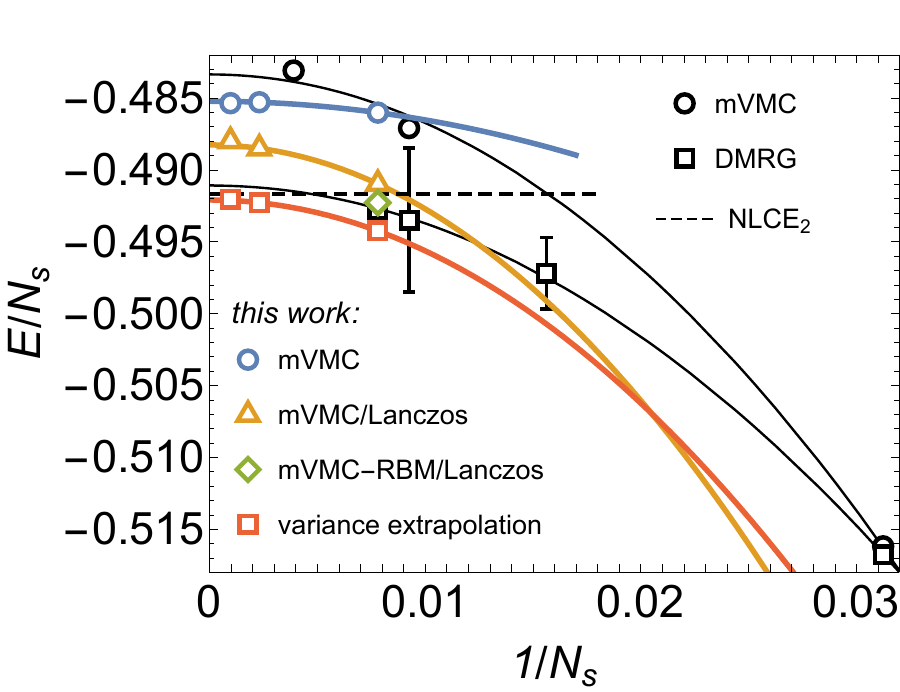}
	\llap{ \footnotesize\parbox[b]{0.9cm}{\cite{Astrakhantsev2021} \\\rule{0ex}{5.1cm}} }
	\llap{ \footnotesize\parbox[b]{1.05cm}{\cite{Hagymasi2021} \\\rule{0ex}{4.62cm}} }
	\llap{ \footnotesize\parbox[b]{1.2cm}{\cite{Schafer2023} \\\rule{0ex}{4.08cm}} }
	\caption{
	Comparison of the normalized ground-state energy $E/N_s$ of ${\mathscr H}$ in 
        Eq.~\eqref{eq:Ham.iso}, as function of inverse system size $1/N_s$. 
	Here, we set the Heisenberg exchange coupling to be the unit of the energy: $J=1$.
  	There are three categories of the numerical results:
  	strictly variational ground-state energy, extrapolated energy from the variational ones,
 	and energy by asymptotic series expansions. 
	For details of the latter two, see SM Sec.V~\cite{SM}.
	The data for mVMC (blue circles), mVMC with the first-step Lanczos (orange triangles),
  	and the RBM and Lanczos (green diamonds), are strictly variational results obtained in the present work.
	We also plot the case of the latter two categories, where the data by the variance 
	extrapolation from the present mVMC results are shown by red squares. 
	The present results are compared to variational mVMC results from 
	Astrakhantsev {\it et al.} [\onlinecite{Astrakhantsev2021}] (black circles),
 	 the results of 3D DMRG results from Hagym\'asi {\it et al.} after bond-dimension 
	 extrapolations [\onlinecite{Hagymasi2021}] (black squares), and the numerical 
	 linked cluster expansion of order two (NLCE$_2$) by Sch{\"a}fer and Placke 
	 {\it et al.} [\onlinecite{Schafer2023}] (black dashed line).
	Data with size dependency have been fitted with a quadratic function (solid curves)
	to give energy estimates in the thermodynamic limit ($1/N_s \to 0$).
	Our best energy estimate from variance extrapolation in the thermodynamic limit
	is $(1/N_s) E_0|_{N_s \to \infty} = -0.4921(4)$.
    	Explicit numerical values are given in SM, Sec.V~\cite{SM}.
	}
	\label{fig:scaling.GSenergy}
\end{figure}

In Fig.~\ref{fig:scaling.GSenergy}, we compare the variational ground-state energy per site, $\mi{E/N_s}$, 
for the Heisenberg Hamiltonian in Eq.~\eqref{eq:Ham.iso} with the previous cutting edge 
studies by mVMC~\cite{Astrakhantsev2021}, DMRG~\cite{Hagymasi2021}, and numerical linked 
cluster expansion~\cite{Schafer2023} methods with the system size dependence as a function 
of $1/N_s$.

When we compare the present results with the previous ones, we note that there are three 
categories of the numerical results:
strictly variational ground-state energy, extrapolated energy from the variational ones,
and energy by asymptotic series expansions.
While the present mVMC results (with or without the restricted Boltzmann machine projection and
the first Lanczos step) and the results by Ref.~[\onlinecite{Astrakhantsev2021}] are strictly variational,
the main results by the 3D DMRG reported in Ref.~[\onlinecite{Hagymasi2021}] are obtained after 
bond-dimension extrapolations, which does not necessarily satisfy the variational principle.
Here, we also show results after variance extrapolation, $\mi{E_0/N_s}$, calculated 
in the procedure described in SM~\cite{SM}, together with the estimate in the 
thermodynamic limit for reference.
We show results of unprecedentedly large systems as well, which certainly makes the 
extrapolation to the thermodynamic limit easier.

The accuracy of the ground-state wave function is measured from the strict variational 
estimate without the variance extrapolation, where the lower energy is better.
The benchmark results for other categories which do not necessarily
follow the strict variational principles are detailed in SM Sec.V~\cite{SM}.
%
%

Among the strictly variational results on the pyrochlore Heisenberg antiferromagnet
in the literature, as summarized in SM Tables SI-SIII~\cite{SM}, we obtained a series of the strictly 
variational energy by mVMC, mVMC-RBM, mVMC/Lanczos, and mVMC-RBM/Lanczos with
spin-parity projection~\cite{Mizusaki2004}.
The best variational energy, $E/N_s = -0.49229(7)$,
is given by the spin-parity mVMC-RBM/Lanczos wave function
while the best variational energy by the 3D DMRG at the finite bond dimension
for $N_s = 128$ is nearly $- 0.49220$, which is 
comparable but slightly higher
than the best variational energy by the present study
at the same size and same boundary condition.
For the benchmark comparison for the latter two categories see SM Sec. V~\cite{SM}.

\section{Definition of Dzyaloshinskii-Moriya interactions}			
\label{app:DM.def}
%

\begin{figure}[b!]
	\centering
	\includegraphics[width=0.55\columnwidth]{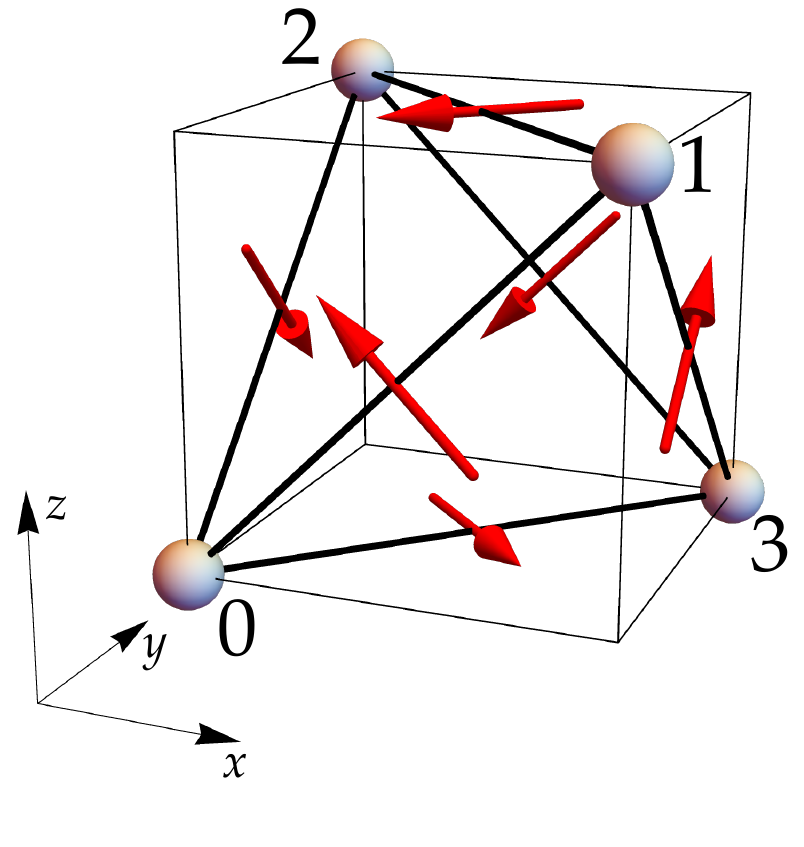}
        \\
	\def\arraystretch{2.5}
	\begin{tabular}{ l  @{\hspace{1cm}}  l} 
		\hhline{==}		
		${\bf e}_{01} = \frac{1}{\sqrt{2}} \left(-1,0,1 \right)$     &  ${\bf e}_{23} = \frac{1}{\sqrt{2}} \left(-1,0,-1 \right)$	\\
		${\bf e}_{02} = \frac{1}{\sqrt{2}} \left(0,1,-1 \right)$	&  ${\bf e}_{13} = \frac{1}{\sqrt{2}} \left(0,1,1 \right)$	\\
		${\bf e}_{03} = \frac{1}{\sqrt{2}} \left(1,-1,0 \right)$	&  ${\bf e}_{12} = \frac{1}{\sqrt{2}} \left(-1,-1,0 \right)$	\\
		\hhline{==}
	\end{tabular}
  	\caption{
	Directions of Dzyaloshinskii–Moriya (DM) vectors [see Eq.~\eqref{eq:DM.def}] 
        on one tetrahedron of the pyrochlore lattice.
        DM vectors are chosen to respect the cross product 
        ${\bf D}_{ij}({\bf S}_i \times {\bf S}_j)$ for site indices
        $ j > i$, and are chosen in their ``indirect'' definition~\cite{Elhajal2005, Canals2008},
        with explicit definitions shown in the lower table.
	}
	\label{fig:DM}
\end{figure}
%

In Fig.~\ref{fig:PhaseDiagram.DM}, we have shown the phase diagram of 
${\mathscr H}_{\sf DM}$ [see Eq.~\eqref{eq:Ham}]
as function of spin-anisotropic Dzyaloshinskii–Moriya (DM) interactions
%
\begin{equation}
   {\bf D}_{ij} = D \ {\bf e}_{ij}      \, ,
\label{eq:DM.def}
\end{equation}
%
with $D$ being the DM interaction strength, and ${\bf e}_{ij}$ their 
unit-vectors defined on bonds between sites $i$ and $j$.
The pyrochlore lattice allows for only two types of DM interactions, 
which are referred to as ``direct'' and ``indirect'' 
cases~\cite{Elhajal2005, Canals2008}.
Here, we used the indirect case with explicit values given for a 
single tetrahedron in Fig.~\ref{fig:DM}.
All remaining DM vectors for the whole pyrochlore lattice are 
uniquely determined by symmetry.

%
\section{Choice of trial wave functions} 			
\label{app:initial.trial}  
%

\subsection{Comparison of initial wave functions} 
%

\begin{figure}[t]
	\centering
	\includegraphics[width=0.48 \textwidth]{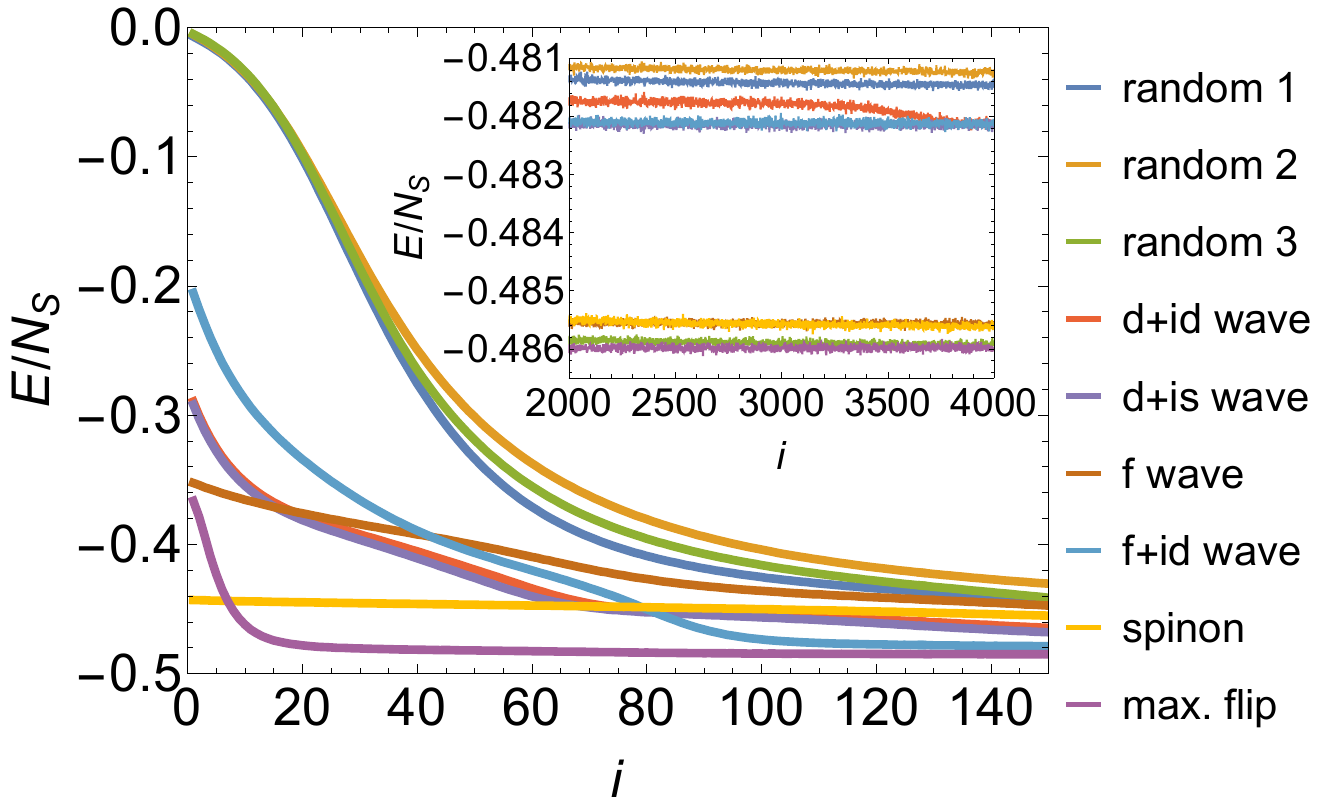}
	\caption{
	Optimization processes of the variational energy for ${\mathscr H}$, in 
        Eq.~\eqref{eq:Ham.iso}, shown as functions 
        of optimization steps $i$, for a $L=2$, $N_s=128$ site cubic cluster with 
        periodic boundary conditions.
        Simulations were initiated from different trial wave functions, 
        namely three random initial states, Gutzwiller projected HFB mean-field states with 
        pairing symmetries of $d+is$, $d+id$, $f$, and $f+id$ wave, 
        the monopole flux spinon mean field (spinon MF) ansatz,
        and the maximally flippable dimer state (max. flip)
        on super-tetrahedra  (see Fig.~\ref{fig:largeN.trial}).
        The inset shows the optimization for $2000 \leq i \leq 4000$.
        The optimization, initiated from the max. flip state 
        gave the fastest convergence to the lowest variational energy state 
        in the mVMC wave function.  
        Simulation details and explicit forms of HBF and spinon mean-field states 
        are given in SM~\cite{SM}). 
        }
	\label{fig:mVMC.opt.trial}
\end{figure}
%

%
\begin{figure}[t]
	\centering
	\includegraphics[width=0.48 \textwidth]{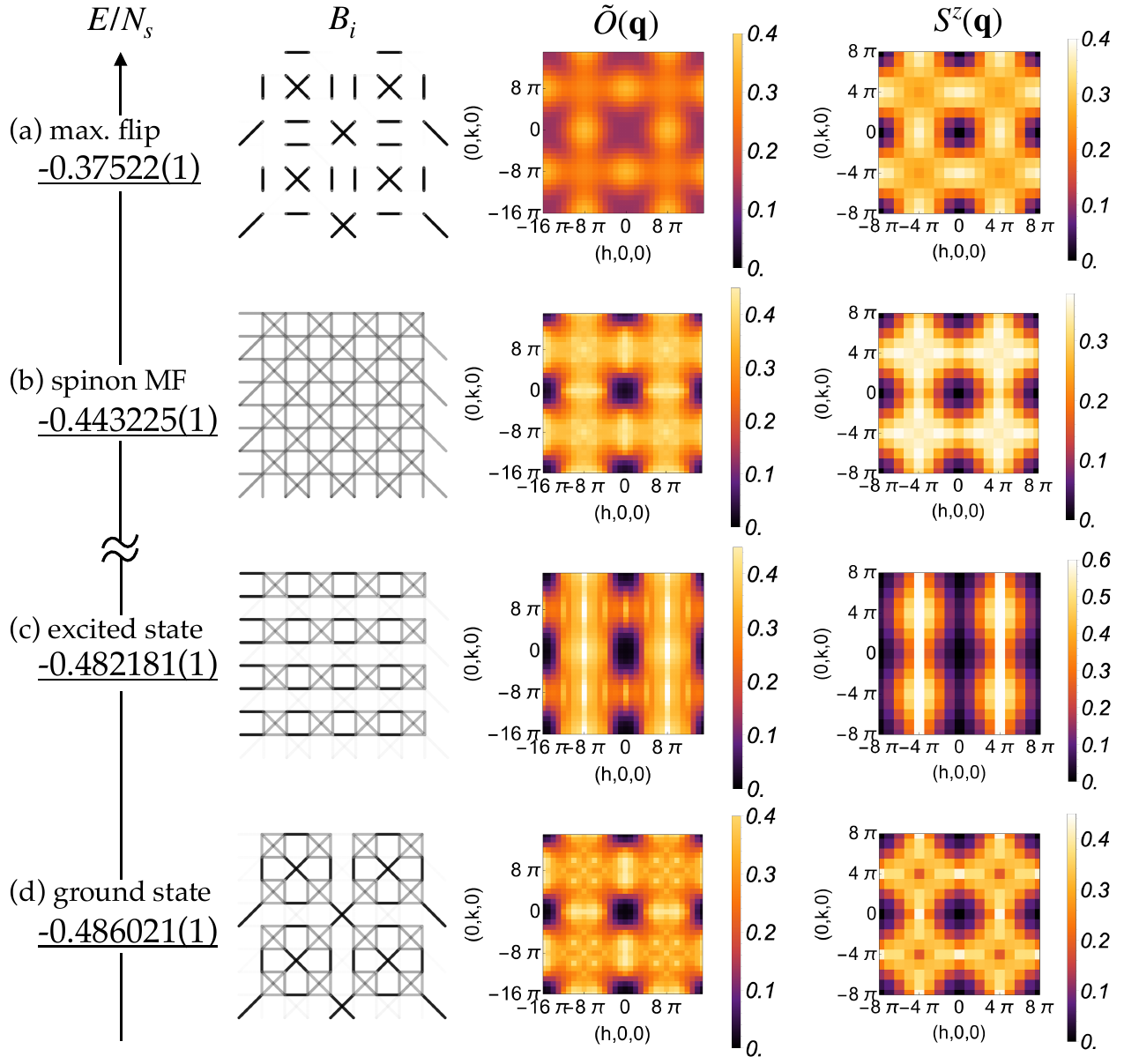}
	\caption{
        Comparison between properties of typical initial and optimized wave functions from  
        Fig.~\ref{fig:mVMC.opt.trial}.
        The columns from the left to the right show the normalized energy, $E/N_s$, 
        the top view of real-space singlet arrangement on bonds, $B_m$ [Eq.~\eqref{eq:B_corr}]
       	(black = strong, white = weak),
        the singlet correlation function in momentum space,
        $\tilde{O}({\bf q})$ [Eq.~\eqref{eq:Oq.no.static}], and 
        the equal-time spin structure factor $S^z({\bf q})$ [Eq.~\eqref{eq:Sq}].
        Each quantity is shown for 
        (a) the initial maximally flippable dimer state (max. flip.),
        (b) the initial monopole flux spinon mean-field (MF) ansatz,
        (c) the optimized HFB mean-field ($d+id$ wave) state, and 
        (d) the lowest-energy state by the mVMC wave function, obtained after 
        optimization of the max. flip. state.
	}
	\label{fig:init.opt.observables}
\end{figure}
%

Variational Monte Carlo methods are based on energy-minimization techniques, where the optimized 
wave function is ideally desirable not to depend on the choice of the initial wave function. 
However, in practice, it could depend if competing states are separated by a large energy barrier in the Hilbert space.
In this case, to avoid getting trapped within a local energy minimum, the optimization should 
start from different choices of the initial trial wave function to reach the global minimum after 
comparing the optimized energy with each other.
A better choice of the initial state also helps to foster our intuition about the nature of the true ground state.

In this section, we investigate the impact of the initial guess on the quality of
optimized wave functions. 
This helps to narrow down the choice of the initial wave function for our comprehensive 
study, and to save computational cost. 
We have performed an initial screening through a simple optimization, omitting 
elaborated optimization with RBM projection and Lanczos steps.

In Fig.~\ref{fig:mVMC.opt.trial} we show variational energies of ${\mathscr H}$ 
in Eq.~\eqref{eq:Ham.iso} as function of
optimization steps $i$, obtained for a $L=2$, $N_s=128$ site cubic cluster 
on the pyrochlore lattice with periodic boundary conditions
(see simulation details in SM~\cite{SM}). 
We compare the optimization processes for various initial 
trial-wave functions, namely three different random initial states, 
a selection of Gutzwiller projected Hartree-Fock-Bogoliubov (HFB)
mean-field states, a state from the monopole flux spinon 
mean-field ansatz~\cite{Burnell2009}, and the maximally flippable dimer 
state on super-tetrahedra (see Appendix~\ref{app:max.flip.state}).
Explicit initial and optimized energies with singlet and spin observables for
typical choices of variational parameters are shown in Fig.~\ref{fig:init.opt.observables}.
We find that the optimization initiated from the ``random 3" and maximally flippable dimer state
reaches the same lowest-energy state, indicating their stable convergence to the same global minimum. 
Additionally, optimization initiated from the maximally flippable dimer state
gives the fastest convergence to the lowest variational energy state for the 
mVMC wave function, $| \Psi \rangle =  {\mathcal L} {\mathcal P}\left | \psi_{\sf pair} \right \rangle$.
Therefore, although we carefully consider other choices for the initial 
wave functions, we prioritize the maximally flippable dimer state as the initial wave function 
for all simulations within the available computer resources in our study, unless stated otherwise.
Below, we show further details of the maximally flippable dimer state and show 
the cases of other initial states including the random initial states in SM~\cite{SM}.

\subsection{Maximally flippable dimers on super-tetrahedra} 			
\label{app:max.flip.state}  
%

\begin{figure*}[t]
	\centering
	\subfloat[][ view along ${(111)}$ axis ]{
		\includegraphics[height=0.35\textheight]{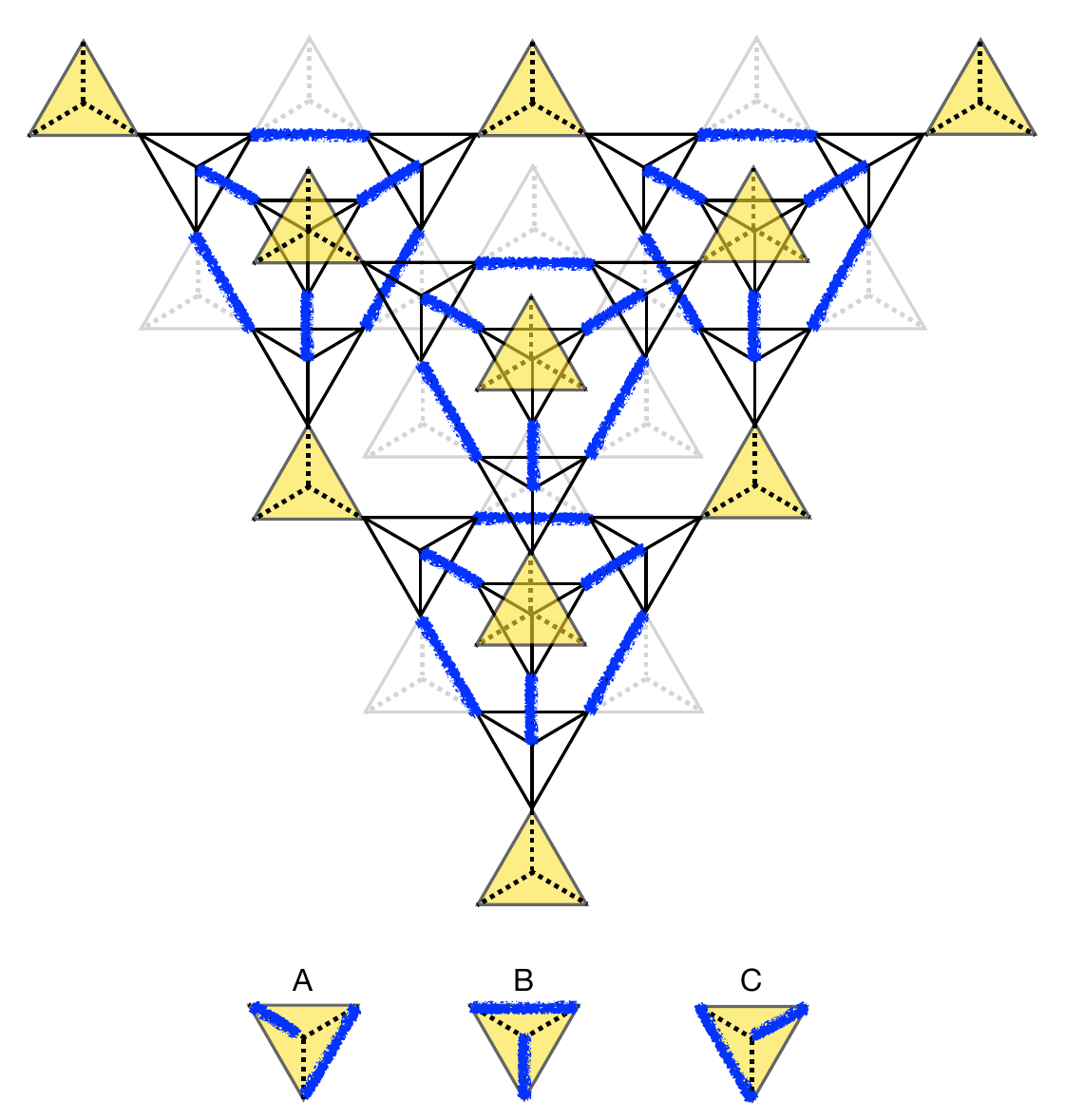}} 
        \hspace{1cm}
	\subfloat[][ view along ${(001)}$ axis ]{
		\includegraphics[height=0.35\textheight]{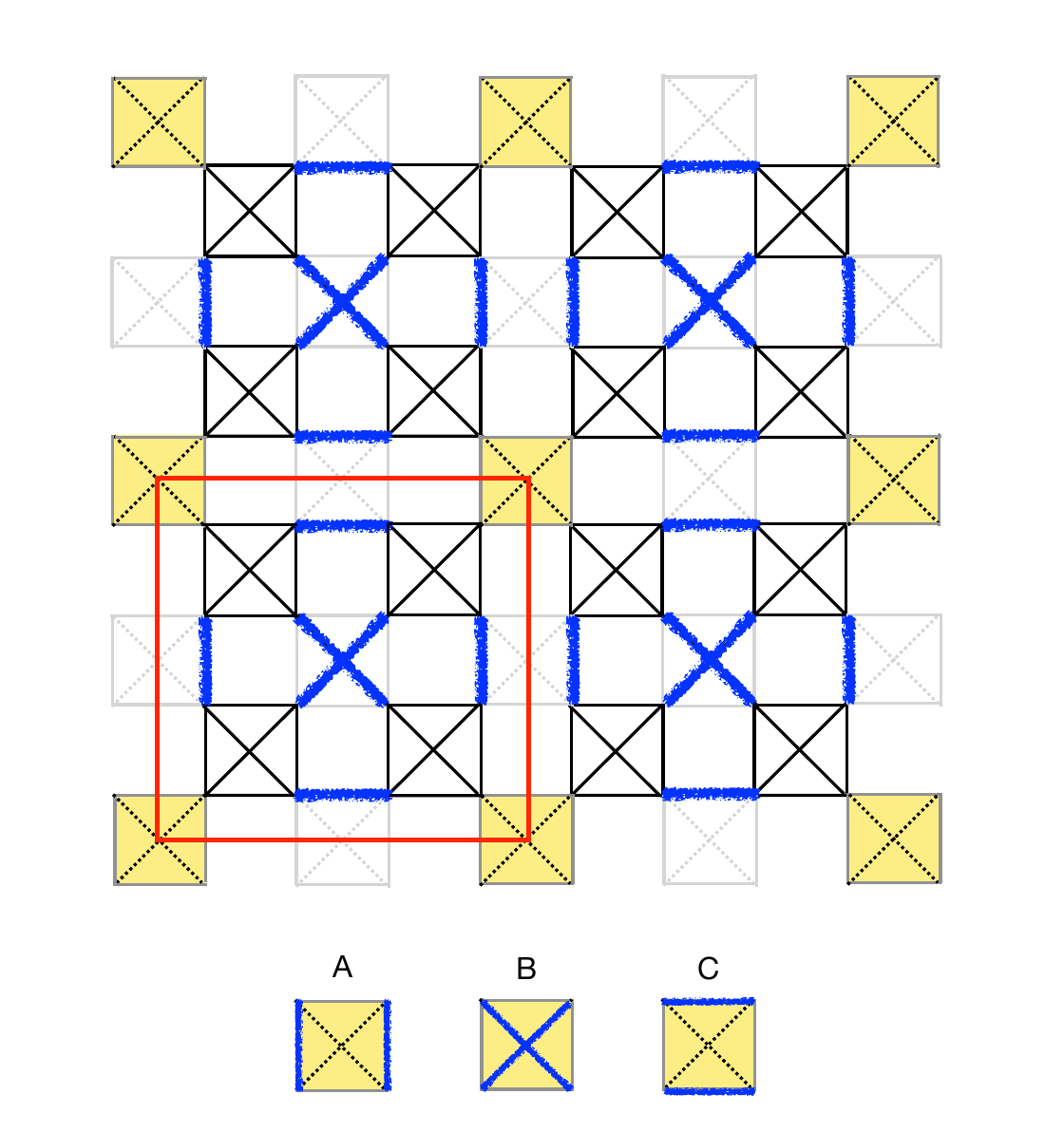}} 	
	\caption{
	Trial wave function for the maximally flippable dimer state on super-tetrahedra. 
        We arrange dimers on blue bonds, connecting site $i$ and $j$,
        by setting $f_{ij} = f_{ji} = 1$ [see Eq.~\eqref{eq:psi.pair}], with an additional 
        overall random offset $-0.1< \epsilon < 0.1$ for all $f_{ij}$.
	(a) View along the $(111)$ axis, perpendicular to the kagome planes.
	(b) View along the $(001)$ axis, perpendicular to the checkerboard planes, with 
	enlarged unit cell shown in red. 
	Super-tetrahedra are connected via ``inter-tetrahedra'' (shaded in yellow), which allow 
	for 3 possible dimer arrangements: A, B and C, as shown in the bottom of (a) and (b).
	}
	\label{fig:largeN.trial}
\end{figure*}
%

The maximally flippable dimer state has connections to the solution of the large-$N$ 
quantum dimer model on the pyrochlore lattice, as presented by Moessner {\it et al.} [\onlinecite{Moessner2006}].
The ground state in this model breaks inversion and translations by covering 
one sublattice of tetrahedra with hardcore dimers in their maximally flippable 
configuration.
As shown in Fig.~\ref{fig:largeN.trial}, a maximally flippable state arranges 
six dimers (colored as blue bonds) such that they cover the hexagonal plaquettes
within the four kagome planes in the pyrochlore lattice.
This dimer arrangement involves 16 sites (4 tetrahedra) of 
the cubic unit cell, which we shall call a ``super-tetrahedron'', in analogy to 
the terminology used in Ref.~[\onlinecite{Moessner2006}].
Dimers on every super-tetrahedron are connected by a four-site 
``inter-tetrahedron'', which are allowed to host three possible dimer coverings,
A, B and C, as shaded in yellow in Fig.~\ref{fig:largeN.trial}.
Moessner {\it et al.} provided an interesting scenario for hard-core dimer coverings on 
the full 3D lattice, by extending the concept of maximally flippable dimers of 
length exceeding the size of a single super-tetrahedron. 
To maximize the number of flippable loops, inter-tetrahedra A, B and C would arrange 
aperiodically throughout the whole lattice,
breaking inversion, translation and rotation symmetries of the lattice.

As we observed from the optimization of the random initial ansatz,
the ground state of the quantum $S$=1/2 nearest-neighbor antiferromagnet in 
Eq.~\eqref{eq:Ham.iso}  seems to also arrange strong 
singlets in their maximally flippable configuration on super-tetrahedra (see SM~\cite{SM}), 
consistently with the result from the large-$N$ quantum dimer model~\cite{Moessner2006}.
However, singlets on inter-tetrahedra do not select states A, B or C to form longer loops 
of maximally flippable dimers extending over a super-tetrahedron by the large-$N$ dimer state itself. 
In reality after thorough optimization of the mVMC wave function, we find that the 
lowest energy state statically selects either state A, B or C by globally correlating all 
inter-tetrahedra [see dark blue bonds in Fig.~\ref{fig:ground.state.3Dpyrochlore}(c)--(e)] instead 
of resonating A, B and C as a locally entangled state by linear combination.
This eventually results in the formation of a decoupled, two-dimensional layered network of singlets 
in the ground state (see detailed discussion in Sec.~\ref{sec:dim.reduction}).
mVMC wave functions can accommodate any kind of entangled singlets by the 
structure of the variational parameters $f_{ij}$, as, for example, shown for the spin liquid without 
symmetry breaking on the \mbox{$J_1$-$J_2$} square lattice~\cite{Nomura2021}. 
However, the optimization on the pyrochlore lattice yields the symmetry broken 
state at this stage.

Motivated by those observations, we prepare the maximally flippable dimer state
for our pair-wave function in Eq.~\eqref{eq:psi.pair} as an initial state by setting 
%
\begin{equation}
	f_{ij} = f_{ji} = 1  \, ,
\end{equation}
%
for pairs of sites $i, j$ on the blue bonds in Fig.~\ref{fig:largeN.trial}.
To allow more flexibility for mVMC to optimize variational parameters we introduced an 
additional overall random offset $-0.1< \epsilon < 0.1$ for all $f_{ij}$.
Since the number of ground states in the dimensionally reduced ground state 
is countable and equivalent to each other (see Sec.~\ref{sec:sym.breaking}), we choose, without loosing 
generality, singlets on inter-tetrahedra to be in the state B of 
Fig.~\ref{fig:largeN.trial}, forming a 2D network of singlets in the $xy$ plane.

Our energy estimate for the initial Gutzwiller projected maximally flippable state is
$E/N_s = -0.37475(1)$, with its corresponding singlet covering on 
one tetrahedral sublattice shown in Fig.~\ref{fig:init.opt.observables}(a).  
The singlet correlations $\tilde{O}({\bf q})$ [Eq.~\eqref{eq:Oq.no.static}] 
and spin correlations $S^z({\bf q})$ [Eq.~\eqref{eq:Sq}] in momentum 
space are very diffuse without any singular structure.

The energy optimization from this initial state with mVMC, as shown in 
Fig.~\ref{fig:mVMC.opt.trial}, shows a quick convergence to the lowest-energy state with
$E/N_s \approx -0.486021(1)$.
Even though optimized for complex $f_{ij}$ parameters, the energy 
matches within error bars to the energy obtained with real $f_{ij}$ 
parameters (see Table~SI of SM~\cite{SM}).
Figure~\ref{fig:init.opt.observables}(d) shows the arrangement of singlets in this ground 
state, which preserves the initial covering of strong singlets on one tetrahedral sublattice,
from the max. flip. state in Fig.~\ref{fig:init.opt.observables}(a).
The energy could be reduced by introducing  weak singlets on the other tetrahedral sublattice,
which introduces more structure in the $\tilde{O}({\bf q})$ and $S^z({\bf q})$ as 
compared to Fig.~\ref{fig:init.opt.observables}(a).

The lowest energy state after the optimization of the variational parameters is obtained from 
the initial wave function of the maximally flippable dimer state. 
However, the state optimized from the random initial condition (see ``random 3'' in Fig.~\ref{fig:mVMC.opt.trial})
exhibits nearly the same energy,  as shown in SM Sec.V (Tables SI-SIII). 
After the variance extrapolation, the max. flip. initial state converges to the energy 
$E_0/N_s=-0.49434(5)$ while $E_0/N_s=-0.4943(1)$ for the random initial state for $N_s=128$. 
For $N_s=432$, they are $E_0/N_s=-0.4924(2)$ and $E_0/N_s=-0.4923(2)$, respectively. 
Physical properties are also essentially the same. 
Therefore, we may start from either of the initial wave function to reach the global minimum.
However, the max. flip. dimer initial state  gives the fastest convergence to the lowest energy
state. 
Therefore, we employ the max. flip. dimer as the trial wave function for all simulation 
results, except where it is stated otherwise.

\section{Spin and singlet correlations on the pyrochlore lattice}	
\label{sec:3D.pyro.correlations}

\begin{figure*}[t]
	\centering
	\includegraphics[width=0.98\textwidth]{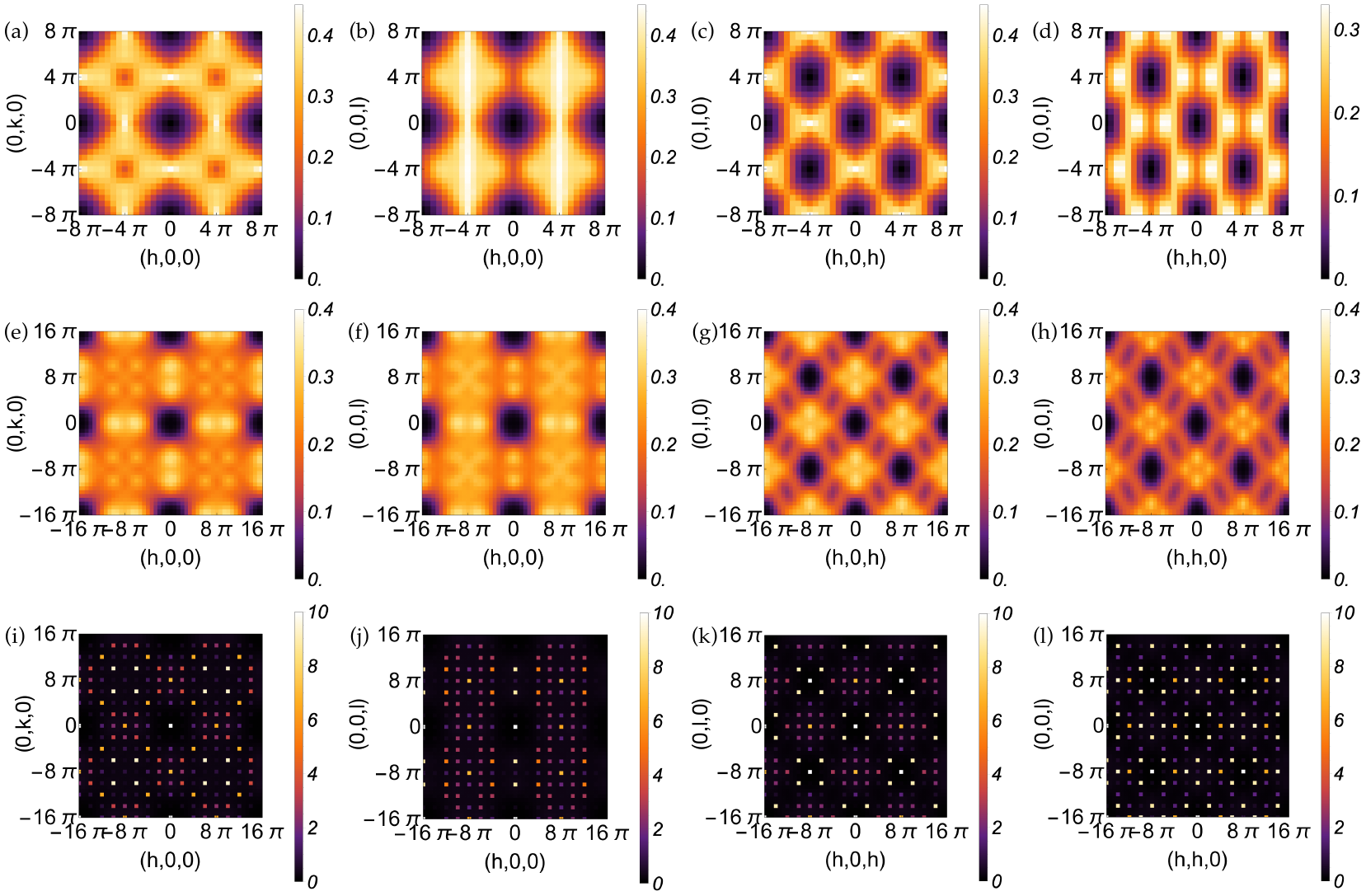}
  	\caption{
	Equal-time structure factors for spins and singlets in the ground state of 
	${\mathscr H}$ [see Eq.~\eqref{eq:Ham.iso}] on the pyrochlore lattice for a 
	$L=3$, $N=432$ site cubic cluster under periodic boundary conditions. 
	(a)--(d) The spin structure factor, $S^z({\bf q})$ [see Eq.~\eqref{eq:Sq}], shows diffuse 
	pattern, indicating the absence of magnetic order. 
	(e)--(h) The singlet structure factor without static contribution,
	$\tilde{O}({\bf q})$ [see Eq.~\eqref{eq:DB_no.static}], shows very rich and diffuse scattering 
	intensities ranging across the Brillouin zone. 
	(i)--(l) The singlet structure factor with static contributions,
	$O({\bf q})$ [see Eq.~\eqref{eq:Oq.with.static}],
	shows the high-intensity Bragg peaks at Brillouin zone centers related to the 
	singlet order as discussed in Sec.~\ref{sec:dim.reduction} of the main text. 
	}
	\label{fig:3D.pyrochlore.spin.singlet}
\end{figure*}
%

As discussed in Sec.~\ref{sec:dim.reduction}, the ground state of the $S$=1/2 pyrochlore 
Heisenberg antiferromagnet breaks the symmetry of the lattice by arranging 
singlets on a complex bond network within the pyrochlore lattice. 
As a supplement to Fig.~\ref{fig:ground.state.3Dpyrochlore} we show in Fig.~\ref{fig:3D.pyrochlore.spin.singlet} 
the equal-time structure factor for spins and singlets in the ground state of ${\mathscr H}$
[see Eq.~\eqref{eq:Ham.iso}] within the (h,k,0), (h,0,l), (h,l,0) and (h,h,l) crystallographic planes.

The ground state has been obtained after optimization from the maximally flippable state, 
aligned within the $xy$ plane.
Figures~\ref{fig:3D.pyrochlore.spin.singlet} (a)--(d) show a very diffuse signal in the $S^z({\bf q})$ 
[see Eq.~\eqref{eq:Sq}] within all four crystallographic planes.
The scattering structure shows a strong global anisotropy between symmetrically equivalent  
planes, with elongated lines of constant intensity along the (0,0,l) direction [see 
Figs.~\ref{fig:3D.pyrochlore.spin.singlet}(b) and (d)].
Similar effects have been reported for spin ice thin films in 
Refs.~[\onlinecite{Jaubert2017}] and [\onlinecite{Hurtubise2018}], 
further supporting our claim of the formation of decoupled 2D layers in the ground state.
In Fig.~\ref{fig:3D.pyrochlore.spin.singlet}(e)--(h) we show the singlet correlation function, 
$\tilde{O}({\bf q})$ [see Eq.~\eqref{eq:DB_no.static}], after subtracting static contributions 
from singlet order.
The structure factor is very diffuse with scattering intensity spread over the whole Brillouin 
zone.
The correlations within the (h,k,0) plane in Fig.~\ref{fig:3D.pyrochlore.spin.singlet}(e) match 
well with the measured signal from the STSL in Fig.~\ref{fig:singlet.correlations}(b), further supporting 
our claim that the layered STSL is a valid choice to investigate dominant correlation effects for the 
ground state in the full 3D model. 
In Fig.~\ref{fig:3D.pyrochlore.spin.singlet}(i)--(l) we show the singlet correlation function, 
$O({\bf q})$ [see Eq.~\eqref{eq:Oq.with.static}], including static contributions from singlet order. 
High-intensity points correspond to Bragg peaks, as discussed in Fig.~\ref{fig:ground.state.3Dpyrochlore}, 
and relate to the complicated singlet order involving 16 sites in the super-tetrahedron unit cell.

In principle, the exact ground state of the finite-size systems should not break the 
rotational symmetry of the original 3D pyrochlore lattice. 
However, since the matrix elements between different symmetry broken states are 
expected to be extremely small, the symmetry breaking with nematic order seems 
to have taken place in the VMC wave function, as is often observed for large system sizes. 
Of course, the exact symmetry could be restored by the quantum number projection, 
but practically the physical properties are not different.

%
\section{Site coordinates in the STSL}			
\label{app:STS.coord}
%

%
\begin{table}
	\newcolumntype{C}{>{}c<{}} 
	\renewcommand{\arraystretch}{1.2}
	\setlength{\tabcolsep}{0.05cm} 
	\begin{tabular}{ C  C  C  C  C}
		\hhline{=====}		
		site index $i$		& position		& 		& site index $i$ 	& position		\\
		\hhline{-----}
		0	& $\frac{1}{8} \left(0, -3, -3 / \sqrt{2} \right)$	& \ 		& 8	& $\frac{1}{8} \left(-2, -1, 1 / \sqrt{2} \right)$	\\
		1	& $\frac{1}{8} \left(1, -2, -1 / \sqrt{2} \right)$	& \ 		& 9	& $\frac{1}{8} \left(-1, 0,  3 / \sqrt{2} \right)$	\\
		2	& $\frac{1}{8} \left(-1, -2,  -1 / \sqrt{2} \right)$	& \ 		&10	& $\frac{1}{8} \left(-3, 0, 3/ \sqrt{2} \right)$	\\
		3	& $\frac{1}{8} \left(0, -1, -3 / \sqrt{2} \right)$	& \ 		&11	& $\frac{1}{8} \left(-2, 1,  1 / \sqrt{2} \right)$	\\
		4	& $\frac{1}{8} \left(2, -1, 1 / \sqrt{2} \right)$		& \ 		&12	& $\frac{1}{8} \left(0, 1, -3 / \sqrt{2} \right)$	\\
		5	& $\frac{1}{8} \left(3, 0,  3 / \sqrt{2} \right)$		& \ 		&13	& $\frac{1}{8} \left(1, 2, -1 / \sqrt{2} \right)$	\\
		6	& $\frac{1}{8} \left(1, 0,  3 / \sqrt{2} \right)$		& \ 		&14	& $\frac{1}{8} \left(-1, 2, -1 / \sqrt{2} \right)$	\\
		7	& $\frac{1}{8} \left(2, 1,  1 / \sqrt{2} \right)$		& \ 		&15	& $\frac{1}{8}\left (0, 3, -3 / \sqrt{2} \right)$	\\
		\hhline{=====}
	\end{tabular}
	\caption{
	Real-space coordinates for the unit cell of the STSL, as visualized in Fig.~\ref{fig:STSL.BCS}.
	}
\label{tab:STSL.real-space.coordinates}
\end{table}
%

\begin{figure}[t]
	\centering
	\includegraphics[width=0.89\columnwidth]{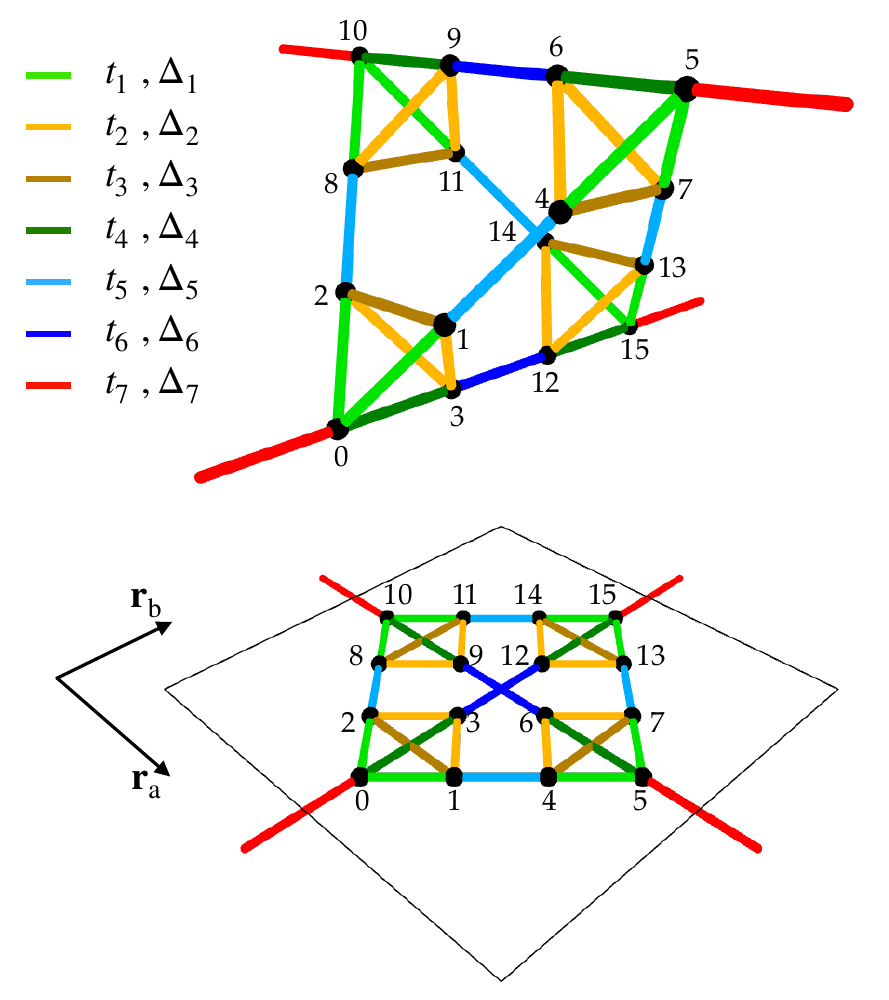}
  	\caption{
	The 16 site unit cell of the STSL with primitive lattice vectors ${\bf r}_{\rm a}$ and 
	${\bf r}_{\rm b}$ [see Eq.~\eqref{eq:Prim.vec.STS}] and site indices $i$, as explicitly 
	given in Table~\ref{tab:STSL.real-space.coordinates}, and its projection onto the $xy$ plane.
	Colored bonds indicate the arrangement of HFB fitting parameters 
	$t_i$ and $\Delta_i$, by satisfying the ${\bf D_{2d}}$ point group symmetry of 
	the lattice.
	}
	\label{fig:STSL.BCS}
\end{figure}
%

As shown in Sec.~\ref{sec:dim.reduction}, the STSL
is a minimal lattice, which captures dominant correlations 
in the ground state of the 3D pyrochlore Heisenberg antiferromagnet.
The STSL is not necessarily intuitive and somewhat different from commonly known
forms of layered pyrochlore systems~\cite{Bergholtz2015, Jaubert2017, Hurtubise2018}.
Therefore, we visualize in Fig.~\ref{fig:STSL.BCS} the 16 sites of the STSL unit cell with 
their projection onto the $xy$ plane, and provide their explicit real-space coordinates in 
Table~\ref{tab:STSL.real-space.coordinates}.
The primitive lattice vectors form a square lattice in real-space with 
%
\begin{align}
	{\bf r}_{\rm a} =   \begin{pmatrix} 1  \\ 0 \\ 0	 \end{pmatrix} \, , \quad
	{\bf r}_{\rm b} =   \begin{pmatrix} 0  \\ 1 \\ 0	 \end{pmatrix} \, ,
\label{eq:Prim.vec.STS}
\end{align}
%
and have a set of vectors ${\pmb \delta}_{\mu, \nu}$ which connect nearest-neighbor 
sublattice sites $\mu$ and $\nu$
%
\begin{equation}
	\begin{aligned}
 		{\pmb \delta}_{\rm ab} &= {\pmb \delta}_{0,1} = {\pmb \delta}_{4,5} = {\pmb \delta}_{8,9} = {\pmb \delta}_{12,13} = \frac{1}{8}  \left(1,1, \sqrt{2} \right)  \, ,	\\
		{\pmb \delta}_{\rm cd} &= {\pmb \delta}_{2,3} = {\pmb \delta}_{6,7} = {\pmb \delta}_{10,11} = {\pmb \delta}_{14,15} = \frac{1}{8}  \left(1,1, -\sqrt{2} \right)  \, ,	\\
		{\pmb \delta}_{\rm ac} &= {\pmb \delta}_{0,2} = {\pmb \delta}_{4,6} =  {\pmb \delta}_{8,10} =  {\pmb \delta}_{12,14}  = \frac{1}{8}  \left(-1,1, \sqrt{2} \right)  \, ,	\\
		{\pmb \delta}_{\rm bd} &= {\pmb \delta}_{1,3} = {\pmb \delta}_{5,7} = {\pmb \delta}_{9,11} = {\pmb \delta}_{13,15} = \frac{1}{8}  \left(-1,1, - \sqrt{2} \right)  \, ,	\\
		{\pmb \delta}_{\rm ad} &= {\pmb \delta}_{0,3} = {\pmb \delta}_{4,7} = {\pmb \delta}_{8,11} = {\pmb \delta}_{12,15} = \frac{1}{8}   \left(0,2, 0 \right)  \, ,	\\	
		{\pmb \delta}_{\rm bc} &= {\pmb \delta}_{1,2} = {\pmb \delta}_{5,6} = {\pmb \delta}_{9,10} = {\pmb \delta}_{13,14} = \frac{1}{8}   \left(-2,0, 0 \right)  \, .	\\
	\label{eq:connection.vector.STSL}
	\end{aligned}
\end{equation}
%
The symmetry of the four tetrahedra within the unit cell allows to combine equal vectors by indices 
$a, b, c$ and $d$, with reciprocal lattice vectors in momentum-space given by
%
\begin{align}
	{\bf k}_a =   2\pi \begin{pmatrix} 1  \\ 0 \\ 0	 \end{pmatrix} \, , \quad
	{\bf k}_b =   2\pi \begin{pmatrix} 0  \\ 1 \\ 0	 \end{pmatrix} \, .
\label{eq:momentum.STSL}
\end{align}
%

\clearpage
\newpage
%
\section{Hartree-Fock-Bogolyubov Hamiltonian for spinons}		
\label{app:HFB}

To perform the Bogolyubov transformation of ${\mathscr H}_{\rm HFB}$, we write  
Eq.~\eqref{eq:BCS1} in the Nambu representation by incorporating electron and hole degree of 
freedom by effectively doubling the size of the Hilbert space

%
\begin{widetext}
\begin{align}
	 {\mathscr H}_{\rm HFB} = \sum_{\bf k} 
	\begin{pmatrix}
		\hat{c}_{{\bf k}, 0, \uparrow}^{\dagger} 	\\
		\vdots \\
		\hat{c}_{{\bf k}, 15, \uparrow}^{\dagger} 	\\
		\hat{c}_{-{\bf k}, 0, \downarrow} 	\\
		\vdots \\
		\hat{c}_{-{\bf k},15 \downarrow}		\\
	\end{pmatrix}^{\intercal}
	\left(
    \renewcommand{\arraystretch}{1.2}
    \begin{array}{c c c  | c c c  }
		\epsilon_{0, 0}{(\bf k)}	& \cdots 		&\epsilon_{0, 15}{(\bf k)} 		&  \Delta_{0, 0}{(\bf k)} 	&  \cdots 		&  \Delta_{0, 15}{(\bf k)}  	 \\
		\vdots  				& \ddots		& \vdots					& \vdots  				& \ddots		& \vdots				 \\
		\epsilon_{15, 0}{(\bf k)}	&  \cdots 		& \epsilon_{15, 15}{(\bf k)} 	&  \Delta_{15, 0}{(\bf k)} 	&  \cdots 		&  \Delta_{15, 15}{(\bf k)}  	 \\
		\hline
		\Delta_{0, 0}{(\bf k)} 		&  \cdots 		& \Delta_{0, 15}{(\bf k)}  		& -\epsilon_{0, 0}{(\bf k)}	& \cdots 		&-\epsilon_{0, 15}{(\bf k)} 		 \\
		\vdots  				&  \ddots		& \vdots					& \vdots  				& \ddots		& \vdots					 \\
		\Delta_{15, 0}{(\bf k)} 	&  \cdots 		&  \Delta_{15, 15}{(\bf k)} 		& -\epsilon_{15, 0}{(\bf k)}&  \cdots 		& -\epsilon_{15, 15}{(\bf k)}  	 \\
    \end{array}    \right)  
    \begin{pmatrix}
		\hat{c}_{{\bf k}, 0, \uparrow}	\\
		\vdots \\
		\hat{c}_{{\bf k}, 15, \uparrow}	\\
		\hat{c}_{-{\bf k}, 0, \downarrow}^{\dagger}  	\\
		\vdots \\
		\hat{c}_{-{\bf k},15 \downarrow}^{\dagger} 		\\
	\end{pmatrix}	\, ,
\label{eq:Hbcs_Nambu}
\end{align}
\end{widetext}
%
where the explicit form of \yy{$\epsilon_{\mu, \nu}{(\bf k)} = \left(\boldsymbol{\epsilon}{(\bf k)}\right)_{\mu, \nu}$} is given by
%
\begin{widetext}
\begin{equation}
    \boldsymbol{\epsilon}{(\bf k)}
    = \left(
    \def\arraystretch{1.2}
    \begin{array}{c c c c | c c c c | c c c c | c c c c }
		0 					& t_{1}\gamma_{ab} 		& t_{1}\gamma_{ac} 		&  t_{4}\gamma_{ad} 	&  0 	&  0 	&  0 	&  0 					&  0 	&  0	 &  0 &  0 						&  0 &  0 &  0 	&  t_{7}\gamma^*_{ad} \\
		t_{1}\gamma^*_{ab}  	& 0					& t_{3}\gamma_{bc}		&  t_{2}\gamma_{bd} 	&  t_{5}\gamma_{ab} &  0 &  0 &  0 		&  0 	&  0	&  0 &  0 						&  0 	&  0 	&  0 	&  0 \\
		t_{1}\gamma^*_{ac} 		&  t_{3}\gamma^*_{bc} 	& 0					&  t_{2}\gamma_{cd} 	&  0 	&  0 	&  0 	&  0 					&  t_{5}\gamma_{ac}	&  0	&  0 	&  0 			&  0 	&  0 	&  0 	&  0 \\
		t_{4}\gamma^*_{ad}		& t_{2}\gamma^*_{bd}	& t_{2}\gamma^*_{cd}	&  0				&  0 	&  0 	&  0 	&  0 					&  0 	&  0	 &  0 &  0 						&  t_{6}\gamma_{ad} &  0 &  0 &  0 \\	
		\hline
		0 &   t_{5}\gamma^*_{ab} 	&  0	&  0 		& 0					& t_{1}\gamma_{ab} 		& t_{2}\gamma_{ac} 		&  t_{3}\gamma_{ad} 	&  0 	&  0	 &  0 &  0 					&  0 	&  0 &  0 	&  0 \\
		0 &  0 &  0 &  0 						& t_{1}\gamma^*_{ab}  	& 0					& t_{4}\gamma_{bc}		&  t_{1}\gamma_{bd} 	&  0 	&  0	 &  t_{7}\gamma^*_{bc}  &  0 	&  0  &  0 	&  0 	&  0 \\
		0 &  0 &  0 &  0 						& t_{2}\gamma^*_{ac} 	&  t_{4}\gamma^*_{bc} 	& 0					&  t_{2}\gamma_{cd} 	&  0 	& t_{6}\gamma_{bc} &  0 &  0		&  0 	&  0 &  0 	&  0 \\
		0 &  0 &  0 &  0 						& t_{3}\gamma^*_{ad}	& t_{1}\gamma^*_{bd}	& t_{2}\gamma^*_{cd} 	&  0					&  0 	&  0	 &  0 &  0 					&  0 	&  t_{5}\gamma_{bd}  &  0 &  0 \\
		\hline
		0 &  0 &  t_{5}\gamma^*_{ac} 	&  0 		&  0 	&  0	&  0 	&  0 						& 0					& t_{2}\gamma_{ab} 		& t_{1}\gamma_{ac} 		&  t_{3}\gamma_{ad} 	&  0 	&  0 	&  0 	&  0 \\
		0 &  0 &  0 &  0						&  0 	&  0	 & t_{6}\gamma^*_{bc} & 0 		& t_{2}\gamma^*_{ab}  	& 0					& t_{4}\gamma_{bc}		&  t_{2}\gamma_{bd} 	&  0 	&  0 	&  0 	&  0 \\
		0 &  0 &  0 &  0						&  0	&  t_{7}\gamma_{bc} &  0 &  0 			& t_{1}\gamma^*_{ac} 	&  t_{4}\gamma^*_{bc} 	& 0					&  t_{1}\gamma_{cd}		&  0 	&  0 	&  0 	&  0 \\
		0 &  0 &  0 &  0 						&  0 	&  0	 &  0 &  0 						& t_{3}\gamma^*_{ad}	& t_{2}\gamma^*_{bd}	& t_{1}\gamma^*_{cd} 	& 0 					&  0 &  0	 &  t_{5}\gamma_{cd}  &  0  \\
		\hline
		0 &  0 &  0 &   t_{6}\gamma^*_{ad} 	&  0  &  0  &  0 &  0 					&  0 	&  0 	&  0 	&  0					& 0					& t_{2}\gamma_{ab} 		& t_{2}\gamma_{ac} 		&  t_{4}\gamma_{ad} 	 \\
		0 &  0 &  0 &  0 					&  0	&  0	 &  0 &  t_{5}\gamma^*_{bd} 	&  0 	&  0 	&  0 	&  0					& t_{2}\gamma^*_{ab}  	& 0					& t_{3}\gamma_{bc}		&  t_{1}\gamma_{bd} 	 \\
		0 &  0 &  0 &  0 					&  0 	&  0	 &  0 &  0 					&  0 	&  0 	&  0 	&  t_{5}\gamma^*_{cd} 	& t_{2}\gamma^*_{ac} 	&  t_{3}\gamma^*_{bc} 	& 0					&  t_{1}\gamma_{cd} 	 \\
		t_{7}\gamma_{ad} &  0 &  0 &  0	&  0 	&  0	 &  0 &  0 					&  0 	&  0 	&  0 	&  0					& t_{4}\gamma^*_{ad}	& t_{1}\gamma^*_{bd}	& t_{1}\gamma^*_{cd}	&  0	 \\
    \end{array}
    \right)  \, .
    \label{eq:BCS-hopping.explicit}
\end{equation}
\end{widetext}
%
The information of the momentum-dependent sublattice structure is incorporated in the phase factor
%
\begin{equation}
	 \gamma_{\mu \nu}({\bf k)} = e^{i ({\pmb \delta}_{\mu \nu} \cdot {\bf k}) }		\, ,
\end{equation}
%
with real-space vectors ${\pmb \delta}$ [Eq.~\eqref{eq:connection.vector.STSL}], connecting nearest-neighbor 
sublattices $\mu$ with $\nu$ and allowed momenta ${\bf k}$, as defined in Eq.~\eqref{eq:momentum.STSL}.
The matrix for the pairing term, $\Delta_{\mu \nu}{(\bf k)}$, shows the same form as $\epsilon_{\mu \nu}{(\bf k)}$ 
[in Eq.~\eqref{eq:BCS-hopping.explicit}] and is obtained by exchanging hopping strengths $t_{i}$ with pairing 
amplitudes $\Delta_i$.

Taking into account the ${\bf D_{2d}}$ point-group symmetry of the STSL, we consider a 
unit cell with 7 inequivalent nearest-neighbor bonds. 
Consequently, this symmetry allows for independent hopping, $t_1, \cdots, t_7$, 
and pairing amplitudes, $\Delta_1, \cdots, \Delta_7$, as illustrated in  Fig.~\ref{fig:STSL.BCS},
which will be subject to optimization of \rp{${\mathscr H}_{\rm HFB}$} 
[Eq.~\eqref{eq:BCS1}].

Our fitting procedure is the following: We fully optimize the variational wave function $|\Psi\rangle$, 
for a finite-size cluster of $L=2$ under periodic boundary conditions by our mVMC calculation
in absence of any additional projection operators, except the Gutzwiller projection.
After successful optimization of ${\mathscr H}_{\sf J_1 J_2}$, in Eq.~\eqref{eq:Ham.J1J2}, for $J_2/J_1=1$,
we Fourier transform the numerically obtained variational parameters 
$f_{ij}$ [Eq.~\eqref{eq:psi.pair}] to obtain the $\bf k$-dependent pair-amplitude $f_{\bf k}^{\rm mVMC}$. 
We then minimize the loss function $\chi^2$, as given in Eq.~\eqref{eq:BCS.fij.loss.function}, by optimizing 
the 14 independent fitting parameters, $t_1, \cdots, t_7$ and  $\Delta_1, \cdots, \Delta_7$.
Note that $\bf u$ and $\bf v$ in Eq.~\eqref{eq:BCS.to.fij}, and hence  $f_{\bf k}^{\rm HFB}$ are 
represented by the fitting parameters $t_i$ and $\Delta_i$ through the Bogolyubov transformation.

Since this $\chi^2$ fitting with 14 variational parameters is not simple, 
we make use of the powerful machine learning library JAX~\cite{Jax2018}, using the gradient processing 
and optimization library “Optax”, with optimizer “Adam”~\cite{Optax2020}.
The $\bf k$ summation in Eq.(\ref{eq:BCS.fij.loss.function}) runs over $N_k$ available symmetrically 
inequivalent points in momentum space, while we stop the optimization 
after the loss function reached values below \mbox{$\chi^2 \leq 10^{-3}$}.
In Table~\ref{tab:BCS.fitted} we present our optimized HFB parameters
and show their corresponding energy dispersion of ${\mathscr H}_{\rm HFB}$ [see Eq.~\eqref{eq:Hbcs_Nambu}]
in Fig.~\ref{fig:BCS.mean-field}.

\begin{table}
	\def\arraystretch{1.5}
	\begin{tabular}{ l  @{\hspace{1cm}}  l} 
		\hhline{==}		
		hopping			&	pairing		\\
		\hhline{--}
		$t_1$ = 0.36532244	& 	$\Delta_1$ = -1.054411		\\
		$t_2$ = 1.714636 	&	$\Delta_2$ = 0.36639872  	\\
		$t_3$ = -0.22435574	&	$\Delta_3$ = 0.19859077  	\\
		$t_4$ = 1.3559698 	&	$\Delta_4$ = 0.4197492   		\\
		$t_5$ = 1.6474987   	&	$\Delta_5$ = 0.28696743 		\\
		$t_6$ = -1.0879616 	&	$\Delta_6$ = 0.38542387		\\
		$t_7$ = 0.89354455	&	$\Delta_7$ = 0.27466503		\\
		\hhline{==}
	\end{tabular}  
	\caption{
	HFB parameters after minimization of the loss function in Eq.~\eqref{eq:BCS.fij.loss.function}.
	}
\label{tab:BCS.fitted}
\end{table}
%

%
\section{Dynamical structure factors of HFB states}			
\label{app:DSF_BCS}

The HFB mean-field Hamiltonian on the STSL obtained in Eq.~(\ref{eq:BCS1}) provides us with insights into  
the power-law spin-spin correlation and the dynamical properties of the present spin liquid state.
As shown in the following, the static spin structure factor for the non-interacting
spinon approximation shows a scaling property that is consistent with the power-law decay of the 
spin-spin correlation $\sim 1/r^{\alpha}$ while the dynamical spin structure factor
shows an essentially gapless and quadratic dispersion relation.

To calculate the dynamical spin structure factor,
we perform the Bogoliubov transformation of ${\mathscr H}_{\rm HFB}$ and obtain the following diagonalized form,
\eqsa{
 {\mathscr H}_{\rm HFB} =
\sum_{{\bf k},n}
|E_{n}({\bf k})|\left(
\hat{\alpha}_{{\bf k}n+}^{\dagger}
\hat{\alpha}_{{\bf k}n+}
-
\hat{\alpha}_{{\bf k}n-}^{\dagger}
\hat{\alpha}_{{\bf k}n-}
\right),
}
where
$\hat{\alpha}_{{\bf k}n\pm}^{\dagger}$
($\hat{\alpha}_{{\bf k}n\pm}^{\ }$)
is the creation (annihilation) operator of
the quasiparticle with the $n$th positive/negative energy eigenvalue, $\pm |E_n ({\bf k})|$.
Here, we use the following unitary transformation
between spinon creation/annihilation operators and
the quasiparticle operators,
\eqsa{
\hat{c}_{{\bf k}, \mu, \uparrow}
&=&
\sum_n
\left[
\left({\bf u}({\bf k})\right)_{\mu,n}
\hat{\alpha}_{{\bf k}n+}
-\left({\bf v}({\bf k})\right)_{\mu,n}
\hat{\alpha}_{{\bf k}n-}
\right],\nonumber\\
\\
\hat{c}_{-{\bf k}, \mu, \downarrow}^{\dagger}
&=&
\sum_n
\left[
\left({\bf v}({\bf k})\right)_{\mu,n}
\hat{\alpha}_{{\bf k}n+}
+
\left({\bf u}({\bf k})\right)_{\mu,n}
\hat{\alpha}_{{\bf k}n-}
\right].\nonumber\\
}
Then, the ground-state
wave function of the HFB Hamiltonian
is written as
\eqsa{
\overline{|\phi_{\rm HFB} \rangle}
=
\left(
\prod_{{\bf k}}
\prod_{n}
\hat{\alpha}_{{\bf k}n-}^{\dagger}
\right)
\prod_{{\bf k}'}
\prod_{\mu}
\hat{c}_{-{\bf k}',\mu, \downarrow}^{\dagger}
|0\rangle.
}

The spin excitation spectra at the non-interacting spinon approximation
are given by bare polarization functions of spinons.
Here, the $z$-component of the polarization function, $\chi^{zz} ({\bf q},\omega)$,
is defined by
the following formula,
\eqsa{
\chi^{zz} ({\bf q},\omega)&=&
\left(\frac{1}{2}\right)^2
\frac{1}{N_{k}}
\sum_{{\bf k},{\bf p}}
\sum_{\mu,\nu}
{e^{+i{\bf q}\cdot {\bf R}_{\mu}-i{\bf q}\cdot {\bf R}_{\nu}}}
\overline{\langle \phi_{\rm HFB}|}
\nonumber\\
&&\times
\left(
\hat{c}_{{\bf p}, \mu, \uparrow}^{\dagger}
\hat{c}_{{\bf p}+{\bf q}, \mu, \uparrow}
-
\hat{c}_{{\bf p}, \mu, \downarrow}^{\dagger}
\hat{c}_{{\bf p}+{\bf q}, \mu, \downarrow}
\right)
\nonumber\\
&&\times
\frac{1}{\omega + i\delta - \mathscr{H}_{\rm HFB} + E_0}
\nonumber\\
&&\times
\left(
\hat{c}_{{\bf k}+{\bf q}, \nu, \uparrow}^{\dagger}
\hat{c}_{{\bf k}, \nu, \uparrow}
-
\hat{c}_{{\bf k}+{\bf q}, \nu, \downarrow}^{\dagger}
\hat{c}_{{\bf k}, \nu, \downarrow}
\right)
\overline{|\phi_{\rm HFB} \rangle},\nonumber\\
\label{eq:bare.polarization.HFB}
}
where $N_k$ is the number of ${\bf k}$ points and $E_0= - \sum_{n}\sum_{{\bf k}}|E_n ({\bf k})|$
is the mean-field ground state energy.
We introduce the real-space coordinates, ${\bf R}_{\mu}$,
of the sites within the unit cell
(see Table \ref{tab:STSL.real-space.coordinates}).
The formula for the bare polarization function using the Bogoliubov transformation is given in the bottom of this section.

The dynamical spin structure factor of the $z$-component
is then given by the imaginary part of $\chi^{zz}$ as
\eqsa{
S^z
({\bf q},\omega)
=
-\frac{1}{\pi}
{\rm Im}\left[
\chi^{zz}
({\bf q},\omega)
\right].
\label{eq:Sqw.spinon}
}
In Fig.~\ref{Fig:SQomega_spinon_MF}, $S^z ({\bf q},\omega)$ along symmetry lines is shown for $J_2/J_1 =1$ and $\omega < 0.5$.
The low-energy spin excitation spectrum shows an
essentially gapless nature and quadratic ${\bf q}$ dependence.

\begin{figure}[tbh]
	\begin{center}
	\includegraphics[width=0.4\textwidth]{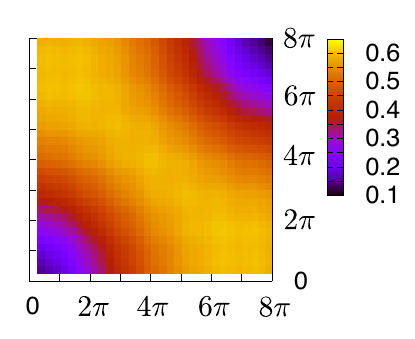}
	\end{center}
	\caption{
	Spin structure factor $S^z ({\bf q})$ [Eq.~\eqref{eq:Sq.spinon}] is shown for $L=8$.
	}
\label{Fig:SQ_8x8_spinon_MF}
\end{figure}

By integrating the $\omega$-dependence,
we obtain the static spin structure factor as
\eqsa{
S^z ({\bf q})
=\int_{-\infty}^{+\infty} d\omega 
S^{z}
({\bf q},\omega),
\label{eq:Sq.spinon}
}
as well, which is shown in Fig.~\ref{Fig:SQ_8x8_spinon_MF}.

%
%

To clarify the spin-spin correlation function at the long-wave-length limit,
we will examine the non-analytical ${\bf q}$ dependence of spin structure factors.
The present mVMC results show the non-analytical behaviors around
{${\bf Q}=(4m\pi, 4m\pi)$} ($m\in \mathbb{Z}$) or equivalent ${\bf q}$ points
(see Sec.~\ref{sec:STS.correlations.spin}).
Below,
the system-size dependence of $S^z({\bf q})$
is explicitly noted as $S^z({\bf q},L)$
to elucidate the non-analytical ${\bf q}$ dependence of the correlation functions.

To analyze the $L$ dependence of
$S^z (\delta {\bf q}+{\bf Q},L)$,
we assume a power-law tail of the real-space spin-spin correlation function
$\sim C/r^{\alpha}$
and perform the Fourier transformation of $C/r^p$ as
\eqsa{
I(|\delta {\bf q}|,L)&=&\int_{a_r}^{c_r L}
r dr \int_{0}^{2\pi} d \theta \frac{C}{r^{\alpha}}e^{-i |\delta {\bf q}| r \cos \theta}
\nonumber \\
&=&
2\pi 
|\delta {\bf q}|^{\alpha-2}
\int_{a_r |\delta {\bf q}|}^{c_r L |\delta {\bf q}|}
dx \frac{C}{x^{\alpha-1}}
J_{0}(x), \label{Eq:scaling_spinon}
}
where $c_r L$ is the long-range cutoff length proportional to $L$, $a_r$ is the short-range 
cutoff of the order of the lattice constant, and $J_0$ is a Bessel function of the first kind.

When we assume $|\delta {\bf q}| \propto 1/L$,
we can estimate Eq.~(\ref{Eq:scaling_spinon}) as
\eqsa{
I(|\delta {\bf q}|,L)
&\simeq&
2\pi C \frac{a_r^{-\alpha+2}}{\alpha-2}
-\frac{\pi Ca_r^{-\alpha+4}}{2} |\delta {\bf q}|^{\alpha-2} \frac{|\delta {\bf q}|^{-\alpha+4}}{4-\alpha}
\nonumber\\
&&
+
{\rm const.} \times |\delta {\bf q}|^{\alpha-2}
+ \mathcal{O}(|\delta {\bf q}|^2).
\label{eq:scaling_spinon_MF}
}
Here, we assume that $I(|\delta {\bf q}|,L)$ is finite, and, thus, $p>2$.
For $2<p<4$, the following scaling relation is obtained,
\eqsa{
I(2\sqrt{2}\pi/L,L)\sim c_0 + c_1/L^{\alpha-2} + c_2/L^2,\label{Eq:fitting_spinon_MF}
}
where the second term, $c_1/L^{\alpha-2}$, in the right hand side corresponds
to a non-analytical ${\bf q}$ dependence
of $S^z({\bf q})$.
The ${\bf q}$ independent term $c_0$ may be affected by the short-range correlations
absent in the power-law tail [Eq.~(\ref{eq:scaling_spinon_MF})]
thus, the first term in Eq.~(\ref{eq:scaling_spinon_MF}) will strongly depend on
${\bf Q}$.

\begin{figure}[tbh]
	\begin{center}
	\includegraphics[width=0.4\textwidth]{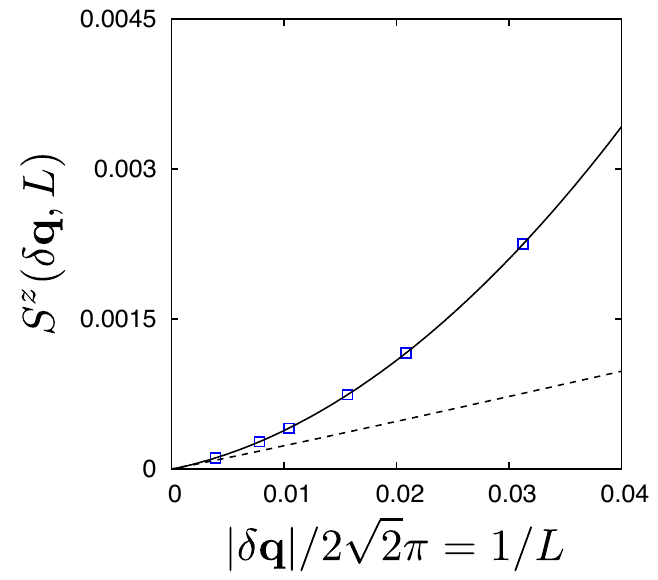}
	\end{center}
	\caption{
	Spin structure factor \textcolor{black}{$S^z (\delta {\bf q},L)$}
	at $\delta {\bf q}=(2\pi/L,2\pi/L)$ is shown
	for $L=32,48, 64, 96, 128,$ and $256$.
	The open squares show $S^z(\delta {\bf q},L)$ and
	the solid line shows a result of the least square fitting by $ c_1/L^{\alpha-2}+c_2/L^2$
	given in Eq.~(\ref{Eq:fitting_spinon_MF}).
	Here, the fitting parameters are determined as 
	$c_1=0.44\pm 0.02$, $c_2 = 24.5 \pm 0.1$, and $\alpha=3.038\pm 0.007$.
}
\label{Fig:fit_square_eta}
\end{figure}

The spin structure factors obtained by the present mVMC calculations show
non-analytical behaviors at ${\bf Q}=(0,0)$, $(4\pi,4\pi)$, and equivalent ${\bf q}$ points.
At the long-wave-length limit, ${\bf q}$ dependence with the smallest exponent $\alpha$ will dominate the power-law decay
of the spin correlation.
Here, we focus on ${\bf Q}=(0,0)$ where no system size dependence appears
since $S^z ({\bf 0},L)=0$ due to the spin conservation.
The spin structure factor $S(\delta {\bf q},L)$ at $\delta {\bf q}=(2\pi/L,2\pi/L)$
is indeed well fitted by $I(2\sqrt{2}\pi/L,L)$ with the fitting parameters,
$c_1=0.44\pm 0.02$, $c_2 = 24.5 \pm 0.1$
and $\alpha=3.038\pm 0.007$
with $c_0=0$, as shown in Fig.~\ref{Fig:fit_square_eta}.
The exponent $\alpha\sim 3$ is consistent with the present mVMC result for the ground state.

In single-orbital systems, $S^z(\delta {\bf q}) \propto |\delta {\bf q}|$ at the limit $|\delta {\bf q}|\rightarrow 0$
has been evidence of the gapless spin excitation as examined in Ref.~[\onlinecite{PhysRevLett.94.026406}].
It is similar to the results of the present multi-orbital system while a linear dispersion of the spin excitation spectrum has been
expected in the previous study~\cite{PhysRevLett.94.026406}.

%
%

For the practical calculation, we rewrite the bare polarization function for the HFB wave 
function [Eq.~(\ref{eq:bare.polarization.HFB})] by using the Bogoliubov transformation 
coefficient matrices ${\bf u}$ and ${\bf v}$ as,
\begin{widetext}
\eqsa{
\chi^{zz} ({\bf q},\omega)&=&
\left(\frac{1}{2}\right)^2
\frac{1}{N_{k}}
\sum_{{\bf k},{\bf p}}
\sum_{\mu,\nu}
\overline{\langle \phi_{\rm HFB}|}
\left(
\hat{c}_{{\bf p}, \mu, \uparrow}^{\dagger}
\hat{c}_{{\bf p}+{\bf q}, \mu, \uparrow}
-
\hat{c}_{{\bf p}, \mu, \downarrow}^{\dagger}
\hat{c}_{{\bf p}+{\bf q}, \mu, \downarrow}
\right)
{e^{+i{\bf q}\cdot {\bf R}_{\mu}}}
\nonumber\\
&&\times
\frac{1}{\omega + i\delta - \mathscr{H}_{\rm HFB} + E_0}
\left(
\hat{c}_{{\bf k}+{\bf q}, \nu, \uparrow}^{\dagger}
\hat{c}_{{\bf k}, \nu, \uparrow}
-
\hat{c}_{{\bf k}+{\bf q}, \nu, \downarrow}^{\dagger}
\hat{c}_{{\bf k}, \nu, \downarrow}
\right)
{e^{-i{\bf q}\cdot {\bf R}_{\nu}}}
\overline{|\phi_{\rm HFB} \rangle}\nonumber
\\
&=&
\left(\frac{1}{2}\right)^2
\frac{1}{N_{k}}
\sum_{{\bf k}}
\sum_{\mu,\nu}
\sum_{m,n}
{e^{+i{\bf q}\cdot ({\bf R}_{\mu}-{\bf R}_{\nu})}}
\left[
\frac{
\left({\bf v}^{\ast}({\bf k})\right)_{\mu,n}
\left({\bf u}({\bf k}+{\bf q})\right)_{\mu,m}
\left({\bf u}^{\ast}({\bf k}+{\bf q})\right)_{\nu,m}
\left({\bf v}({\bf k})\right)_{\nu,n}
}{\omega + i\delta - |E_{m}({\bf k}+{\bf q})| - |E_{n}({\bf k})|}
\right.
\nonumber\\
&&
-
\frac{
\left({\bf v}^{\ast}(-{\bf k}-{\bf q})\right)_{\mu,n}
\left({\bf u}(-{\bf k})\right)_{\mu,m}
\left({\bf v}^{\ast}(-{\bf k})\right)_{\nu,m}
\left({\bf u}(-{\bf k}-{\bf q})\right)_{\nu,n}
}{\omega + i\delta - |E_{m}(-{\bf k})| - |E_{n}(-{\bf k}-{\bf q})|}
\nonumber\\
&&
-
\frac{
\left({\bf u}^{\ast}({\bf k})\right)_{\mu,n}
\left({\bf v}({\bf k}+{\bf q})\right)_{\mu,m}
\left({\bf u}^{\ast}({\bf k}+{\bf q})\right)_{\nu,m}
\left({\bf v}({\bf k})\right)_{\nu,n}
}{\omega + i\delta - |E_{m}({\bf k}+{\bf q})| - |E_{n}({\bf k})|}
\nonumber\\
&&+
\left.
\frac{
\left({\bf u}^{\ast}(-{\bf k}-{\bf q})\right)_{\mu,n}
\left({\bf v}(-{\bf k})\right)_{\mu,m}
\left({\bf v}^{\ast}(-{\bf k})\right)_{\nu,m}
\left({\bf u}(-{\bf k}-{\bf q})\right)_{\nu,n}
}{\omega + i\delta -  |E_{m}(-{\bf k})| - |E_{n}(-{\bf k}-{\bf q})|}
\right],
}
where $N_k$ is the number of ${\bf k}$ points and $\delta$ is a small and positive broadening factor.
The above dynamical spin susceptibility is given by the particle-hole excitations in the 32 spinon bands
generated by diagonalizing the HFB Hamiltonian.
\end{widetext}

%
%
\bibliography{Bibliography}
%

\clearpage
\widetext


\section*{Supplementary material}

Rico Pohle,[1,2,3] Youhei Yamaji,[4] and Masatoshi Imada[1,5,6]

[1]*Waseda Research Institute for Science and Engineering,*
*Waseda University, 3-4-1 Okubo, Shinjuku-ku, Tokyo, 169-8555, Japan*

[2]*Department of Applied Physics, University of Tokyo, Hongo, Bunkyo-ku, Tokyo, 113-8656, Japan*

[3]*Graduate School of Science and Technology, Keio University, Yokohama 223-8522, Japan*

[4]*Center for Green Research on Energy and Environmental Materials (GREEN),*
*and Research Center for Materials Nanoarchitectonics (MANA),*

*National Institute for Materials Science (NIMS), Namiki, Tsukuba-shi, Ibaraki 305-0044, Japan*

[5]*Toyota Physical and Chemical Research Institute, Yokomichi, Nagakute, Aichi, 480-1192, Japan*

[6]*Sophia University, Kioicho, Chiyoda-ku, Tokyo, Japan*

(Dated: November 19, 2023)

## I. DETAILS OF MANY-VARIABLE VARIATIONAL MONTE CARLO

The variational Monte Carlo method is widely used to approximate a quantum mechanical many-body wave function by using statistical sampling over a restricted number of phase-space parameters. In the present work, we make full use of the open-source software named many-variable variational Monte Carlo (mVMC), which demonstrates the state-of-the-art performance among conventional variational methods, thanks to its highly generalized form and large number of variational parameters [1, 2]. An open access version of mVMC is available at the GitHub repository [3] and is so far constantly maintained to achieve a user-friendly environment. We review some details of the algorithm in the following paragraphs of this section.

### A. Overview of variational wave function

Within the mVMC method, we evaluate the system Hamiltonian $\mathscr{H}$, by measuring the expectation value of the energy $\langle E \rangle$, for a variational wave function $|\Psi\rangle$

$$
\begin{aligned}
\langle E \rangle &= \frac{\langle \Psi | \mathscr{H} | \Psi \rangle}{\langle \Psi | \Psi \rangle} \\
&= \sum_x \frac{\langle \Psi | \mathscr{H} | x \rangle \langle x | \Psi \rangle}{\langle \Psi | \Psi \rangle} \\
&= \sum_x \rho(x) \frac{\langle \Psi | \mathscr{H} | x \rangle}{\langle \Psi | x \rangle},
\end{aligned}
\tag{S1}
$$

which is represented by the summation over real-space configurations $x$ to expand the complete set of the Hilbert space. Since the full summation is not tractable, the summation is replaced by a sampling with the probability $\rho(x)$ represented by

$$
\rho(x) = \frac{|\langle x | \Psi \rangle|^2}{\langle \Psi | \Psi \rangle}.
\tag{S2}
$$

In the present case, $|x\rangle$ is a component of the real-space spin configuration in the basis diagonal in the $z$ component of the

spin $S^z$. $\langle E \rangle$ is obtained by averaging over such generated statistically independent Monte Carlo samples. Each real-space configuration is usually updated via two-spin exchange updates for the Heisenberg model [1, 2].

We write our variational wave function in the form

$$
|\Psi\rangle = \mathcal{Z}\mathcal{L}\mathcal{N}\mathcal{P} |\psi_{\mathsf{pair}}\rangle,
\tag{S3}
$$

with the correlation factor $\mathcal{P}$, the artificial neural-network projector $\mathcal{N}$, the quantum-number projector $\mathcal{L}$, and the first-step Lanczos projector $\mathcal{Z}$. $\mathcal{P}$ in the present work is the Gutzwiller factor to strictly keep the constraint of one electron per site in the fermionic representation of spins, which we introduce below. The neural-network projector $\mathcal{N}$ plays the role of taking into account correlations among many spins, which can be regarded as an extension of the Jastrow projector [4]. To obtain the lowest energy state with given quantum numbers, we use the quantum-number projector $\mathcal{L}$.

### B. Pair product wave function

In Eq. (S3), we introduce the pair-product form of the many-body wave function in its real-space representation using Pfaffian matrices (an extension of the general Slater determinant)

$$
|\psi_{\mathsf{pair}}\rangle = \left( \sum_{ij}^{N_s} \sum_{\sigma\sigma'} f_{ij}^{\sigma\sigma'} c_{i\sigma}^\dagger c_{j\sigma'}^\dagger \right)^{N_e/2} |0\rangle,
\tag{S4}
$$

where the amplitude $f_{ij}^{\sigma\sigma'}$ of an electron pair serves as the variational parameter which will be optimized. $N_s$ and $N_e$ represent the site and electron number, respectively. The operator $c_{i\sigma}^\dagger$ creates one electron at site $i$ with corresponding spin $\sigma = \uparrow$ or $\downarrow$ by acting on the vacuum $|0\rangle$. This form accommodates two-particle bound states, as needed to represent singlet resonating-valence-bond (RVB) states as well as Bardeen-Cooper-Schrieffer (BCS) type (or in other words, Hartree-Fock-Bogolyubov type) wave functions [5].

The variational parameters $f_{ij}^{\sigma\sigma'}$ are crucial in our simulations. For simulations of SU(2) symmetric Hamiltonians (see

Eqs. (1) and (3) of the main text), we restrict the $f_{ij}^{\sigma\sigma'}$ parameters to be real numbers in the form $f_{ij}^{\uparrow\downarrow}$ (called from now on $f_{ij}$ for simplicity, as also used in the main text), where each pair contains one spin up and one spin down electron. Eq. (S4) is then rewritten as

$$|\psi_{\mathsf{pair}}\rangle = \left( \sum_{ij}^{N_s} f_{ij} c_{i\uparrow}^{\dagger} c_{j\downarrow}^{\dagger} \right)^{N_e/2} |0\rangle , \qquad (S5)$$

which contains only states with conserved $S^z = \sum_i S_i^z = 0$.

To further reduce the computational cost, we impose full translational invariance on the $f_{ij}$ parameters by considering a $f_{ij}$ sublattice. When this sublattice contains $M$ spins, the number of independent variational parameters $f_{ij}$ is reduced to $M \times N_s \sim O(N_s)$, whereas without assuming translational symmetry, the number of $f_{ij}$ is $(N_s \times N_s) \sim O(N_s^2)$.

The size and shape of the $f_{ij}$ sublattice must be compatible with the magnetic unit cell of the optimized wave function. As discussed in Sec. III A of the main text, the arrangement of singlets in the ground state of the pyrochlore Heisenberg antiferromagnet cover 16 sites on a super-tetrahedron. We therefore take $M = 16$ in simulations of the 3D pyrochlore lattice, while $M = 2 \times 2 \times 16 = 64$ for simulations of the quasi-2D super-tetrahedron square lattice (STSL), which also allows us to resolve the $2 \times 2$ total momenta (see Sec. I F) of the desired wave function.

### C. Correlation factor

Since our work exclusively focuses on the Heisenberg model, we need to exclude doubly-occupied and empty sites of electrons. This can be achieved by keeping the number of electrons $N_e$ to be the same as the number of sites $N_s$, namely $N_s = N_e$, together with the Gutzwiller correlation factor [6]

$$\mathcal{P}_G \equiv \mathcal{P} = \prod_i (1 - n_{i\uparrow} n_{j\downarrow}) . \qquad (S6)$$

### D. Restricted Boltzmann machine

To improve the quality of our variational wave function, we combine mVMC with a neural network, namely the restricted Boltzmann machine (RBM) [7–9]. This implementation for the Hubbard and Heisenberg models [9] has been successfully applied to understand the nature of spin liquid states in 2D systems [10, 11].

For Eq. (S3), we write the neural-network projector:

$$\mathcal{N} = \sum_{\{h_k\}} \exp \left( \sum_i a_i \sigma_i + \sum_{i,k} W_{ik} \sigma_i h_k + \sum_k b_k h_k \right) , \qquad (S7)$$

where $\{a_i, b_k, W_{ik}\}$ respectively represent complex variational parameters for physical units, hidden units, and the "neurons" connecting them. The summation over $h_k$ is the trace summation with respect to the hidden variable $h_k$ only. For the $S = 1/2$ Heisenberg model we write $\sigma_i = 2S_i^z = n_{i,\uparrow} - n_{i,\downarrow} = \pm 1$, counting the number of electrons with spin $\sigma$ at the physical site $i$. Since the hidden variables can be traced out, we can simplify Eq. (S7) into

$$\mathcal{N} = \prod_k 2 \cosh \left( b_k + \sum_i W_{ik} \sigma_i \right) e^{\sum_i a_i \sigma_i} . \qquad (S8)$$

### E. Optimization of variational parameters

Our mVMC+RBM method optimizes both $f_{ij}$ in $|\psi_{\mathsf{pair}}\rangle$ [see Eq. (S5)] and $\{a_i, b_k, W_{ik}\}$ in $\mathcal{N}$ [see Eq. (S8)] to reach a reliable variational wave function of high quality. These parameters are optimized by the natural gradient (or equivalently stochastic reconfiguration) method [12–14] with the help of the conjugate gradient method [12, 15]. This allows us to access system sizes as large as $N_s = 1024$ sites by reducing memory and computational cost [2]. In this paper, optimizations were carried out by starting from a general, unbiased random initial state, as well as from the maximally flippable dimer initial state, as discussed in Appendix C of the main text. Details of other choices of initial states are provided in Sec. IV of this Supplemental Material.

### F. quantum-number projection

The quantum-number projection technique enables us to restore the symmetry of the Hamiltonian, which has to be satisfied in finite-size clusters [16]. This technique convincingly showed substantial improvement of variational wave functions for itinerant and strongly correlated electron systems [1, 16–19] and also spin models [10, 20].

The quantum-number projector $\mathcal{L}$ superimposes elements of the symmetry transformation $T_n$ as

$$\mathcal{L}|\psi\rangle = \sum_n w_n |\psi_n\rangle ,$$
$$|\psi_n\rangle = T_n |\psi\rangle , \qquad (S9)$$

to restore the symmetry. In this paper we impose momentum projection $\mathcal{L}_K$, point-group projection $\mathcal{L}_\lambda$, and spin-projection $\mathcal{L}_{S_{\mathsf{tot}}}$ [1, 2, 16]:

$$\mathcal{L} = \mathcal{L}_{S_{\mathsf{tot}}} \mathcal{L}_\lambda \mathcal{L}_K , \qquad (S10)$$

with

$$\mathcal{L}_{\mathbf{K}} = \frac{1}{N} \sum_{\mathbf{R}} e^{i\mathbf{K}\cdot\mathbf{R}} \, T_{\mathbf{R}} , \qquad (S11)$$

$$\mathcal{L}_\lambda = \frac{1}{N_\lambda} \sum_n \chi_n(\lambda) \, T_n , \qquad (S12)$$

$$\mathcal{L}_{S_{\mathsf{tot}}} = \frac{2S_{\mathsf{tot}} + 1}{8\pi^2} \int d\Omega P_{S_{\mathsf{tot}}}(\cos\beta) R(\Omega) . \qquad (S13)$$

The momentum projector $\mathcal{L}_{\mathbf{K}}$ [Eq. (S11)] selects wave functions with specific total momentum $\mathbf{K}$, by acting with the translational operator $T_{\mathbf{R}}$ between all sites within the $f_{ij}$ sublattice. Wave functions with particular point-group symmetry can be selected with $\mathcal{L}_{\lambda}$ [Eq. (S12)] by acting with the symmetry transformation $T_n$ and its corresponding character $\chi_n(\lambda)$ of the selected irreducible representation $\lambda$. We sum over $n$ sufficient symmetry transformations to separate states of different point group. As an example, we write $\mathcal{L}_{\lambda}$ for the symmetric irrep $A_1$ of the $\mathbf{D_{2d}}$ group (see Table I of the main text) on the STSL

$$
\begin{aligned}
\mathcal{L}_{A_1} \propto &\, T_E + T_{S_4} + T_{C_{2(z)}} + T_{C_2'} \\
&+ T_{S_4} * T_{C_{2(z)}} + T_{S_4} * T_{C_2'} + T_{C_{2(z)}} * T_{C_2'} \\
&+ T_{S_4} * T_{C_{2(z)}} * T_{C_2'} ,
\end{aligned}
\tag{S14}
$$

which requires in total eight projection operators. To reduce computational cost, in most cases we do not evaluate all available point-group symmetries, but rather a minimal set of symmetry operations to only separate states of character $\{AB\} = \{A_1, A_2, B_1, B_2\}$ from $\{E\}$

$$
\mathcal{L}_{AB} \propto T_E + T_{C_{2(z)}} ,
\tag{S15}
$$

$$
\mathcal{L}_E \propto T_E - T_{C_{2(z)}} ,
\tag{S16}
$$

which requires only two projection operators.

The spin-projection operator $\mathcal{L}_{S_{\text{tot}}}$ [Eq. (S13)] restores the $SU(2)$ spin-rotation symmetry by superimposing rotated states for Euler angles $\Omega$ over the whole sphere

$$
R(\Omega) = R^z(\alpha) R^y(\beta) R^z(\gamma) = e^{i\alpha S^z} e^{i\beta S^y} e^{i\gamma S^z} ,
\tag{S17}
$$

with $\Omega = \alpha, \beta, \gamma$ and $S^y$ and $S^z$ spin operators for $y$ and $z$ directions. Hereby, the weight $w_n$ in Eq. (S9) is set to the $S_{\text{tot}}$-th Legendre polynomial $P_{S_{\text{tot}}}(\cos\beta)$. In the case of conserved spin-quantum numbers $S^z = 0$ the integration over $\alpha$ and $\gamma$ can be done analytically, thus Eq. (S13) reduces to

$$
\mathcal{L}_{S_{\text{tot}}} = \frac{2S_{\text{tot}}+1}{2} \int_0^{\pi} d\beta \sin\beta P_{S_{\text{tot}}}(\cos\beta) R^y(\beta) .
\tag{S18}
$$

The integration over $\beta$ is evaluated efficiently by the Gauss-Legendre quadrature in numerical calculations, by using a discrete number of mesh points (usually $\geq 10$).

Unfortunately, taking a large numbers of mesh-points requires large numerical cost. In that case, we introduce the spin-parity projector [16] when we only need to distinguish between odd and even total spin $S_{\text{tot}}$. The parity operator flips a spin with

$$
P = e^{i\pi S^y} = -i S^y ,
\tag{S19}
$$

with

$$
\langle \Psi | e^{i\pi S^y} | \Psi \rangle = P_{S_{\text{tot}}}(\cos\pi) = (-1)^{S_{\text{tot}}} ,
\tag{S20}
$$

telling us that "$+$" parity corresponds to a wave function with even total spin $S_{\text{tot}} = 0, 2, 4, \cdots$, and "$-$" parity to wave functions with odd total spin $S_{\text{tot}} = 1, 3, 5, \cdots$. The spin-parity projection operator can therefore be written as

$$
\mathcal{L}_{\pm} = (1 \pm P)/2 ,
\tag{S21}
$$

which corresponds to only 2 mesh points in Eq. (S18) and therefore reduces the numerical cost significantly.

### G. First-step Lanczos method

To improve the accuracy of the optimized variational wave function, we apply the first-step Lanczos method by multiplying the projector $\mathcal{Z}$ in Eq. (S3)

$$
\mathcal{Z} = (1 + \alpha_L \mathcal{H}) ,
\tag{S22}
$$

where $\alpha_L$ is a variational parameter, which is used to minimize the energy in Eq. (S1) [2, 21]. The first-step Lanczos method remains a variational method and provides an improved upper bound to the ground state energy. In principle, higher-power Lanczos steps are possible and will improve the energy even further, however are very costly, which is the reason why we consider only the first-step Lanczos method in this work.

### H. Observables

Correlation functions are generally measured in the form of Green's functions

$$
G(\{i, j, \sigma_i, \sigma_j\}) = \langle \psi | \prod_{\{i,j,\sigma,\sigma'\}} c_{i,\sigma}^{\dagger} c_{j,\sigma'} | \psi \rangle ,
\tag{S23}
$$

with $\{i, j, \sigma, \sigma'\}$ defining the set of correlations of interest. To analyze spin and dimer structure factors, we measure two-body and four-body Green's functions, respectively. Two-body Green's functions are measured as

$$
\begin{aligned}
G(i, j, &\sigma_1, \sigma_2, \sigma_3, \sigma_4) \\
&= \langle \psi | c_{i,\sigma_1}^{\dagger} c_{i,\sigma_2} c_{j,\sigma_3}^{\dagger} c_{j,\sigma_4} | \psi \rangle \\
&= \sum_x \langle \psi | c_{i,\sigma_1}^{\dagger} c_{i,\sigma_2} c_{j,\sigma_3}^{\dagger} c_{j,\sigma_4} | x \rangle \langle x | \psi \rangle ,
\end{aligned}
\tag{S24}
$$

while four-body Green's functions are measured as

$$
\begin{aligned}
G(m, n, \sigma_1, \sigma_2, \sigma_3, \sigma_4, \sigma_5, \sigma_6, \sigma_7, \sigma_8) &= \langle \psi | c_{m_1,\sigma_1}^{\dagger} c_{m_1,\sigma_2} c_{m_2,\sigma_3}^{\dagger} c_{m_2,\sigma_4} c_{n_1,\sigma_5}^{\dagger} c_{n_1,\sigma_6} c_{n_2,\sigma_7}^{\dagger} c_{n_2,\sigma_8} | \psi \rangle \\
&= \sum_x \langle \psi | c_{m_1,\sigma_1}^{\dagger} c_{m_1,\sigma_2} c_{m_2,\sigma_3}^{\dagger} c_{m_2,\sigma_4} | x \rangle \langle x | c_{n_1,\sigma_5}^{\dagger} c_{n_1,\sigma_6} c_{n_2,\sigma_7}^{\dagger} c_{n_2,\sigma_8} | \psi \rangle ,
\end{aligned}
\tag{S25}
$$

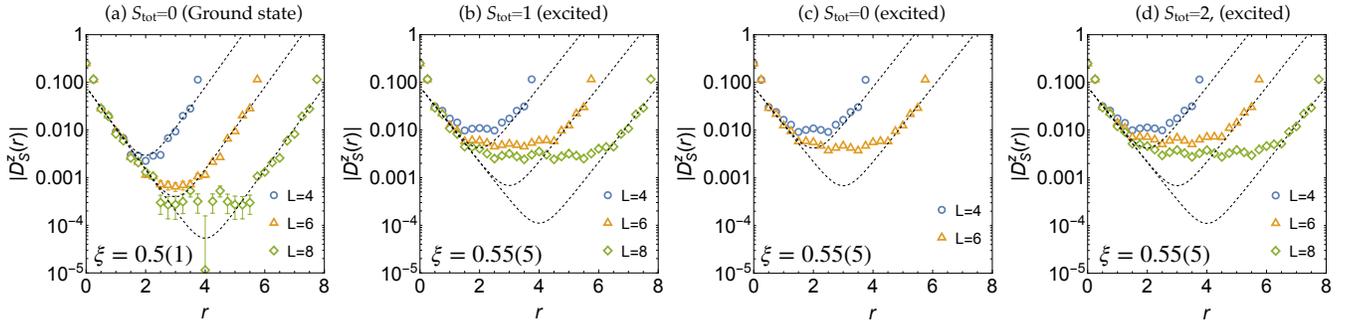

Figure S1. Spin-spin correlations in real space for the anisotropic STSL model [see Eq. (3) of the main text] with $J_2 = 0.6$. Data points are the same as shown in Fig.6 of the main text, however plotted on a semi-log plot. Correlations decay exponentially [dashed black fit from Eq. (S28)] only at short distances, $\sim e^{-r/\xi}$, with correlation length (a) $\xi = 0.5(1)$ in the ground state and (b)–(d) $\xi = 0.55(5)$ for excited states.

with $m_{1,2}$ and $n_{1,2}$ indicating the pair of spins living on bonds $m$ and $n$ in the lattice. For the four-body part, we reduce computational cost and memory consumption by calculating the product of two-body Green's functions, which, however results in lower statistical accuracy as a trade-off.

## I. Variance extrapolation

To estimate the exact energy for a variational wave function we employ the technique of variance extrapolation [22]. By computing the energy variance

$$\sigma^2 = \frac{\langle E^2 \rangle - \langle E \rangle^2}{\langle E \rangle^2}, \tag{S26}$$

one can directly estimate the quality of the optimized variational wave function because an exact eigenstate must satisfy $\sigma^2 = 0$. Usually, larger number of variational parameters allow for higher accuracy of wave functions with smaller variance. By calculating pairs of energy and variance $\{E, \sigma^2\}$ for states of different quality (mVMC with or without RBM or first step Lanczos), one can linearly extrapolate the wave function energy to the exact result in the limit of zero variance

$$E_0 \equiv \lim_{\sigma^2 \to 0} E. \tag{S27}$$

This method showed great success in estimating exact energy eigenvalues on various models hosting superconducting or quantum spin liquid ground states [11, 23]. The energy variance has also proven to offer a useful criterion to assess the accuracy of variational solutions and expose challenges of strongly interacting quantum many-body systems [24].

## II. SPIN-SPIN CORRELATIONS ON THE SUPER-TETRAHEDRON-SQUARE LATTICE

In the present study, the magnetic unit cell is large, involving 16 sites in one super-tetrahedron, which makes a convincing finite-size scaling to the thermodynamic limit computationally demanding. Within the mVMC method, used in

this work, we were able to access finite size clusters of up to $N_s = 1024$ sites, which allows us to perform a scaling analysis of the super-tetrahedron square lattice (STSL) for up to $L = 8$, since $8^2 \times 16 = 1024$.

In Fig. 6 of the main text, we show the spin-spin correlation function $|D_{\vec{\xi}}^z(\mathbf{r})|$ for the $S_{\text{tot}} = 0$ ground state and its excited states in $\mathscr{H}_{J_1 J_2}$ [Eq. (3) of the main text, for $J_2 = 0.6$], on a log-log plot to demonstrate that they are well fit by power-law decay of long-ranged correlations. Here, in comparison, we plot the same dataset on a semilog plot in Fig. S1, by assuming exponential decay in the form

$$|D_{\vec{\xi}}^z(\mathbf{r})| = A \left( e^{-r/\xi} + e^{-(L-r)/\xi} \right). \tag{S28}$$

We see that the short range part could be fit by exponential decay but it clearly deviates at long distance and the exponential decay is numerically excluded on a firm basis. This deviation is rather easily confirmed in the excited states even at $L = 4$, but requires cluster sizes of $L \geq 6$ for the ground state.

We also confirmed the algebraic decay of correlations from the equal-time structure factor $S^z(\mathbf{q})$ in Fig. 7 of the main text. In Fig. S2 we additionally plot the intensity of the $S^z(\mathbf{q})$ along the vertical line from $\mathbf{q} = (4\pi, -4\pi)$ to $(4\pi, 4\pi)$. The intensity function has finite values over the full momentum path and shows, next to the singular cusp at $|\mathbf{q}| = 4\pi$ (as discussed in the main text), also additional singular points at multiples of $2\pi$, but with a weaker singularity.

## III. SIMULATION DETAILS

In this Section we provide technical details of simulations done with mVMC. In Fig. 1 of the main text, we perform simulations for the full spin space with complex variational parameters $f_{ij}^{\sigma\sigma'}$ of Eq. (S4) and 16 site $f_{ij}$ sublattice structure. We did not use any additional projection operators and solved for $|\Psi\rangle = \mathcal{P}_{\text{G}}|\psi_{\text{pair}}\rangle$, by optimizing from an initial maximally flippable state (green diamonds), the all-in/ all-out state (yellow triangles) and a coplanar antiferromagnetic state (blue circles).

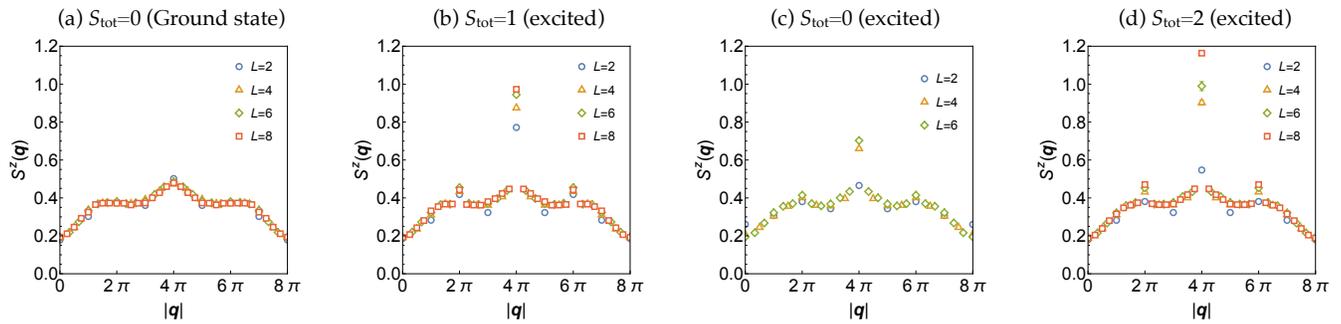

Figure S2. Equal-time structure factor $S^z(\mathbf{q})$ (see Eq. (A4) of the main text) in Fig. 7 of the main text, plotted along the vertical line from $\mathbf{q} = (4\pi, -4\pi)$ to $(4\pi, 4\pi)$.

In Fig. 3, and Fig. 17 of the main text, we performed simulations for a restricted subspace ($S^z = 0$) with real variational parameters $f_{ij}$ of Eq. (S5) and 16 site $f_{ij}$ sublattice structure. To obtain higher accuracy for the ground-state wave function we used the parity-even spin-projection $|\Psi\rangle = \mathcal{L}_+ \mathcal{P}_G |\psi_{\mathrm{pair}}\rangle$ and optimized from an initial maximally flippable state.

In Fig. 4 of the main text, we performed simulations for a restricted subspace ($S^z = 0$) with real variational parameters $f_{ij}$ of Eq. (S5) and 16 site $f_{ij}$ sublattice structure. We used the projection operators, in (a) $|\Psi\rangle = \mathcal{L}_0 \mathcal{P}_G |\psi_{\mathrm{pair}}\rangle$, in (b) $|\Psi\rangle = \mathcal{L}_0 \mathcal{P}_G |\psi_{\mathrm{pair}}\rangle$, and in (c) $|\Psi\rangle = \mathcal{L}_{A_1} \mathcal{L}_0 \mathcal{P}_G |\psi_{\mathrm{pair}}\rangle$, with 12 Gauss-Legendre mesh points for the spin projection $\mathcal{L}_0$ to the $S_{\mathrm{tot}} = 0$ ground state. All simulations were initiated from a maximally flippable trial wave function.

In Fig. 5 of the main text, we performed simulations for a restricted subspace ($S^z = 0$) with real variational parameters $f_{ij}$ of Eq. (S5). We used a $2 \times 2 \times 16$ site $f_{ij}$ sublattice structure to access wave functions with different momenta and solved for $|\Psi\rangle = \mathcal{L}_{\mathbf{K}} \mathcal{L}_{S_{\mathrm{tot}}} \mathcal{P}_G |\psi_{\mathrm{pair}}\rangle$. To distinguish between spin states with $S_{\mathrm{tot}} = 0$ and $S_{\mathrm{tot}} = 2$ quantum numbers, we performed simulations using the total spin-projection operator with 12 Gauss-Legendre mesh points. For the ground state ($S_{\mathrm{tot}} = 0$) and excited states ($S_{\mathrm{tot}} = 1, 2$) we projected onto the $\mathbf{K} = (0,0)$, while for the excited state with $S_{\mathrm{tot}} = 0$, we used $\mathbf{K} = (\pi, 0)$. All simulations were initiated from the maximally flippable trial wave function.

In Figs. 6–9 of the main text and Figs. S1–S2 we performed simulations for a restricted subspace ($S^z = 0$) with real variational parameters $f_{ij}$ of Eq. (S5). For system sizes $L = 2, 4$ and 6, we used the same simulation procedure as explained for Fig. 5. To contain the required accuracy for $L = 8$ cluster sizes, we excluded the momentum projector and solved for $|\Psi\rangle = \mathcal{L}_{S_{\mathrm{tot}}} \mathcal{P}_G |\psi_{\mathrm{pair}}\rangle$.

In Figs. 14 and 15 we performed simulations for a restricted subspace ($S^z = 0$) with complex variational parameters $f_{ij}$ of Eq. (S5). We solved for $|\Psi\rangle = \mathcal{L}_+ \mathcal{P}_G |\psi_{\mathrm{pair}}\rangle$ by using a 16 site $f_{ij}$ sublattice structure for all initial states, except for the spinon mean-field case, where we used a $2 \times 2 \times 2 \times 16$ $f_{ij}$ sublattice.

## IV. CHOICES OF INITIAL WAVE FUNCTIONS AND RELATIONS TO PREVIOUS WORK

Here, we supplement Appendix D of the main text and outline the choices of initial variational wave functions different from the maximally flippable dimer state.

### A. Random initial state

A simple first choice of an initial trial state is to take $f_{ij}$ at random. Previous studies have demonstrated that employing $f_{ij}$, which are random but decay in amplitude with a power law based on the distance between $i$ and $j$, helps to rapidly convergence to a QSL ground state [10, 25]. However, in this study, we prepared $f_{ij}$ by random with equal amplitudes, independent of indices $i$ and $j$. In Fig. 14 of the main text we show the optimization process for three different initial random trial states. Their initial Gutzwiller projected energies are essentially zero. However, after sufficiently long optimization we obtain energies of $E/N_s = -0.481515(1)$, $E/N_s = -0.481292(1)$, $E/N_s = -0.485961(1)$, for the random initial state 1, 2 and 3, respectively. While "random 1" and "random 2" seem to be trapped in a local minimum, "random 3" converges to essentially the lowest-energy state with a value well comparable to the state obtained from the max. flip. initial state (see discussion in Appendix D 2, of the main text). Since simulations were performed using complex $f_{ij}$ parameters, the energy of "random 3" is slightly lower than for the random state in Table SI, which has been obtained using real $f_{ij}$ parameters. We observe that the optimized wave function for "random 3" is a non-magnetic, symmetry-broken state, which shows a singlet arrangement, essential identical to the state obtained after optimization from a max. flip state (see Fig. 15(d) of the main text). Based on this comparison, we conclude that optimization from a maximally flippable state does not bias our simulations. In fact, it aids in reaching the lowest energy state quickly.

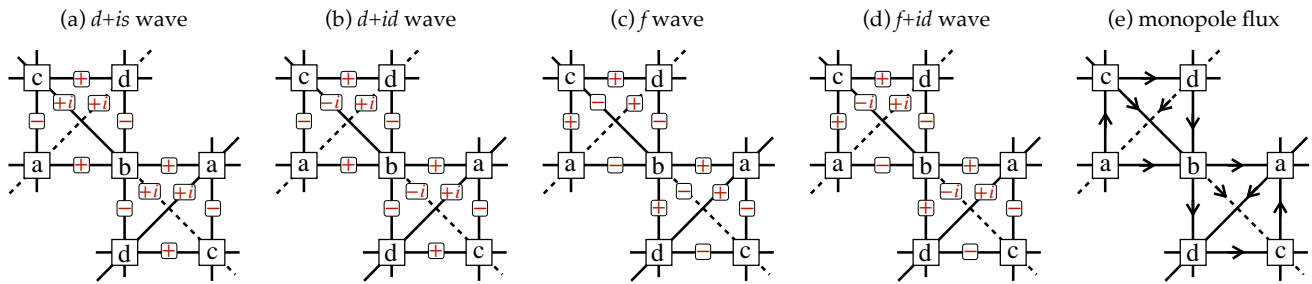

| (a) $d+is$ wave | (b) $d+id$ wave | (c) $f$ wave | (d) $f+id$ wave | (e) monopole flux |

Figure S3. Selected mean-field ansätze, defined on the primitive unit cell of the pyrochlore lattice (sublattice sites a, b, c and d), used as trial wave functions in Fig. 14 of the main text. (a)–(d) HFB mean field ansätze, with pairing symmetry of $d+is$, $d+id$, $f$, and $f+id$, respectively. Symbols on the bonds correspond to the pairing amplitudes $\Delta_{\mu\nu}$. (e) Monopole flux spinon mean-field ansatz as proposed in [26] and [27]. Arrows correspond to the direction of spinon hopping with complex amplitudes $e^{i\theta_{ij}}$, for $\theta_{ij} = \pi/2$ along the direction of the arrows, and $\theta_{ij} = -\pi/2$ against it.

## B. HFB-RVB wavefunction

A very useful approach to prepare good initial wave functions for finding QSL ground states is given by the Bardeen-Cooper-Schrieffer (BCS), or more general Hartree-Fock-Bogoliubov (HFB), mean-field theory. As highlighted by Anderson [28], a resonating valence bond (RVB)-type wave function can be approximated by utilizing a Gutzwiller projected BCS state. Such a concept has found successful application in the exploration of QSL candidates [11, 29–31], and shall help us to formulate a trial-initial wave function for our mVMC simulations.

We solve $\mathscr{H}_{\mathrm{HFB}}$ in Eqs. (9)–(11) of the main text, where we set the fermionic hopping amplitude to unity and the chemical potential to -2. In Fig. S3(a)–(d) we provide a selection of HFB mean-field wave functions, with pairing symmetries of $d+is$, $d+id$, $f$ and $f+id$, where symbols on the bonds correspond to the normalized pairing amplitudes $\Delta_{\mu\nu}$, which can be real (+, −) or imaginary ($+i$, $-i$). Here, we note that the real-space configurations of the pairings are denoted by the symmetry labels $s$, $d$, and $f$. However, since the symmetry of the STSL differs from simple tetragonal ones, the labels are not following the strict tetragonal symmetry notation as one sees the notation in Fig. S3.

As visualized in Fig. 14 of the main text, we obtain initial Gutzwiller projected energies of $E/N_s \approx -0.290(1)$ for $d + is$ and $d + id$, and $E/N_s = -0.3527(1)$ for $f$ and $E/N_s = -0.2055(1)$ for $f + id$ wave functions. After sufficiently long optimization (see inset in Fig. 14 of the main text), energies for $d + is$, $d + id$, and $f + id$ states converge to $E/N_s = -0.4821(2)$, while the $f$-wave function converges to almost to the ground state with $E/N_s = -0.4856(2)$.

## C. Spinon mean-field state

We also took into account another suggestive initial choice from spinon mean-field theory, as shown in Fig. S3(e), the monopole flux state by Burnell, Chakravarty and Sondhi [27], which is equivalent to the $[\pi/2, \pi/2, 0]$ flux state by Kim and Han [26]. By setting the pairing term to zero, and giving the

hopping term a complex phase $t_{ij} = e^{i\theta_{ij}}$ for all nearest-neighbor pairs $(i, j)$, and $\theta_{ij} = \pm\pi/2$ depending on the direction of arrows on the bonds, we obtain a Gutzwiller projected initial energy of $E/N_s = -0.443264(1)$, which is well comparable to the value in Ref. [27]. From Fig. 15(b) of the main text we see that the the arrangement of singlets in this state does not break any symmetry of the lattice, producing a diffuse signal in $\tilde{O}(\mathbf{q})$ and $S^z(\mathbf{q})$ structure factors.

After sufficiently long optimization, the variational state reaches $E/N_s = -0.48572(1)$, which is very close, however still higher in energy, than the optimized state obtained from a maximally flippable dimer state (see inset in Fig. 14 of the main text). We find that the optimized wave function is a non-magnetic, symmetry-broken state, where singlets are arranged on super-tetrahedra, as seen in the lowest-energy wave function obtained from the max. flip. state (see Fig. 16 of the main text). However, singlets on inter-tetrahedra do not select states A, B or C globally (as discussed in Appendix D 2. of the main text), but rather aperiodically throughout the lattice, which, we believe, causes the increase in energy.

## V. SUPPLEMENTAL BENCHMARK DATA FOR ACCURACY OF THE PRESENT VARIATIONAL WAVE FUNCTION IN COMPARISON TO OTHER METHODS

Here we present supplementary data to benchmark the present mVMC calculations including cases that do not necessarily satisfy the variational principle.

### A. Results of variance extrapolations

Figure S4 shows the plot of variational energies extrapolated by the energy variance for the isotropic Heisenberg model in Eq. (1) of the main text on the pyrochlore lattice with periodic boundary conditions. Here, the energy variance is defined by Eq.(S26). To respect the full symmetry of the lattice, we performed calculations on cubic cluster shapes, with 16 sites per unit cell, and linear dimensions $L = 2, 3, 4$ ($N_s = 128, 432, 1024$). Wave functions have been optimized from a

(a) L = 2 (N=128)  (b) L = 3 (N=432)  (c) L = 4 (N=1024)

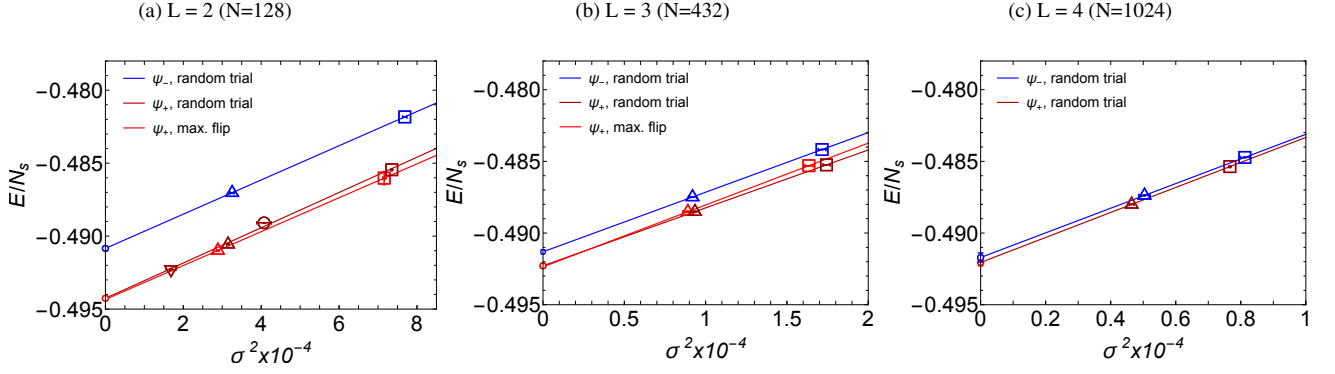

Figure S4. Variational energies and variance extrapolation for spin-parity even ($\psi_+$) and spin-parity odd ($\psi_-$) sectors of Eq. (1) in the main text on the pyrochlore lattice with cubic unit cell and periodic boundary conditions for linear dimension $L = 2, 3, 4$ ($N_s = 128, 432, 1024$). The optimization was initiated from random and maximally flippable dimer (max. flip) trial states (see Appendix D in the main text, where energy values were obtained from statistically independent measurements on mVMC (□), mVMC-RBM (○), mVMC/Lanczos (△), and mVMC-RBM/Lanczos (▽) optimized variational wave functions. The ground state $\psi_+$ is a singlet state, showing a size-dependent finite gap to triplet excitations, $\psi_-$. Explicit values for individual energies of $\psi_+$ and $\psi_-$ states are given in Tables SI–SIII.

random initial state and a maximally flippable dimer (max. flip.) trial wave function [see Appendix D 2 of the main text and Sec. IV A of this SM) for the spin-parity even, $\psi_+$, singlet sector of the ground state, and the spin-parity odd, $\psi_-$, sector of its triplet excitations. Linear extrapolation to zero variance has been done for 20 to 40 statistically independent measurements of the energy-eigenvalues from a singly optimized wave function within mVMC and mVMC-RBM, in combination with the first Lanczos step (see Eqs. (S3) and (S22) and technical details in Sec II). Data points for the same state, obtained using mVMC, mVMC-RBM and Lanczos methods align along an almost identical linear fitting line, even when the number of data points becomes larger than 2, as highlighted in [22]. Although the slope of the line slightly depends on the initial wave function, the extrapolation to zero variance converges to excellently close values between optimization from random and max. flip. states. Since the different initial conditions, including the random one, converge to essentially the same state if the symmetry is the same, our optimization procedure has likely succeeded in reaching the global minimum without falling into local excited states, as discussed in detail in Appendices B and D of the main text.

Energy values of our variational wave functions are summarized in Tables SI–SIII, where the first line in each row corresponds to the normalized energy $E/N_s$, and the second line to its corresponding variance $\sigma^2$. We show in the bottom row of each table the estimated ground state energy $E_0/N_s$ after variance extrapolation by fitting the variational energies obtained from different methods linearly to $\sigma^2 \to 0$. By com-

| $L=2$ ($N_s=128$) | $\psi_+$ (random) | $\psi_+$ (max. flip.) | $\psi_-$ (random) |
|---|---|---|---|
| mVMC | $-0.485451(3)$ | $-0.486024(3)$ | $-0.481829(4)$ |
| | $0.0007353(14)$ | $0.000716(4)$ | $0.000769(5)$ |
| mVMC-RBM | $-0.489089(2)$ | | |
| | $0.000407(19)$ | | |
| mVMC/ Lanczos | $-0.490325(5)$ | $-0.49098(1)$ | $-0.48703(1)$ |
| | $0.0003150(34)$ | $0.000289(4)$ | $0.000325(1)$ |
| mVMC-RBM/ Lanczos | $-0.49229(7)$ | | |
| | $0.000169(12)$ | | |
| $E_0 \equiv \lim_{\sigma \to 0} E$ | $-0.4943(1)$ | $-0.49434(5)$ | $-0.4908(1)$ |

Table SI. Variational energies for the spin-parity even, $\psi_+$, singlet ground state and its excited spin-parity odd, $\psi_-$, triplet state for $L = 2$ ($N_s = 128$), as obtained by mVMC, mVMC-RBM, mVMC with first step Lanczos and mVMC-RBM with first step Lanczos calculations of $\mathcal{H}$ [Eq. (1)]. The normalized energies $E/N_s$ (first line) are shown with their corresponding variance $\sigma^2$ [see Eq. (S26)] (second line), while the variance-extrapolated energies $E_0$ are shown in the last row. Optimizations were initiated from a maximally flippable dimer trial wave function (see Appendix D 2 of the main text) and a random wave function.

| $L=3$ ($N_s=432$) | $\psi_+$ (random) | $\psi_+$ (max. flip.) | $\psi_-$ (random) |
|---|---|---|---|
| mVMC | $-0.485243(7)$ | $-0.485302(8)$ | $-0.484176(9)$ |
| | $0.0001743(22)$ | $0.0001635(24)$ | $0.0001716(24)$ |
| mVMC/ Lanczos | $-0.48851(3)$ | $-0.48851(3)$ | $-0.48748(2)$ |
| | $0.0000935(27)$ | $0.0000891(30)$ | $0.0000919(25)$ |
| $E_0 \equiv \lim_{\sigma \to 0} E$ | $-0.4923(2)$ | $-0.4924(2)$ | $-0.4913(2)$ |

Table SII. Variational energies for the spin-parity even, $\psi_+$, singlet ground state and its excited spin-parity odd, $\psi_-$, triplet state for $L = 3$ ($N_s = 432$). The normalized energies $E/N_s$ (first line) are shown with their corresponding variance $\sigma^2$ [see Eq. (S26)] (second line), while the variance-extrapolated energies $E_0$ are shown in the last row. Optimizations were initiated for $\psi_+$ from a maximally flippable dimer trial wave function (see Appendix D 2 of the main text), and for $\psi_-$ from a random initial wave function.

| $L=4$ ($N_s=1024$) | $\psi_+$ (random) | $\psi_-$ (random) |
|---|---|---|
| mVMC | $-0.48537(1)$ | $-0.48473(2)$ |
| | $0.00007664(27)$ | $0.0000812(15)$ |
| mVMC/ Lanczos | $-0.48800(1)$ | $-0.48738(2)$ |
| | $0.0000464(7)$ | $0.0000504(17)$ |
| $E_0 \equiv \lim_{\sigma \to 0} E$ | $-0.4921(2)$ | $-0.4917(3)$ |

Table SIII. Variational energies for the spin-parity even, $\psi_+$, singlet ground state and its excited spin-parity odd, $\psi_-$, triplet state for $L=4$ ($N_s=1024$). The normalized energies $E/N_s$ (first line) are shown with their corresponding variance $\sigma^2$ [see Eq. (S26)] (second line), while the variance-extrapolated energies $E_0$ are shown in the last row. Optimizations were initiated from random initial wave functions with imposed $C_3$ point-group symmetry projections.

puting the eigenvalue of the $\mathbf{S}^2$ operator

$$\langle \mathbf{S}^2 \rangle = S_{\text{tot}}(S_{\text{tot}}+1) = \sum_{ij} \langle \mathbf{S}_i \cdot \mathbf{S}_j \rangle, \quad (S29)$$

we confirmed that the ground state $\psi_+$ is a singlet ($S_{\text{tot}}=0$), which shows a small but finite size-dependent energy gap $\Delta E = E(\psi_-) - E(\psi_+)$ to $\psi_-$ triplet ($S_{\text{tot}}=1$) excitations.

### B. Comparison in thermodynamic limit

In Fig. 12 of the main text, we plot our variational energies obtained by mVMC and mVMC/Lanczos together with their variance-extrapolated ground state energy $E_0$, as a function of inverse system size $1/N_s$. By fitting the three data points for $N_s = 128, 432$ and $1024$ with a quadratic function we estimate the following energies in their thermodynamic limit:

$$(1/N_s)E^{\text{mVMC}}|_{N_s \to \infty} = -0.4853(1), \quad (S30)$$

$$(1/N_s)E^{\text{mVMC}}_{\text{Lanczos}}|_{N_s \to \infty} = -0.4881(3), \quad (S31)$$

$$(1/N_s)E_0|_{N_s \to \infty} = -0.4921(4). \quad (S32)$$

Our variance-extrapolated ground state energy $E_0/N_s$ is very competitive, nevertheless slightly lower than the estimated ground state energy from numerical linked cluster expansion of order two (NLCE$_2$), which estimates $E/N_s = -0.4917(5)$ [32].

We compare our energies to other state-of-the-art numerical methods from the literature. Astrakhantsev et al. [33] treated $\mathcal{H}$ in Eq. (1) with the same mVMC method as we used, for finite-size clusters respecting the tetrahedral symmetry of the pyrochlore lattice (4 site unit cell) with periodic boundary conditions. Energies are $E/N_s = -0.5162(1)$, $E/N_s = -0.4871(1)$, and $E/N_s = -0.4831(1)$, for cluster sizes of $N_s = 4 \times 2^3 = 32$, $N_s = 4 \times 3^3 = 108$ and $N_s = 4 \times 4^3 = 256$, respectively. By fitting with a quadratic function, we extrapolate those energies in Fig. 12 to the thermodynamic limit and obtain $(1/N_s)E|_{N_s \to \infty} = -0.4834(1)$.

In comparison, our best mVMC wave functions for cluster shapes respecting the cubic symmetry yield ground state energies of $E/N_s = -0.486024(3)$, $E/N_s = -0.485302(8)$, and $E/N_s = -0.48537(1)$, for cluster sizes of $N_s = 16 \times 2^3 = 128$, $N_s = 16 \times 3^3 = 432$ and $N_s = 16 \times 4^3 = 1024$, respectively. Fitting our values with a quadratic function gives an estimated variational ground state energy for mVMC of $(1/N_s)E^{\text{mVMC}}|_{N_s \to \infty} = -0.4853(1)$ for the thermodynamic limit. We believe that the reduction of energy, compared to [33] is coming from the choice of cluster shape. As discussed in Sec. III A, the QSL ground state favors dominant correlations within the quasi-2D STSL with an enlarged unit cell containing 16 lattice sites. In order to capture those correlations, the cluster shape within mVMC simulations and the $f_{ij}$ sublattice (details in Sec. I) must be chosen commensurate to this unit cell.



\end{document}